\documentclass[11pt]{book}
\usepackage{a4}

\usepackage[english]{babel}
\usepackage{latexsym}
\usepackage{amsmath}
\usepackage{dsfont}
\usepackage{bbm}
\usepackage{amssymb}
\usepackage{amsfonts}
\usepackage{eufrak}
\usepackage{euscript}
\usepackage[dvips]{graphicx}
\usepackage{theorem}
\newtheorem{defi}{Definition}[chapter]

\newtheorem{theo}[defi]{Theorem}
\newtheorem{lem}[defi]{Lemma}
\newtheorem{cor}[defi]{Corollary}
\newtheorem{prop}[defi]{Proposition}
\newtheorem{assump}[defi]{Assumption on the massive generator}

\newtheorem{examp}[defi]{Example}
\newtheorem{examps}[defi]{Examples}
\newtheorem{rem}[defi]{Remark}
\def\R{\mathbb{R}}
\def\Mi{\mathbb{M}}

\def\C{\mathbb{C}}

\def\Z{\mathbb{Z}}
\def\1{\mathbbm{1}}

\def\Aa{\mathfrak{A}}
\def\Ba{\mathfrak{B}}
\def\Fa{\mathfrak{F}}
\def\Ma{\mathfrak{M}}
\def\Na{\mathfrak{N}}
\def\Wa{\mathfrak{W}}
\def\Va{\mathfrak{V}}
\def\Da{\mathfrak{D}}
\def\Za{\mathfrak{Z}}

\def\A{\mathcal{A}}

\def\Cg{\mathcal{C}^\infty}

\def\D{\mathcal{D}}
\def\E{\mathcal{E}}
\def\F{\mathcal{F}}

\def\H{\mathcal{H}}
\def\L{\mathcal{L}}
\def\M{\mathcal{M}}
\def\N{\mathcal{N}}
\def\O{\mathcal{O}}
\def\P{\mathcal{P}}
\def\S{\mathcal{S}}
\def\U{\mathcal{U}}
\def\V{\mathcal{V}}
\def\W{\mathcal{W}}
\def\X{\mathcal{X}}
\def\Y{\mathcal{Y}}

\def\Gr{\mathbf{G}}
\def\k{\mathbf{k}}
\def\p{\mathbf{p}}
\def\x{\mathbf{x}}
\def\y{\mathbf{y}}
\def\o{\omega}
\def\Hs{\mathbf{H}}

\def\BH{\mathcal{B}(\H)}
\def\LH{\mathcal{L}(\H)}
\def\LCH{\mathcal{LC}(\H)}

\def\EA{\mathbf{E}_{\mathcal{\Aa}}}

\def\*{{}^{*}}
\def\t{($\sigma,\beta$)-KMS state }
\def\Re{{\text{Re}}}
\def\Im{{\text{Im}}}
\def\const{{\text{const}}}
\def\grad{{\text{grad}}}
\def\supp{{\text{supp}}}
\def\spec{{\textbf{Spec\,}}} 

\def\Tr{\text{Tr}}

\def\singsupp{{\text{singsupp}}}

\def\proof{{\hspace{-0.7cm} Proof: }}
\def\sproof{{\hspace{-0.7cm} Sketch of the Proof: }}
\def\qed{{\begin{flushright}$\Box$\end{flushright}}}

\usepackage{fancyhdr}
\begin{document}

\begin{center}

\thispagestyle{empty}

\vspace*{3cm}
{\huge \textbf{Modular Action \\ \vspace*{1cm} on the Massive Algebra}}\\
\vspace{4cm} 
{\LARGE Dissertation}\\
\vspace{0.5cm}
{\Large zur Erlangung des Doktorgrades\\
des Fachbereichs Physik\\ \vspace{0.1cm}
der Universit\"at Hamburg}\\

\vspace{2,2cm}

{\large vorgelegt von}\\
\vspace{1,5cm}
{\LARGE Timor Saffary}\\ 
\vspace{0,2cm}
%{\large aus }\\
%\vspace{0,2cm}
{\large aus Kabul, Afghanistan}\\
\vspace{2cm}
{\Large  Hamburg 2005}\\
%\vspace{0.2cm}
%{\Large November 2005}

\end{center}

\newpage

\thispagestyle{empty}
\vspace*{15cm}
\begin{flushleft}
\hspace*{-0.8cm}
\begin{tabular}{ll}
\textbf{Gutachter der Dissertation:} & Prof. Dr. K. Fredenhagen\\
                                     & Prof. Dr. G. Mack\\
\\
\textbf{Gutachter der Disputation:} & Prof. Dr. K. Fredenhagen\\
                                    &  Dr. V. Schomerus\\ 
\\
\textbf{Datum der Disputation:} & 12. Dezember 2005\\
\\
\textbf{Vorsitzender des Pr\"ufungsausschusses:} & Prof. Dr. J. Barthels\\
\\
\textbf{Vorsitzender des Promotionsausschusses:} & Prof. Dr. G. Huber\\
\\
\textbf{Dekan des Fachbereichs Physik:} & Prof. Dr. G. Huber\\
\end{tabular}
\end{flushleft}

\newpage

\thispagestyle{empty}
\vspace*{8cm}
\begin{center}
For my parents\\ 
\vspace{0,5cm}
Shafika and Sohrab Ali Saffary
\end{center}

\newpage

%%%%%%%%%%%%%%%%%%%%%%%%%%%%%%%%%%%%%%%%%%%%%%%%%%%%%%%%%%%%%%%%%%%%%%%%%%%%

\thispagestyle{empty}

\begin{center}
{\large\textbf{Modular Action on the Massive Algebra}}
\end{center}

\vspace{2cm}

\begin{center}
\textbf{Abstract:}\\
\end{center}

The subject of this thesis is the modular group of automorphisms $\big(\sigma_m^t\big)_{t\in\R}$, $m>0$, acting on the massive algebra of local observables $\Ma_m(\O)$ having their support in $\O\subset\R^4$. After a compact introduction to micro-local analysis and the theory of one-parameter groups of automorphisms, which are used exensively throughout the investigation, we are concerned with modular theory and its consequences in mathematics, e.g., Connes' cocycle theorem and classification of type $III$ factors and Jones' index theory, as well as in physics, e.g., the determination of local von Neumann algebras to be hyperfinite factors of type $III_1$, the formulation of thermodynamic equilibrium states for infinite-dimensional quantum systems (KMS states) and the discovery of modular action as geometric transformations. However, our main focus are its applications in physics, in particular the modular action as Lorentz boosts on the Rindler wedge, as dilations on the forward light cone and as conformal mappings on the double cone. Subsequently, their most important implications in local quantum physics are discussed. 

The purpose of this thesis is to shed more light on the transition from the known massless modular action to the wanted massive one in the case of double cones. First of all the infinitesimal generator $\delta_m$ of the group $\big(\sigma_m^t\big)_{t\in\R}$ is investigated, especially some assumptions on its structure are verified explicitly for the first time for two concrete examples. Then, two strategies for the calculation of $\sigma_m^t$ itself are discussed. Some formalisms and results from operator theory and the method of second quantisation used in this thesis are made available in the appendix.

\newpage

%%%%%%%%%%%%%%%%%%%%%%%%%%%%%%%%%%%%%%%%%%%%%%%%%%%%%%%%%%%%%%%%%%%%%%%%%%%%

\thispagestyle{empty}

\begin{center}
{\large\textbf{Modulare Wirkung auf der Massiven Algebra}}
\end{center}

\vspace{2cm}

\begin{center}
\textbf{Zusammenfassung:} 
\end{center}

Gegenstand dieser Dissertation ist die modulare Automorphismengruppe $\big(\sigma_m^t\big)_{t\in\R}$, $m>0$, auf der massiven Algebra der lokale Observablen $\Ma_m(\O)$ mit Tr\"ager in $\O\subset\R^4$. Nach einer kompakten Einf\"uhrung in die mikrolokale Analysis und die Theorie einparametriger Automorphismengruppen, von denen in dieser Arbeit ausgiebig Gebrauch gemacht wird, behandeln wir die modulare Theorie und ihre Konsequenzen sowohl in der Mathematik, z.B. das Kozykel-Theorem und die Klassifizierung von Faktoren vom Typ $III$ von Connes und die Indextheorie von Jones, als auch in der Physik, als da sind die Bestimmung der lokalen von Neumann Algebren als hyperfinite Faktoren vom Typ $III_1$, die Formulierung von thermodynamischen Zust\"anden in unendlichdimensionalen Quantensystemen (KMS-Zust\"ande) und die Entdeckung der modularen Wirkung als geometrische Transformation. Unser Hauptaugenmerk sind jedoch die physikalischen Anwendungen und hier ganz besonders die modulare Wirkung als Lorentz-Boosts auf dem Rindler-Keil, als Dilatationen auf dem Vorw\"artslichtkegel und als konforme Abbildungen auf dem Doppelkegel. Ihre wichtigsten Folgerungen in der lokalen Quantenphysik werden anschlie\ss end besprochen. 

Ziel dieser Arbeit ist es, im Falle des Doppelkegels mehr Licht auf den \"Ubergang von der bekannten masselosen modularen Wirkung auf die noch zu berechnende massive zu werfen. Zun\"achst wird der infinitesimale Generator $\delta_m$ der Gruppe $\big(\sigma_m^t\big)_{t\in\R}$ analysiert, insbesondere werden einige Vermutungen \"uber seine Struktur zum ersten Mal f\"ur zwei konkrete Beispiele explizit best\"atigt. Danach diskutieren wir zwei Strategien f\"ur die Berechnung von $\sigma_m^t$ selbst. Die in dieser Arbeit verwendeten Formalismen und Resultate aus der Operatortheorie und der zweiten Quantisierungsmethode werden im Anhang zur Verf\"ugung gestellt.

\newpage

\setcounter{page}{1}
\setcounter{tocdepth}{2}
\tableofcontents

\setcounter{secnumdepth}{-1}
\setcounter{secnumdepth}{2}

\chapter{Introduction}

\begin{flushright}
\emph{Problems worthy of attack\\
prove their worth by hitting back.}\\
\vspace{0,5cm}
P. H. Grooks
\end{flushright}
\vspace{0,5cm}

Although the Lagrangian formulation of relativistic quantum mechanics, especially the perturbation theory, has yielded some spectacular agreement with experiment, its theoretical structure is not consistent since the singularities and divergencies appearing there are handled insufficiently leaving the approach in an unsatisfactory state. For a deeper understanding and a better mastering of quantum field theory one has reclaimed the very fundamental concepts and returned to mathematically more rigorous approaches, as there are, above all, the Lehmann-Symanzik-Zimmermann theory (LSZ) \cite{Lehmann:1954rq}, the Wightman theory \cite{Wightman:1957} and the Haag-Kastler-Araki theory, the so-called algebraic quantum field theory or local quantum physics \cite{Haag:1963dh}, emphasising special aspects. While the LSZ formalism is suited for the calculation of the $S$-matrix from the time ordered correlation functions, the Wightman ansatz reflects the relation between locality and the spectrum condition. In the Wightman theory one faces, in contrast to the algebraic formulation, domain problems as a consequence of the appearence of unbounded operators and one has to restrict the causal structure by hand. Unfortunately, bounded operators, which are used in local quantum physics, do not get along with strict locality of states, a disadvantage of local quantum physics. One expects these three approaches to be more or less physically equivalent, but the transition from one theory into another is not fully understood yet.

In algebraic quantum field theory, the setting of this thesis, the main objects are $C^*$-algebras or von Neumann algebras, to be more precise. Its core is the assignment to each open subset $\O\subset\M$ of a spacetime $\M$ a $C^*$-algebra $\Aa(\O)$ generated by local observables,
\begin{gather}\label{net}
\O\mapsto\Aa(\O).
\end{gather}
Under some physically reasonable conditions this mapping is assumed to contain in principle all physical information. The quasi-local algebra is defined as the $C^*$-inductive limit of the net $\{\Aa(\O)\}_{\O\subset\M}$, and the global algebra of observables is introduced as its bicommutant $\Ma:=\Aa''$. The states are represented by normalised and positive linear functionals,
\begin{gather*}
\o:\Aa\longrightarrow\C.
\end{gather*} 
The ``usual'' approach and the algebraic formulation of quantum field theory is then connected through the GNS representation.

The choice of an algebra is motivated by the facts that, first, the $S$-matrix depends only on large classes of fields, the so-called Borchers' classes, and not on a special field system from the class, and second, quantum field theories, i.e., quantum systems with infinitely many degrees of freedom, have a host of inequivalent irreducible representations describing classes of states for which the superposition principle is not valid. Algebraic quantum field theory entails the conceptual separation of the physical system (algebra) and the possible states of the system (representations).

Last but not least, the algebraic language admits the entry of modular theory with its powerful tools into quantum field theory. Modular theory or Tomita-Takesaki theory is the generalisation of the modular function, which constitutes the difference between the left and right Haar measure, to non-commutative algebras. Although the prerequisite for this theory is only the specification of an underlying von Neumann algebra $\Ma$ and a cyclic and separating vector $\Omega\in\H$ or, equivalently, a faithful and normal state $\o$, it provides a deep insight into the most complex structure of von Neumann algebras. The main properties of the modular objects are addressed in Tomita's theorem \cite{Takesaki:1970}, i.e., the anti-unitary modular conjugation $J$ relates $\Ma$ to its commutant $\Ma'$, 
\begin{gather*}
J\Ma J=\Ma',
\end{gather*}
and the positive, selfadjoint modular operator $\Delta$ ensures the existence of an automorphism group, 
\begin{align*}
\sigma_\o^t:\Ma&\longrightarrow\Ma\\
A&\mapsto\sigma_\o^t(A):=\pi_\o^{-1}\big(\Delta^{it}\pi_\o(A)\Delta^{-it}\big),\end{align*}
for all $t\in\R$, where $\pi_\o$ is the cyclic GNS representation of $\Ma$ with respect to the faithful state $\o$. These statements, in particular that a state already determines the dynamics of a system, have far-reaching consequences in mathematics as well as in physics. 

To start with, Connes shows that modular groups are equivalent up to inner automorphisms, i.e., two arbitrary groups $\sigma_{\o_1}^t$ and $\sigma_{\o_2}^t$ with respect to the states $\o_1$ and $\o_2$, respectively, are linked via a one-parameter family of unitaries $\Gamma_t$, the so-called cocycle,
\begin{gather*}
\sigma_{\o_2}^t(A)=\Gamma_t\sigma_{\o_1}^t(A)\Gamma_t^*,\quad\forall A\in\Ma,t\in\R.
\end{gather*}
This suggests the introduction of the modular spectrum $S(\Ma):=\bigcap_\o\mathbf{Spec}\Delta_\o$ by means of which Connes gives a complete classification of factors \cite{Connes:1973}, i.e., von Neumann algebras with $\Ma\cap\Ma'=\C\1$: 
\begin{itemize}
\item $\mathfrak{M}$ is of type $I$ or type $II$, if $S(\mathfrak{M})=\{1\}$;
\item $\mathfrak{M}$ is of type $III_0$, if $S(\mathfrak{M})=\{0,1\}$;
\item $\mathfrak{M}$ is of type $III_\lambda$, if $S(\mathfrak{M})=\{0\}\cup\{\lambda^{n}|\;0<\lambda<1,n\in \mathbb{Z}\}$;
\item $\mathfrak{M}$ is of type $III_1$, if $S(\mathfrak{M})=\R_+$.
\end{itemize}
The next development of paramount transboundary importance is Jones' classification of type $II_1$ subfactors \cite{Jones:1983kv}. He shows, contrary to everyone's expectation, that for the (global) index $[\Ma:\Na]$ not all positive real numbers are realised, but
\begin{gather*}
[\Ma:\Na]\in\Big\{4\cos^2\frac{\pi}{n}|\;n\in\N,n\geq3\Big\}\cup\big[4,\infty\big].
\end{gather*}
This result is extended by Kosaki to arbitrary factors \cite{Kosaki:1986}. Jones' index theory on his part connects widely separated areas, such as parts of statistical mechanics with exactly solvable models, and leads to some groundbreaking developments, e.g., a new polynomial invariant for knots and links in $\R^3$.

The interplay of modular theory and quantum field theory is most naturally apparent in the algebraic formulation since here the requirements of modular theory are already fulfilled: an underlying von Neumann algebra $\Ma(\O)$ is given and, due to the Reeh-Schlieder theorem, a cyclic and separating vacuum vector. 
The first physical application of modular theory is proved by Takesaki who recognises that the equilibrium dynamics is determined by the modular groups, since their infinitesimal generator is the thermal Hamiltonian and, due to the property
\begin{align*}
\big(\Delta^{1/2}\pi_\o(A)\Omega_\o,\Delta^{1/2}\pi_\o(B)\Omega_\o\big)&=\big(J\pi_\o(A^*)\Omega_\o,J\pi_\o(B^*)\Omega_\o\big)\\
&=\big(\pi_\o(B^*)\Omega_\o,\pi_\o(A^*)\Omega_\o\big),
\end{align*}
they satisfy the KMS condition, the generalisation of Gibbs' notion of equilibrium to systems with infinitely many degrees of freedom, 
\begin{gather*}
\o\big(A\sigma^{i\beta}(B)\big)=\o(BA), 
\end{gather*}
where $\beta$ is the inverse of the temperature. 

The classification theory is not less important in physics, in fact, the quest for decomposition of quantum systems has been one of von Neumann's most important reasons for the investigation of operator algebras. In local quantum physics, one is interested in the structure of the von Neumann algebra of local observables $\Ma(\O)$. The analysis, which has been undertaken by a colloboration of many persons, discovered $\Ma(\O)$ as a hyperfinite factor of type $III_1$. The substructure of $\Ma(\O)$, which is of utmost significance for decoding the physical information contained in the mapping \eqref{net}, is determined only for conformal local nets with central charge $c<1$ yet.

The third main application for modular theory in local quantum physics is the modular action as a geometric transformation on the local algebra for special spacetimes. For the local algebra generated by Wightman fields with mass $m\geq0$ and which are localised in the right wedge, Bisognano and Wichmann identify  the modular action with the Lorentz boosts $\Lambda$ and the modular conjugation with the TCP operator $\Theta$ \cite{Bisognano:1975ih},\cite{Bisognano:1976za}:
\begin{gather*}
J_{\W_R}=\Theta U\big(R_1(\pi)\big),\\
\sigma_{\W_R}^t\big(\varphi[f]\big)=\varphi[f_s],\quad f_s(x):=f\big(\Lambda_{s}(x)\big),\quad x\in\W_R,\; s:=2\pi t,
\end{gather*}
where $R_1$ denotes the spatial rotation around the $x^1$-axis. For massless theories this result has been transferred via conformal transformations to other spacetime regions. While the r\^ole of $J$ remains unchanged, for forward light cones $\V_+$ the modular action conicides with dilations as shown by Buchholz \cite{Buchholz:1977ze}, and for double cones $\D$ they are conformal transformations as proved by Hislop and Longo \cite{Hislop-Longo:1981uh}:
\begin{gather*}
x_\pm(s)=\frac{1+x_\pm-e^{-s}(1-x_\pm)}{1+x_\pm+e^{-s}(1-x_\pm)}\\
\text{with}\quad x_+:=x^0+|\mathbf{x}|\quad\text{and}\quad x_-:=x^0-|\mathbf{x}|,\quad x\in\D,s\in\R.
\end{gather*}
The geometric interpretation of the modular group is of paramount importance in further applications. The result of Bisognano and Wichmann is not only closely related to the Unruh effect and the black hole evaporation, actually, in analogy of the Rindler wedge with a black hole, it implies the Hawking radiation, but has also made possible the derivation of some most fundamental concepts of quantum field theory, as there are the proofs of the PCT theorem by Borchers \cite{Borchers:1991xk} and of the spin-statistics theorem by Guido and Longo \cite{Guido:1995fy}, where modular theory intervenes twice through Jones' index theory, the construction of the Poincar\'e group by Brunetti, Guido and Longo \cite{Brunetti:1992zf}, and the introduction of modular nuclearity condition by Buchholz, D'Antoni and Longo \cite{Buchholz:1989bj}, nuclearity as a tool to single out models with decent phase space properties. Moreover, Schroer and Wiesbrock's investigation gives a hint that modular theory plays a decisive r\^ole in the construction of field theories with interaction \cite{Schroer:1998wp}. 

As aforementioned, the modular group for massive theories, $\sigma_m^t$, is known only for wedge regions. In fact, the transfer of Bisognano and Wichmann's result via conformal mappings to forward light cones and double cones does not work in massive theories. If one assumes the modular group to act locally, then, as shown by Trebels \cite{Trebels:1997}, the action can be determined as the ones of Bisognano-Wichmann, Buchholz and Hislop-Longo up to a scaling factor. But in general the modular action has to be non local and does not act as a  geometric transformation anymore. This is mainly due to the fact that the massive scalar field is not invariant under conformal transformations. 

Since the discovery of Bisognano and Wichmann, there have been many attempts to derive the massive modular group in the case of double cones, the most important spacetime regions, but no progress has been made so far. What should be mentioned are some assumptions on its nature, to be more precise, on the structure of its infinitesimal generator $\delta_m$, where $m$ denotes the mass. It is well known that the generators $\delta_0$ of the massless groups are all ordinary differential operators of order one. Because of the non local action and a result of Figliolini and Guido, who prove that $\delta_m$ is depending on $m$ in the strong generalised sense \cite{Figliolini:1989vf}, one assumes that the generator $\delta_m$ has to be of pseudo-differential nature, the generalisation of differential operators. One expects exactly the following structure,
\begin{gather}\label{assumption-gen}
\delta_m=\delta_0+\delta_r,
\end{gather}
where $\delta_0$ constitutes the leading (principal) part of $\delta_m$, i.e., it comprises the term of highest order, namely one. The additional part $\delta_r$ is expected to be a pseudo-differential operator of order less than one and responsible for the non local character of the modular group $\sigma_m^t$. 

The ultimate desire still remains the calculation of the modular group acting on massive fields localised in the double cone, which would allow for many applications of not yet foreseeable significance, since the modular group governs the dynamics of quantum systems.

In this thesis, after an extensive and very detailed overview of the state-of-the-art of modular theory and its applications in mathematics as well as in local quantum physics, we will confirm the assumption \eqref{assumption-gen} on the structure of the infinitesimal generator, in fact, we will verify it explicitly for the first time for two examples of modular groups with non local action, given by Borchers and Yngvason \cite{Yngvason:1994nk},\cite{Borchers:1998ye}. Concerning the derivation of  $\sigma_m^t$ itself, we first discuss the approach of Figliolini and Guido  \cite{Figliolini:1989vf}, second determine some properties of the general massive infinitesimal generator, and third elaborate two approaches of our own. As an intermediate step, we present a modular group with respect to the massless vacuum acting on the massive algebra.\\

The thesis is structured as follows.

In the second chapter we give a compact summary of microlocal analysis, since this discipline has become more and more important in local quantum physics. Especially our investigation will make use of the terminology as well as of some its powerful tools, as there are, for example, the generalisation of differential operators to pseudo-differential or even Fourier integral operators and their mapping properties.

Chapter 3 contains the most important facts on one-parameter groups of automorphisms, the central objects of this thesis, and an introduction to conformal transformations.

In chapter 4 we are concerned with modular theory and its state-of the-art status in mathematics and local quantum physics. While the mathematical consequences, e.g., the classification theories of Connes and Jones, are given in a nutshell, the applications in physics are discussed in more detail. We start with the determination of the type of local algebras $\Ma(\O)$ and the formulation of equilibrium states (KMS states).  Subsequently, the modular action will be analysed for three spacetime regions, the wedges, the forward light cones and the double cones. We close this chapter with the most significant concepts of quantum field theory where modular theory plays a decisive r\^ole, as there are, the Hawking radiation, the PCT theorem, the spin-statistics theorem, the construction of the Poincar\'e group and the modular nuclearity condition. 

The main results of our own are comprised in chapter 5. After a short motivation of our assumption, the very nature of the infinitesimal generator $\delta_m$ and some of its properties, we will analyse the ansatz of Figliolini and Guido, whose investigation of the massive group is based on the second quantisation formalism. Then, we will present two approaches, first via unitary equivalence of free local algebras, and second through the cocycle theorem.

We conclude the thesis with a summary and an appendix containing useful information on operator algebras and free quantum fields.

%%%%%%%%%%%%%%%%%%%%%%%%%%%%%%%%%%%%%%%%%%%%%%%%%%%%%%%%%%%%%%%%%%%%%%%%%%%%

\chapter{Pseudo-Differential Operators}

\begin{flushright}
\emph{au taraf-e auriz mora.\\
Water flows towards the waterfall.}\\
\vspace{0,5cm}
A Hazaragi proverb
\end{flushright}
\vspace{0,5cm}

One of the most fundamental characteristics of physical equations is their being concerned with local measurements, i.e., they represent the interplay of physical quantities at the same time and in the same place or at least in an infinitesimal neighbourhood of a spacetime point. The physical state in a subset $\Omega\subset\R^{3}$ of the configuration space at a particular moment determines via the physical laws governing the evolution in time the future situation in the causal shadow $\tilde{\Omega}$ of $\Omega$. However, effects from the outside of $\Omega$ cannot influence the events in $\tilde{\Omega}$ instantaneously. By shifting this principle of locality, i.e., the independence of events in spacelike separated regions, onto the phase space, one can introduce the principle of micro-locality, i.e., finite velocity of causal effects via equations governing their propagation. The micro-local analysis dealing with this subject can be considered as the local analysis in the cotangent bundle, and it is for two reasons of great help in mathematical physics. First it gives a more precise treatment of the notion of singular points, and second it provides an easy treatment of their propagation. In this chapter we follow mainly \cite{Kumamo-go:1981} and \cite{Treves:1980}. Beside these books there are the classics \cite{Hoermander:1990}, \cite{Hoermander:1971} and \cite{DuistFour:1972}. 

We want to give some more details of these two advantages. In the theory of partial differential equations (PDE) it is possible to extract already from the form of the equation,  
\begin{gather*} 
Pu=f \quad\text{with}\\
P=\sum_{|\alpha|\leq m}a_{\alpha}(x)D^{\alpha},\quad a_{\alpha}\in \Cg(\Omega),\quad D^{\alpha}:=(-i)^{|\alpha|}\partial^{\alpha},\\
u,f\in\mathcal{D}'(\Omega),\quad\Omega\subset\R^{n},
\end{gather*}
information about the regularity of the weak (distributional) solution $u$, i.e., one is able to analyse quite extensively the singularities of $u$ with the knowledge of the differential operator $P$ and the inhomogeneity $f$ only. For this purpose one has to generalise the notion of the singular support $\singsupp(u)$ of $u$, that is the set of all points in $\Omega$ having no neighbourhood in which $u$ could be described as a $\Cg$-function, to the so-called wave front set $WF(u)$ of $u$, see Definition \ref{wavefrontset}. An equivalent characterisation of the wave front set will be given in Proposition. The wave front set is a subset of the cotangent space $\Omega\times\R^{n}$ of $\Omega$ and it provides beside the singularity itself also the direction of the singular `behaviour' of $u$.

The concept of the wave front set can be introduced in two different ways, via the local Fourier transformation and the pseudo-differential operators (PsDO). It is also possible to establish the notion of singularities for PsDOs, and one gets the so-called micro-support $\mu\supp(P)$ of the differential operator $P$. The wave front set and the micro-support are closely connected via the relation

\begin{displaymath}
WF(Pu)\subseteq WF(u)\cap\mu\text{supp}(P),
\end{displaymath}
the more precise form of the `micro-locality property' 
\begin{displaymath}
WF(Pu)\subseteq WF(u). 
\end{displaymath}
In the theory of PsDOs the elliptic ones, see Definition \ref{elliptic}, play a special role since they leave the wave front set of distributions invariant:
\begin{displaymath}
(x_{0},\xi_{0})\in WF(u)\Longleftrightarrow (x_{0},\xi_{0})\in WF(Pu),\quad (x_{0},\xi_{0})\in\Omega\times\R^{n}.
\end{displaymath}
Because $P$ is elliptic in $(x_{0},\xi_{0})$ if and only if the principal symbol $p_{0}$ of $P$, i.e. the leading symbol of $P$, does not vanish in $(x_{0},\xi_{0})$ and the so-called bicharacteristic curve 
\begin{displaymath}
\gamma:\R\supseteq I\longrightarrow \Omega\times\R^{n}\quad\mbox{with}\quad p_{0}\big(\gamma(s)\big)=0\quad\forall s\in I
\end{displaymath} 
runs through each zero point of $p_{0}$, one can prove that under particular conditions on $p_{0}$ the singularities of $u$ must propagate along $\gamma$:
\begin{displaymath}     
\Big(\exists s\in I:\gamma(s)\in WF(u)\Big)\quad\Longrightarrow\quad\Big(\forall s'\in I:\gamma(s')\in WF(u)\Big).
\end{displaymath}

\vspace{1cm} 
In the sequel we give the exact definitions of the concepts mentioned in the preceding paragraph, focussing on PSDOs.

First of all, let us consider a differential operator with variable coordinates
\begin{gather*}
p(x,D):=\sum_{|\alpha|\leq m}a_{\alpha}(x)D_{x}^{\alpha},
\end{gather*}
then one has
\begin{equation*}
\begin{split}
p(x,D)u(x)&=\frac{1}{(2\pi)^{\frac{n}{2}}}p(x,D)\int_{\R^{n}}\tilde{u}(\xi)e^{ix\xi}d^{n}\xi\\
&=\frac{1}{(2\pi)^{\frac{n}{2}}}\int_{\R^{n}}p(x,\xi)\tilde{u}(\xi)e^{ix\xi}d^{n}\xi,
\end{split}
\end{equation*}
where $\tilde{u}$ denotes the Fourier transform of $u$ and $p(x,\xi):=\sum_{|\alpha|\leq m}a_{\alpha}(x)\xi^{\alpha}$. This concept of differential operators with variable coordinates can be generalised by replacing the polynomial $p(x,\xi)$ with the so-called symbols. 

\begin{defi}\label{symbol-definition}
Let $m\in\R$ and $0<\delta\leq\rho\leq 1,\,\delta<1$. Then $p\in\mathcal{C}^{\infty}(\R_{x}^{n}\times\R_{\xi}^{n})$ is said to be a symbol of order $m$ and of type $(\rho,\delta)$ if $p$ satisfies the following condition:
\begin{equation}
\begin{split}
\forall\alpha,\beta\in\N_{0}^{n}:\,\exists C_{\alpha,\beta}\geq 0:\;\Big(\big|\partial_{\xi}^{\alpha}\partial_{x}^{\beta}p(x,\xi)\big|\leq C_{\alpha,\beta}&\big(1+|\xi|\big)^{m+\delta|\beta|-\rho|\alpha|}\\
&\forall(x,\xi)\in\R_x^{n}\times\R_\xi^{n}\Big).
\end{split}
\end{equation}
The vector space generated by the symbols  is denoted by $S_{\rho,\delta}^{m}=S_{\rho,\delta}^{m}(\R^{n}\times\R^{n})$. Furthermore we introduce the spaces
\begin{gather*}
S^{-\infty}:=\bigcap\big\{S_{\rho,\delta}^{m}|\,m\in\R\big\}\quad\text{and}\quad\;S_{\rho,\delta}^{\infty}:=\bigcup\big\{S_{\rho,\delta}^{m}|\,m\in\R\big\}.
\end{gather*}

\end{defi}

\begin{rem}\label{symbol-Hoermander}\emph{
It should be mentioned that some authors use a different notion of symbols $p\in S_{\rho,\delta}^{m}(K\times\R_{\xi}^{n})$, often referred as H\"ormander's version, by demanding the variable $x$ to lie in a compact subset $K\subset\Omega$, where $\Omega$ is an open subset of $\R_x^n$. But be aware that H\"ormander works with both versions, with the latter definition in \cite{Hoermander:1971} and with Definition \ref{symbol-definition} in \cite{Hoermander:1990}. The analysis in this section will be based on Definition \ref{symbol-definition}.}
\end{rem}

The space of all symbols equipped with the semi-norms 
\begin{gather*}
|p|_{S_{\rho,\delta}^{m}}^{(l)}:=\sup\Big\{\big|\partial_{\xi}^{\alpha}\partial_{x}^{\beta}p(x,\xi)\big|\big(1+|\xi|^{2}\big)^{-(m+\delta|\beta|-\rho|\alpha|)}:\\
\hspace{6.5cm}\alpha,\beta\in\N_{0}^{n},\,|\alpha+\beta|\leq l,\,(x,\xi)\in\R^{2n}\Big\}
\end{gather*}
becomes a Fr\' echet space. In this thesis we have to deal only with symbols of type $(1,0)$, i.e., we will consider elements of $S^{m}=S_{1,0}^{m}$. 

\begin{defi}
Let $m'\in\R$, $0<\delta\leq\rho\leq 1,\,\delta<1$, and adopt the remaining notation as given above. Then we denote as $S_{\rho,\delta}^{m,m'}=S_{\rho,\delta}^{m,m'}(\R^{n})$ the vector space of all $p\in\mathcal{C}^{\infty}\big(\R_{x}^{n}\times\R_{\xi}^{n}\times\R_{x'}^{n}\times\R_{\xi'}^{n}\big)$, such that
\begin{gather*}
\forall\alpha,\alpha',\beta,\beta'\in\N_{0}^{n}:\,\exists C_{\alpha,\alpha',\beta,\beta'}\geq 0:\,\forall(x,x',\xi,\xi')\in\R^{2n}\times\R^{2n}:
\end{gather*}
\begin{equation*}
\begin{split}
\Big(\big|\partial_{\xi}^{\alpha}\partial_{\xi'}^{\alpha'}\partial_{x}^{\beta}\partial_{x'}^{\beta'}p(x,x',\xi,\xi')\big|\leq &C_{\alpha,\alpha',\beta,\beta'}\big(1+|\xi|^{2}\big)^{m+\delta|\beta|-\rho|\alpha|}\\
&\big(1+|\xi|^{2}+|\xi'|^{2}\big)^{\delta|\beta'|}(1+|\xi'|^{2})^{m'-\rho|\alpha'|}\Big).
\end{split}
\end{equation*}
The elements of the set $S_{\rho,\delta}^{m,m'}$ are called double symbols.
\end{defi}
The space $S_{\rho,\delta}^{m,m'}$ is also a Fr\' echet space when equipped with the semi-norms
\begin{equation*}
\begin{split}
|p|_{S_{\rho,\delta}^{m,m'}}^{(l)}:=\sup\Big\{&\big|\partial_{\xi}^{\alpha}\partial_{\xi'}^{\alpha'}\partial_{x}^{\beta}\partial_{x'}^{\beta'}p(x,x',\xi,\xi')\big|\big(1+|\xi|^{2}\big)^{-(m+\delta|\beta|-\rho|\alpha|)}\\
&\big(1+|\xi|^{2}+|\xi'|^{2}\big)^{-\delta|\beta'|}\big(1+|\xi'|^{2}\big)^{-(m'-\rho|\alpha'|)}:\;\alpha,\alpha',\beta,\beta'\in\N_{0}^{n},\\
&|\alpha+\alpha'+\beta+\beta'|\leq l;\;(x,x',\xi,\xi')\in\R^{4n}\Big\}.
\end{split}
\end{equation*}
As above we set $S^{m,m'}=S_{1,0}^{m,m'}$.\\
Now we are ready to give the definition of PsDOs.

\begin{defi}\label{PsDO-defi}
To each symbol $p\in S_{\rho, \delta}^{m}$ we associate a pseudo-differential operator (PsDO) 
\begin{equation*}
\begin{split} 
P\equiv p(X,D_{x}):\mathcal{S}(\R^{n})&\longrightarrow\mathcal{S}(\R^{n})\\
u&\mapsto p(X,D_{x})u
\end{split}
\end{equation*}
with
\begin{align}
p(X,D_{x})u(x)&:=\int p(x,\xi)\tilde{u}(\xi)e^{ix\xi}d\xi\notag\\
&=\frac{1}{(2\pi)^{n/2}}\iint p(x,\xi)u(x')e^{i(x-x')\xi}dx'd\xi
\end{align}
for all $u\in\mathcal{S}(\R^{n})$ and $x\in\R^{n}$.
The space of PsDOs associated with symbols from $S_{\rho,\delta}^m$ is denoted by $\EuScript{S}_{\rho,\delta}^{m}(\R^{n})$. We set
\begin{gather*}
\EuScript{S}^{m}:=\EuScript{S}_{1,0}^{m},\;\EuScript{S}^{\infty}:=\bigcup\big\{\EuScript{S}^{m}:\,m\in\R\big\},\\
\text{and}\quad\EuScript{S}^{-\infty}:=\bigcap\big\{\EuScript{S}^{m}:\,m\in\R\big\}.
\end{gather*}
The elements of $\EuScript{S}^{-\infty}$ are called regularising.\\
For the PsDO $P\in\EuScript{S}_{\rho,\delta}^{m}$ we define the formal adjoint  $P^{*}$ and the transpose ${}^{t}P$ of $P$ by
\begin{align*}
\overline{\langle u,P^{*}\overline{v}\rangle}&:=\langle Pu,v\rangle\quad\text{and}\\
\langle u,{}^{t}Pv\rangle&:=\langle Pu,v\rangle\quad\forall u,v\in\mathcal{S}(\R^{n}),
\end{align*}
respectively. $P^{*}$ and ${}^{t}P$ are still elements of $\EuScript{S}_{\rho,\delta}^{m}$.

\end{defi}
PsDOs will be of great interest for us insofar as the infinitesimal generator $\delta_m$ of the modular group acting on the massive algebras localised in a double cone is assumed to be of this kind. One expects that the transfer from the massless algebra, where the infinitesimal generator $\delta_0$ is a differential operator derived by Hislop and Longo \cite{Hislop-Longo:1981uh}, to the massive algebra will conserve $\delta_0$ as the principal part of $\delta_m$, but will also yield an additional part of pseudo-differential nature. 

The mapping
\begin{align*}
S_{\rho,\delta}^{m}(\R^{n})&\longrightarrow\EuScript{S}_{\rho,\delta}^{m}(\R^{n})\\
 p&\mapsto p(X,D_{x})
\end{align*}
is  bijective and its inverse is denoted by $\sigma$. Each PsDO $P$ is a linear and continuous mapping $P:\mathcal{S}(\R^n)\rightarrow\mathcal{S}(\R^n)$, i.e., for each $l\in\N_{0}$ one can find constants $C_{l}\geq 0$ and $l'\in\N_{0}$ such that
\begin{gather*}
|Pu|_{\mathcal{S}}^{(l)}\leq C_{l}|p|_{S_{\delta,\rho}^{m}}^{(l')}|u|_{\mathcal{S}}^{(l')}\quad\forall u\in\mathcal{S}(\R^n),\,P=p(X,D_{x})\in S_{\delta,\rho}^{m},
\end{gather*}
with
\begin{gather*}
|u|_{\mathcal{S}}^{(l')}:=\sup\big\{|x^\alpha D^\beta\ u(x)|\,|\;\alpha,\beta\in\N_0^n,\;|\alpha+\beta|\leq l\big\}.
\end{gather*}
Since we are concerned with the Klein-Gordon equation in this work, the next two examples will appear in our analysis of the modular group, especially in the context of the approach of Figliolini and Guido in Section \ref{Guido-section}.

\begin{examp}\label{Laplace}\emph{
The d'Alembert operator $\Box:=\partial_{x^0}^2-\sum_{i=1}^{n}\partial_{x^i}^2$ is a pseudo-differential operator of order two with symbol $p_\Box(x,\xi)=-\xi_0^2+\sum_{i=1}^{n}\xi_i^2$. The Laplace operator $\Delta:=\sum_{i=1}^{n}\partial_{x^i}^2$ is even an elliptic pseudo-differential operator of order two with symbol $p_\Delta(x,\xi)=-\sum_{i=1}^{n}\xi_i^2$, see Definition \ref{elliptic}.}
\end{examp}

\begin{defi}\label{elliptic}
The PsDO $P$ is said to be elliptic of order $m$ if for every compact subset $K\subset\Omega$, where $\Omega$ is an open region in $\R^n$, there exist constants $C_K$ and $R$ such that
\begin{gather*}
|p(x,\xi)|\geq C_K\big(1+|\xi|\big)^m
\end{gather*}
for all $x\in K$ and $|\xi|\geq R$.
\end{defi} 
Compared with other types of PsDOs, the functional calculus of elliptic PsDOs is better understood, i.e., the quest for the class of all functions $f(\lambda)$ such that $f(P)$ is still a PsDO. Seeley \cite{Seeley:1967ea} investigates the special case of complex powers of elliptic operators on a compact manifold, and his results assure that the energy operator $\o:=(-\Delta+m^2)^{1/2}$ is a PsDO with symbol $p_\o(x,\xi)=(\xi^2+m^2)^{1/2}$.\\   
The following notion concerns the expansion of PsDOs and plays an important role in the construction of PsDOs.

\begin{defi}\label{expansion}
Let $(m_{i})_{i\in\N_{0}},\,m_{i}\in\R$, be a decreasing sequence with $\lim_{i\rightarrow\infty}m_{i}=-\infty,\;p(x,\xi)\in S_{\rho,\delta}^{m}$ and $p_{i}(x,\xi)\in S_{\rho,\delta}^{m_{i}}\;\forall i\in\N_{0}$. Then $p$ is said to have the asymptotic expansion $\sum_{i=0}^{\infty}p_{i}(x,\xi)$, denoted by $p(x,\xi)\sim\sum_{i=0}^{\infty}p_{i}(x,\xi)$, if the following statement is valid:
\begin{gather*}
p(x,\xi)-\sum_{i=0}^{N-1}p_{i}(x,\xi)\;\in  S_{\delta,\rho}^{m_{N}}\qquad\forall N\in\N.
\end{gather*}
In this case $p_0(x,\xi)$ is called the principal symbol of $P$.

\end{defi}

\begin{cor}
Let $(m_{i})_{i\in\N_{0}}$ and $\big(p_{i}(x,\xi)\big)_{i\in\N_{0}}$ be as aforementioned, then there exists a symbol $p(x,\xi)\in S_{\rho,\delta}^{m_{0}}$ such that
\begin{gather*}
p(x,\xi)\sim\sum_{i=0}^{\infty}p_{i}(x,\xi).
\end{gather*}
$p(x,\xi)$ is determined uniquely modulo $S^{-\infty}$.
\end{cor}
In the sequel we summarise some properties of PsDOs and of their products.

\begin{theo}
Let us assume that $p(x,\xi)\in S_{\rho,\delta}^{m},\,q(x,\xi)\in S_{\rho,\delta}^{m'}$, $P:=p(X,D_{x})$ and $Q:=q(X,D_{x})$. Then the following statements are valid:
\begin{itemize}
\item[(i)] The product $PQ$ lies in $\EuScript{S}_{\rho,\delta}^{m,m'}$.
\item[(ii)] $PQ$ is generated by the double symbol
\begin{gather*}
p(x,\xi)\cdot q(x',\xi')=:r(x,x',\xi,\xi')\in S_{\rho,\delta}^{m,m'},
\end{gather*}
i.e., $PQ=r(X,X',D_{x},D_{x'})$. Then
\begin{gather*}
|r|_{S_{\rho,\delta}^{m,m'}}^{(l)}\leq\underset{i+i'=l}{\max}|p|_{S_{\rho,\delta}^{m}}^{(i)}|q|_{S_{\rho,\delta}^{m'}}^{(i')}\quad\forall l\in\N_{0}.
\end{gather*}
\item[(iii)] Let $r\in S_{\rho,\delta}^{m,m'}$ with $r(X,D_{x})=PQ$. Then one has 
\begin{gather*}
r\sim\sum_{|\alpha|\geq 0}\frac{1}{\alpha!}\partial_{\xi}^{\alpha}p(x,\xi)\cdot(-i)^{|\alpha|}\partial_{x}^{\alpha}q(x',\xi').
\end{gather*}
\item[(iv)] For $p\sim\sum p_{i}$ and $q\sim\sum q_{i}$ one obtains
\begin{gather*}
r\sim\sum_{\substack{|\alpha|\geq 0 \\ i,j\in\N_{0}}}\frac{1}{\alpha!}\partial_{\xi}^{\alpha}p_{i}(x,\xi)\cdot (-i)^{|\alpha|}\partial_{x}^{\alpha}q_{j}(x',\xi'),
\end{gather*}
which means that the asymptotic expansion of the product of PsDOs is derived from the asymptotic expansions of their factors.
\end{itemize}
\end{theo} 

\begin{examp}\label{asymptotic0}\emph{
In Example \ref{Laplace} we have seen that the Laplace operator $\Delta$ is a positive and elliptic operator. Due to the investigation of Seeley on complex powers of elliptic operators \cite{Seeley:1967ea}, one can apply the functional calculus to $\Delta$ and show that, since $\Delta$ and $-\Delta+m^2$ are positive, its square root and the energy operator $\o=(-\Delta+m^2)^{1/2}$ are of pseudo-differential nature, too. Moreover, the symbol of $\o$ can be determined as
\begin{gather*}
p_\o(x,\xi)=(\xi^2+m^2)^{1/2},
\end{gather*}
and one obtains the identity
\begin{align*}
\o f(x)&=\int(\xi^2+m^2)^{1/2}\tilde{f}(\xi)e^{ix\xi}d\xi\\
&=\underset{|\xi|<m}{\int}(\xi^2+m^2)^{1/2}\tilde{f}(\xi)e^{ix\xi}d\xi+\underset{|\xi|>m}{\int}(\xi^2+m^2)^{1/2}\tilde{f}(\xi)e^{ix\xi}d\xi.
\end{align*}
In order to get a well-defined expansion of the symbol $p_\o(x,\xi)$ we adopt the approach of L\"ammerzahl \cite{Laemmerzahl:1993xe} and decompose the test function $\tilde{f}$ in $\tilde{f}_{nr}$ and $\tilde{f}_{ur}$ with their support lying in $U_{nr}:=\{\xi\in\R^3|\;|\xi|<m\}$ and $U_{ur}:=\{\xi\in\R^3|\;|\xi|>m\}$, respectively. Then we expand the symbol in the first integral with respect to $\xi$ and in the second one with respect to $m$, i.e.,
\begin{align}
(\xi^2+m^2)^{1/2}&=m+\frac{\xi^2}{2m}+\sum_{k=2}^\infty (-1)^{k+1}\frac{1\cdot 3\cdot 5\cdots(2k-3)}{2\cdot 4\cdot 6\cdots 2k}m^{-2k+1}\xi^{2k},\notag\\
(\xi^2+m^2)^{1/2}&=(\xi^2)^{1/2}+\frac{m^2}{2}(\xi^2)^{-1/2}+\sum_{k=2}^\infty (-1)^{k+1}\frac{1\cdot 3\cdot 5\cdots(2k-3)}{2\cdot 4\cdot 6\cdots 2k}(\xi^2)^{-k+\frac{1}{2}}m^{2k}\notag\\
&=:\sum_{i=0}^\infty p_{i}(x,\xi)\label{asymptotic},
\end{align}
where the former series converges for $|\xi|<m$ and the latter one for $|\xi|>m$. The right-hand side of \eqref{asymptotic} constitutes an asymptotic expansion of the symbol $(\xi^2+m^2)^{1/2}$, because the requirements in Definition \ref{expansion} are fulfilled, i.e.,
\begin{gather*}
p_{i}(x,\xi)\in S_{\delta,\rho}^{m_i}\quad\text{with}\quad m_i\overset{i\rightarrow\infty}{\longrightarrow}-\infty,\\
\text{and}\quad p(x,\xi)-\sum_{i=0}^{N-1}p_{i}(x,\xi)\;\in  S_{\delta,\rho}^{m_{N}}\qquad\forall N\in\N.
\end{gather*}
Thus we obtain for the energy operator:
\begin{align*}
\o f(x)&=\underset{|\xi|<m}{\int}\left(m+\frac{\xi^2}{2m}+\sum_{k=2}^\infty (-1)^{k+1}\frac{1\cdot 3\cdots(2k-3)}{2\cdot 4\cdots 2k}m^{-2k+1}\xi^{2k}\right)\tilde{f}_{nr}(\xi)e^{ix\xi}d\xi\\
&\;+\underset{|\xi|>m}{\int}\Bigg((\xi^2)^{1/2}+\frac{m^2}{2}(\xi^2)^{-1/2}+\\
&\qquad\qquad\qquad+\sum_{k=2}^\infty (-1)^{k+1}\frac{1\cdot 3\cdot 5\cdots(2k-3)}{2\cdot 4\cdots 2k}(\xi^2)^{-k+\frac{1}{2}}m^{2k}\Bigg)\tilde{f}_{ur}(\xi)e^{ix\xi}d\xi\\
&=\left(m-\frac{\Delta}{2m}-\sum_{k=2}^\infty\frac{1\cdot 3\cdots(2k-3)}{2\cdot 4\cdots 2k}m^{-2k+1}\Delta^k\right)f_{nr}(x)\\
&\;+\Bigg((-\Delta)^{1/2}+\frac{m^2}{2}(-\Delta)^{-1/2}\\
&\hspace{4cm}+\sum_{k=2}^\infty\frac{1\cdot 3\cdot 5\cdots(2k-3)}{2\cdot 4\cdots 2k}(-\Delta)^{-k+\frac{1}{2}}m^{2k}\Bigg)f_{ur}(x)\\
&=:\o_{nr}f_{nr}(x)+\o_{ur}f_{ur}(x),
\end{align*}
where we set $\tilde{f}_{nr}(\xi):=\chi_{|\xi|<m}\tilde{f}(\xi)$ and $\tilde{f}_{ur}(\xi):=\chi_{|\xi|>m}\tilde{f}(\xi)$. Via the expansions, converging on $|\xi|<m$ and $|\xi|>m$,
\begin{align*}
(\xi^2+m^2)^{-1/2}&=\frac{1}{m}-\frac{\xi^2}{2m^3}+\sum_{k=2}^\infty (-1)^k\frac{1\cdot 3\cdots(2k-3)}{2\cdot 4\cdot 6\cdots 2k}m^{-2k-1}\xi^{2k},\\
(\xi^2+m^2)^{-1/2}&=(\xi^2)^{-1/2}-\frac{m^2}{2}(\xi^2)^{-3/2}+\sum_{k=2}^\infty (-1)^k\frac{1\cdot 3\cdots(2k-3)}{2\cdot 4\cdot 6\cdots 2k}(\xi^2)^{-k-\frac{1}{2}}m^{2k},
\end{align*}
one calculates in the same manner the inverse of the energy operator:
\begin{align*}
\o^{-1}f(x)&=\left(\frac{1}{m}+\frac{\Delta}{2m^3}-\sum_{k=2}^\infty\frac{1\cdot 3\cdot 5\cdots(2k-3)}{2\cdot 4\cdots 2k}m^{-2k-1}\Delta^k\right)f_{nr}(x)\\
&\quad+\Bigg((-\Delta)^{-1/2}-\frac{m^2}{2}(-\Delta)^{-3/2}\\
&\qquad\qquad\qquad+\sum_{k=2}^\infty\frac{1\cdot 3\cdot 5\cdots(2k-3)}{2\cdot 4\cdots 2k}(-\Delta)^{-k-\frac{1}{2}}m^{2k}\Bigg)f_{ur}(x)\\
&=:\o_{nr}^{(-1)}f_{nr}(x)+\o_{ur}^{(-1)}f_{ur}(x).
\end{align*}
Thus $\o_{ur}$ and $\o_{ur}^{(-1)}$ are asymptotic expansions of $\o$ and $\o^{-1}$, respectively.}
\end{examp}

As aforementioned, one of the most important classes of PsDOs are the elliptic ones.

\begin{defi}
$Q'\in\EuScript{S}_{\delta,\rho}^{m}$ is called left parametrix or right parametrix of $Q\in\EuScript{S}_{\delta,\rho}^{m}$, if 
\begin{gather*}
Id-Q'Q\in\EuScript{S}^{-\infty}\quad\text{or}\quad Id-QQ'\in\EuScript{S}^{-\infty},
\end{gather*}
respectively. If $Q'$ possesses both properties then it is said to be a parametrix of $Q$.
\end{defi}
If the conditions given in Definition \ref{expansion} for the asymptotic expansion are fulfilled, then one can show that the requirement, 
$$
\inf\big\{|p_{0}(x,\xi)|:\;(x,\xi)\in\R^{n}\times S^{n}\big\}>0,
$$
is sufficient for ellipticity. Every elliptic PsDO $P\in\EuScript{S}_{\rho,\delta}^{m}$ has a parametrix $Q\in\EuScript{S}_{\rho,\delta}^{-m}$.

The wave front set, one of the main tools in micro-local analysis which we mentioned in the beginning of this chapter, can also be characterised in terms of PsDOs.
\begin{prop}\label{wavefrontset}
Let $u\in\D'(\R^n)$, then one has for the wave front set,
\begin{gather*}
WF(u)=\underset{\substack{P\in\EuScript{S}^{0}\\ Pu\in\mathcal{C}^{\infty}}}{\bigcap}\text{char}(P),
\end{gather*}
where
\begin{gather*}
\text{char}(P)\equiv \text{char}(p):=\big\{(x,\xi)\in\R^{n}\times\R_{*}^{n}:\;p_0(x,\xi)=0\big\}
\end{gather*}
is the so-called characteristic set of the (properly supported, see Definition \ref{properly-supported}) PsDO $P$ with principal symbol $p_0$. 
\end{prop}
Since Radzikowski \cite{Radzikowski:1996pa} has shown that the so-called Hadamard condition, which extracts the physical relevant states, is encoded in the wave front set of the two-point function, the microlocal analysis plays an increasing role in algebraic quantum field theory. The most important properties of the wave front set is collected in the following

\begin{theo}
Let $u,v\in\mathcal{D}'(\Omega)$ and $\Omega\subset\R^n$ an open subset, then the following statements hold:
\begin{itemize}
\item[(i)] $WF(u)$ is closed in $\Omega\times\R_{*}^{n}$ and conical in $\xi$ for all $u$.
\item[(ii)] For the complex-conjugate $\overline{u}$ one has:
\begin{gather*}
WF(\overline{u})=-WF(u),\quad\text{i.e.,}\\
\Big((x,\xi)\in WF(\overline{u})\Longleftrightarrow (x,-\xi)\in WF(u)\Big). 
\end{gather*}
\item[(iii)] For each closed and conical subset $A\subseteq\Omega\times\R_{*}^{n}$ there exists a distribution $u$ with $WF(u)=A$.
\item[(iv)] $WF(u+v)\subseteq WF(u)\cup WF(v)$.
\item[(v)] For the tensor product one has:
\begin{align*}
WF(u\otimes v)\subseteq \big(WF(u)\times WF(v)\big)&\cup((\supp(u)\times\{0\})\times WF(v))\\
&\cup(WF(u)\times(\supp(v)\times\{0\})).\notag
\end{align*} 
\item[(vi)] Let $\chi:\Omega\longrightarrow\Omega'$ with $\Omega,\Omega'\subset\R^{n}$ be a diffeomorphism. Then 
\begin{gather*}
WF(\chi^{*}u)=\chi^{*}WF(u):=\Big\{\big(x,{}^{t}[D\chi(x)]\xi\big):\;\big(\chi(x),\xi\big)\in WF(u)\Big\},
\end{gather*}
where
\begin{gather*}
\chi^{*}u(\phi):=u\big(\det(D_{\chi})^{-1}\phi\circ\chi^{-1}\big),\;\phi\in\mathcal{D}(\Omega'), 
\end{gather*} 
is the so-called pull-back operator.
\item[(vii)] The wave front set is related to the conventional notion of singularity in the following way:
\begin{gather*}
\pi_{x}\big(WF(u)\big)=\singsupp(u).
\end{gather*}
\end{itemize}

\end{theo}
It is a well-known fact that the product of distributions is not well-defined in general and, therefore, the space of distributions $\D'(\Omega)$, $\Omega\subset\R^n$, is not an associative differential algebra. Nevertheless, for a particular class of distributions the product can be introduced properly.

\begin{defi}
The so-called Fourier product $w\in\D'(\Omega)$ of the distributions $u,v\in\D'(\Omega)$, $\Omega\subset\R^n$, is said to exist, if for all $x\in\Omega$ there is always a test function $f\in\D(\Omega)$, identically one in an arbitrary neighbourhood of $x$, such that the convolution,
\begin{gather*}
\widetilde{f^2w}(\xi)=\frac{1}{(2\pi)^{n/2}}\int\widetilde{fu}(\eta)\widetilde{fv}(\eta)(\xi-\eta)d^n\eta,
\end{gather*} 
converges absolutely.
\end{defi} 
The existence of the Fourier product may be verified by a requirement on the wave front set.

\begin{prop}
If for $u,v\in\D'(\Omega)$, $\Omega\subset\R^n$, the following condition,
\begin{gather*}
(x,\xi)\in WF(u)\Longrightarrow(x,-\xi)\notin WF(v),
\end{gather*}
is satisfied for all $x\in\Omega$, then the Fourier product $u\cdot v$ exists and is uniquely defined. Furthermore one has,
\begin{gather*}
WF(u\cdot v)\subseteq\big(WF(u)\oplus WF(v)\big)\cup WF(u)\cup WF(v),
\end{gather*}
where
\begin{gather*}
WF(u)\oplus WF(v):=\big\{(x,\xi+\eta)|\;(x,\xi)\in WF(u),(x,\eta)\in WF(v)\big\}.
\end{gather*}
\end{prop}

Now, we want to introduce Sobolev spaces and list some of their fundamental properties, because, in Section \ref{Guido-section} in particular, we will make extensive use of them.

\begin{defi}
For each $s\in\R$ the Sobolev space
\begin{gather*}
H^{s}(\R^{n}):=\big\{u\in\mathcal{S}'(\R^{n}):\;\langle D_{x}\rangle^{s}u\in L_{2}(\R^n)\big\},
\end{gather*}
where $\langle D_{x}\rangle^{s}\in\EuScript{S}^{s}$ is the unique PsDO with reference to the symbol $(x,\xi)\mapsto(1+|\xi|^{2})^{s/2}$, equipped with the scalar product
\begin{align*}
(u,v):&=\big(\langle D_x\rangle^su,\langle D_x\rangle^sv\big)_{L_2(\R^n)}\\
&=\int(1+|\xi|^{2})^{2s}\tilde{u}(\xi)\overline{\tilde{v}(\xi)}d\xi,\qquad u,v\in H^s(\R^{n}),
\end{align*}
is a Hilbert space.
\end{defi}
Because of 
\begin{equation}\label{Sobolev1}
s<s'\quad\Longrightarrow\quad H^{s'}(\R^{n})\subset H^{s}(\R^{n}),
\end{equation}
one sets  
\begin{gather*}
H^{\infty}(\R^{n}):=\bigcap\big\{H^{s}(\R^{n}):\;s\in\R\big\}\quad\text{and}\\
H^{-\infty}(\R^{n}):=\bigcup\big\{H^{s}(\R^{n}):\;s\in\R\big\}.
\end{gather*}
Due to Sobolev's lemma on embedding, the inclusion
\begin{gather*}
H^{s}(\R^{n})\subseteq\mathcal{C}^{k}(\R^{n})\quad\text{and}\\
\exists C_{n,s}\geq 0:\;\Big(|u(x)|\leq C_{n,s}\parallel u\parallel_{H_{s}(\R^n)}\quad\forall x\in\R^{n},\,u\in H^{s}(\R^{n})\Big)
\end{gather*}
hold. In particular one obtains $H^{\infty}(\R^{n})\subseteq\mathcal{C}^{\infty}(\R^{n})$.

\begin{prop}\label{propSobolev2}
Let us assume that $P\in\EuScript{S}_{\rho,\delta}^{m}$. Then: 

\begin{itemize}
\item[(i)] $P$ is well defined on $H^{\infty}(\R^{n})$.
\item[(ii)] $P$ has the following property,
\begin{equation}\label{Sobolev2}
P:H^{s}(\R^{n})\rightarrow H^{s-m}(\R^{n}),
\end{equation}
and is continuous. Furthermore,
\begin{gather*}
P\big(H^{-\infty}(\R^{n})\big)\subseteq H^{-\infty}(\R^{n})\quad \text{and}\\
P\in\EuScript{S}^{-\infty}\quad\Longrightarrow\quad P\big(H^{-\infty}(\R^{n})\big)\subseteq H^{\infty}(\R^{n})\subseteq\mathcal{B}(\R^{n})\subseteq\mathcal{C}^{\infty}(\R^{n}).
\end{gather*}
\end{itemize}
\end{prop}
Let us consider the following Sobolev spaces defined on the open subsets $\Omega\subset\R^n$ and with respect to compact subsets $K$ of $\Omega$:
\begin{gather*}
H^s_c(K):=\big\{u\in\D'(K)\big\},\\
H^s_c(\Omega):=\bigcup_{K\subset\Omega} H^s_c(K),\\
H_{loc}^s(\Omega):=\big\{u\in\D'(\Omega)|\;\varphi\, u\in H^s(\R^{n})\text{ for }\varphi\in\D(\Omega)\big\}.
\end{gather*}
The topology of $H_{loc}^s(\Omega)$ is then defined by the semi-norms $u\mapsto\|\varphi u\|_s$, $\varphi\in\D(\Omega)$, with respect to the scalar product in $H^s(\Omega)$. The open sets in $H^s_c(\Omega)$, i.e., the topology on $H^s_c(\Omega)$, can be introduced through open intersections with $H^s_c(K)$. One may verify the following continuous linear injections with dense images:
\begin{gather*}
\S(\R^n)\hookrightarrow H^s(\R^n)\hookrightarrow H^{s'}(\R^n)\hookrightarrow\S'(\R^n)\quad\text{for}\quad s'\leq s,\\ 
\D(\Omega)\hookrightarrow H^s_c(\Omega)\hookrightarrow H^s_{loc}(\Omega)\hookrightarrow\D'(\Omega),
\end{gather*}
and the set-theoretical identifications
\begin{align*}
\mathcal{C}^\infty(\Omega)&=\bigcap_s H_{loc}^s(\Omega),&\D(\Omega)&=\bigcap_s H_c^s(\Omega),\\
\E'(\Omega)&=\bigcup_s H_c^s(\Omega),&\D'_F(\Omega)&=\bigcup_s H_{loc}^s(\Omega),
\end{align*}
where $\D'_F(\Omega)$ denotes the set of distributions of finite order in $\Omega$.

\begin{theo}
Let $\rho>0$, $\delta<1$ and $\delta<\rho$, then any continuous linear operator $P\in\EuScript{S}_{\rho,\delta}^m(\Omega)$ with $P:\E'(\Omega)\longrightarrow\D'(\Omega)$ defines a continuous map from $H^s_c(\Omega)$ onto $H_{loc}^s(\Omega)$ for arbitrary real numbers $m$ and $s$.
\end{theo}

\begin{defi}\label{properly-supported}
The distributional kernel $K_{P}\in\mathcal{D}'(\R_{x}^{n}\times\R_{x'}^{n})$ of $P\in\EuScript{S}_{\rho,\delta}^{m}$ is defined as
\begin{gather*}
\langle K_{P},u\times v\rangle:=\langle Pu,v\rangle\quad\forall u,v\in\mathcal{D}(\R^{n}).
\end{gather*}
A closed set $M\subseteq\R^{n}\times\R^{n}$ is said to be properly supported if\begin{gather*}
(K\times\R^{n})\cap M\sqsubset\R^{n}\times\R^{n}\quad\text{and}\\
(\R^{n}\times K)\cap M\sqsubset\R^{n}\times\R^{n}\quad\forall K\sqsubset\R^{n}
\end{gather*}
hold, where `$\;\sqsubset$' symbolises the inclusion of a compact subset. $P\in\EuScript{S}_{\rho,\delta}^{m}$ is called properly supported if $\supp(K_{P})$ has this property.
\end{defi} 
Due to Schwartz' kernel theorem, the existence of a unique $K_{P}$ is ensured. The PsDO $P$ is properly supported if and only if the following criterion is fulfilled:
\begin{gather*}
\forall K\sqsubset\R^{n}:\,\exists K'\sqsubset\R^{n}:\,\forall u\in\mathcal{D}(\R^{n}):\\
\Big(\supp(u)\subseteq K\quad\Longrightarrow\quad \supp(Pu),\,\supp({}^{t}Pu)\subseteq K'\Big).
\end{gather*}
For each $P\in\EuScript{S}_{\rho,\delta}^{m}$ one can always find a decomposition $P=P_{0}+P'$, where $P_{0}\in\EuScript{S}_{\rho,\delta}^{m}$ is properly supported and $P'$ is regularising, i.e., $P'\in\EuScript{S}^{-\infty}$. A properly supported PsDO $P$ can be uniquely extended to a continuous operator 
\begin{gather*}
P:\mathcal{D}'(\R^{n})\rightarrow\mathcal{D}'(\R^{n}).
\end{gather*}
Beyond this the following three restrictions, 
\begin{align*}
P:\mathcal{E}'(\R^{n})&\rightarrow\mathcal{E}'(\R^{n}),\;\\
P:\mathcal{C}^{\infty}(\R^{n})&\rightarrow\mathcal{C}^{\infty}(\R^{n}),\quad\text{and}\\
P:\mathcal{D}(\R^{n})&\rightarrow\mathcal{D}(\R^{n}),
\end{align*}
are continuous in their respective domains.

At the beginning we had announced that the wave front set of a distribution can be characterised with the help of PsDOs.

In a final step we mention the pseudo-locality property in the following  

\begin{theo}
Let $P\in\EuScript{S}_{\rho,\delta}^{m}$ be given with its distributional kernel $K_{P}\in\mathcal{D}'(\R^{n}\times\R^{n})$. Then
\begin{gather*}
\singsupp(K_{P})\subseteq\Delta:=\big\{(x,x):\;x\in\R^{n}\big\},
\end{gather*} 
where the singular support of the kernel, $\singsupp(K_{P})$, is the complement of the open set on which $K_{P}$ is smooth. There holdds the so-called pseudo-locality property: 
\begin{gather*}
\singsupp(Pu)\subseteq \singsupp(u),
\end{gather*}
for all $u\in\mathcal{E}'(\R^{n})$ or $u\in\mathcal{D}'(\R^{n})$, provided $P$ is properly supported.
\end{theo}
  
One may generalise PsDOs by addmitting general phase functions $\theta(x,\xi)$ in Definition \ref{PsDO-defi} instead of the scalar product $x\xi$ which satisfy the following conditions:
\begin{itemize}
\item[(i)] $\theta(x,\xi)$ is (in general) complex-valued, smooth and homogeneous of degree one. 
\item[(ii)] The gradient $\nabla_x\theta(x,\xi)$ does not vanish on the conic support of the symbol $a(x,\xi)$ for $\xi\neq 0$.
\end{itemize}
This leads to so-called Fourier integral operators.

\begin{defi}\label{FIO}
To each symbol $a\in S_{\rho,\delta}^{m}$ we associate a Fourier integral operator (FIO) 
\begin{equation*}
\begin{split} 
A\equiv a(X,D_{x}):\mathcal{S}(\R^{n})&\longrightarrow\mathcal{S}(\R^{n})\\
u&\mapsto a(X,D_{x})u
\end{split}
\end{equation*}
with
\begin{gather*}
Au(x):=\int a(x,\xi)\tilde{u}(\xi)e^{i\theta(x,\xi)}d\xi
\end{gather*}
for all $u\in\mathcal{S}(\R^{n})$ and $x\in\R^{n}$.
\end{defi}

One of the main differences between PsDOs and FIOs is that, while the distributional kernel of every PsDO is smooth off the diagonal in $\R^n\times\R^n$, a FIO does not have this property in general. 

We want to show in the next example \cite{Egorov:1993} that FOIs appear naturally in concrete problems and are not only of theoretical interest.

\begin{examp}\emph{
Let us consider the Cauchy problem for the wave equation
\begin{gather*}
\frac{\partial^2u}{\partial t^2}-a^2\Delta u=0,\quad t>0,\\
u=u_0,\quad\frac{\partial u}{\partial t}=u_1,\quad t=0,
\end{gather*}
where $a=\const$ and $u_0,u_1\in\D(\R^n)$. This problem can be transformed via 
\begin{gather*}
v(t,\xi):=\int u(t,x)e^{-ix\xi}dx
\end{gather*}
to a Cauchy problem for an ordinary differential equation:
\begin{gather*}
\frac{\partial^2v}{\partial t^2}+a^2|\xi|^2 v=0,\quad t>0,\\
v=\widehat{u}_0,\quad\frac{\partial v}{\partial t}=\widehat{u}_1,\quad t=0.
\end{gather*}
The solution $v$ is given as
\begin{gather*}
v(t,\xi)=\widehat{u}_0(\xi)\cos(at|\xi|)+\widehat{u}_1(\xi)\frac{\sin(at|\xi|)}{at|\xi|},
\end{gather*}
and finally $u$ can be determined as:
\begin{align*}
u(t,x)=\frac{(2\pi)^{-n}}{2}\Bigg\{&\int\Big(\widehat{u}_0(\xi)+\frac{1}{iat|\xi|}\widehat{u}_1(\xi)\Big)e^{i(at|\xi|+x\xi)}d\xi\\
&+\int\Big(\widehat{u}_0(\xi)-\frac{1}{iat|\xi|}\widehat{u}_1(\xi)\Big)e^{i(-at|\xi|+x\xi)}d\xi\Bigg\},
\end{align*}
which is a sum of two FIOs with real-valued phase function $\theta(x,\xi,t):=\pm at|\xi|+x\xi$.}
\end{examp}

%%%%%%%%%%%%%%%%%%%%%%%%%%%%%%%%%%%%%%%%%%%%%%%%%%%%%%%%%%%%%%%%%%%%%%%%%%%%

\chapter{One-Parameter Groups, Conformal Group}

\begin{flushright}
  \emph{Es ist unglaublich, wie unwissend die studirende Jugend \\
auf Universit\"aten kommt, wenn ich nur 10 Minuten\\
rechne oder geometrisire, so schl\"aft 1/4 derselben ein.}\\
\vspace{0,5cm}
G. C. Lichtenberg
\end{flushright}
\vspace{0,5cm}

All theories in physics comprise a kinematical part which contains the elements of the system, e.g., states and observables, and a dynamical part consisting of physical laws which determine the interaction between the elements. The states and observables are represented in classical mechanics by points in a differential manifold and functions defined on the manifold, respectively, in quantum mechanics by rays in a Hilbert space and linear operators acting on the space, and in algebraic quantum field theory by positive linear functionals on a $C^*$-algebra of local observables defined as operator-valued distributions. The dynamical nature of the theory, i.e., the time development of the system, as well as symmetries of the theory are formulated in terms of one-parameter groups of automorphisms, e.g., in classical mechanics one has a group of diffeomorphisms, in quantum mechanics a group of unitaries operating on the Hilbert space, and in quantum field theory a group of automorphisms of the $C^*$-algebra. 

In this thesis we will be concerned with the so-called modular group of automorphisms which will be introduced and investigated in detail in Section \ref{modulartheory}. For this purpose we need some general preparations. First we want to give an introduction to one-parameter groups and summarise some main features, which are excerpted mainly from \cite{Engel:2000} and \cite{Bratteli:1979tw}. In the second subsection we will address the conformal group and the conformal transformation, as they will appear throughout our investigation. This is mainly due to the fact that the modular group of automorphisms acts conformally on the massless algebra.

\section{One-Parameter Groups}

The starting point for the investigation of one-parameter groups is the quest for all maps $\alpha:\Gr\longrightarrow\L(\X)$, where $\Gr$ is a locally compact group, $\X$ a Banach space and $\alpha$ a representation of $\Gr$  in $\L(\X)$, the space of bounded linear operators on $\X$, satisfying the functional equation
\begin{gather}\label{semigroup}
\alpha^{g_1}\alpha^{g_2}=\alpha^{g_1g_2}\quad\text{and}\quad\alpha^e=\1.
\end{gather}
One may formulate the whole theory in a more general setting, namely in terms of semigroups, but, since we will only deal with the case $\Gr=\R$, we restrict ourselves to groups. 

\begin{defi}
A family $(\alpha^t)_{t\in\R}$ of bounded linear operators on a Banach space $\X$ fulfilling the functional equation \eqref{semigroup} is called a one-parameter group on $\X$. 
\end{defi}
In physics one usually refers to a dynamical system $(\X,\Gr,\alpha)$. We will always be concerned with $C^*$-dynamical or even $W^*$-dynamical systems because in our case $\X$ will be a $C^*$-algebra or a von Neumann algebra. But, nonetheless, we will stick to the general formalism. One may require these representations to be continuous with respect to different topologies. The strongest topology, namely the uniform topology, is too restrictive, since many one-parameter groups naturally arising in physics are not uniformly continuous. One counter-example is the group of left translations,
\begin{align*}
\alpha_l:\R&\longrightarrow\mathcal{L}\big(L^\infty(\R)\big)\\
t&\mapsto\alpha^t_l(f)(s):=f(t+s).
\end{align*}
For our purposes strong continuity will suffice, i.e., the family $(\alpha^t)_{t\in\R}$ should satisfy the functional equation \eqref{semigroup}, and the orbits,
\begin{align}\label{orbits}
\xi_X:\R&\longrightarrow\X\notag\\ 
t&\mapsto\xi_X(t):=\alpha^t(X),
\end{align}
are supposed to be continuous for each $X\in\X$, where $\R$ carries the natural topology and $\X$ the uniform topology. The next theorem justifies the restriction to strong continuity, in view of the fact that the central object of this thesis, the modular automorphism group, is (originally) weakly continuous.

\begin{theo}
The group $(\alpha^t)_{t\in\R}$ acting on a Banach space is strongly continuous if and only if it is weakly continuous. 
\end{theo} 
For the analysis of groups on Banach spaces two tools are of great help, namely the infinitesimal generator and the resolvent. We will make extensive use of the former one.

\begin{defi}
Let $(\alpha^t)_{t\in\R}$ be a strongly continuous group acting on a Banach space $\X$. Then its infinitesimal generator is defined as 
\begin{gather*}
\delta:\X\supseteq\mathbf{D}(\delta)\longrightarrow\X,\\
\mathbf{D}(\delta):=\big\{X\in\X|\;\xi_X \text{ in \eqref{orbits} is differentiable}\big\},\\
\delta X:=\frac{d}{dt}\xi_X(0)=\lim_{h\rightarrow 0}\frac{\alpha^h(X)-X}{h}.
\end{gather*}
\end{defi}
The generator is a linear, closed operator, but in general it is not defined on the whole Banach space.

\begin{theo}
Every strongly continuous group is uniquely determined by its densely defined infinitesimal generator.
\end{theo} 
For the domain of the generator one can prove a much stronger statement.

\begin{prop}
For the generator $\delta$ of a strongly continuous group on a Banach space $\X$ even the space $\bigcap_{i=1}^\infty\mathbf{D}(\delta^i)$, where $\delta^{i+1}:=\delta(\delta^i)$, is dense in $\X$. 
\end{prop}
The domain of the closed infinitesimal generator is a dense subspace of $\X$, and it is even $\X$ itself if and only if the group $(\alpha^t)_{t\in\R}$ is uniformly continuous. In all these cases the group can be formulated with the help of the exponential function,
$$
\alpha^t=e^{t\delta}:=\sum_{i=0}^\infty \frac{t^i\delta^i}{i!} \qquad\forall t\in\R.
$$
This is why the functional equation \eqref{semigroup} determining the group can be replaced by the equivalent differential equation
$$
\frac{d}{dt}\alpha^t=\delta\alpha^t\quad\forall t\in\R\quad\text{and}\quad\alpha^0=\1.
$$
If the group is strongly continuous, then the exponential series converges only on a linear subset of $\mathbf{D}(\delta)$, the so-called set of (entire) analytic vectors which is still a dense subset of $\X$. Consequently, for the analytic vectors $X\in\X$ we can describe the one-parameter group as
$$
\alpha^t(X)=e^{t\delta}X:=\sum_{i=0}^\infty \frac{t^i}{i!}\delta^iX \qquad\forall t\in\R.
$$  
For strongly continuous semigroups the analytic vectors do not have to constitute even a dense subset; there are examples for which the exponential series converges for $t=0$ and $X=0$ only.

For the sake of completeness we give the connection between the group $\alpha^t$ and the resolvent $R(\lambda,\delta):=(\lambda\1-\delta)^{-1}$ at $\lambda$ in the resolvent set $\rho(\delta)$.

\begin{theo}
Let $\delta$ be the infinitesimal generator of the strongly continuous group $(\alpha^t)_{t\in\R}$ on the Banach space $\X$ satisfying 
$$
\|\alpha^t\|\leq Me^{Nt},\qquad M\geq 1,N\in\R,
$$
for all $t\in\R$. If $\lambda\in\C$ such that 
$$
R(\lambda)X:=\int_0^\infty e^{-\lambda s}\alpha^s(X)ds
$$
exists for all $X\in\X$, then $\lambda\in\rho(\delta)$ and $R(\lambda)=R(\lambda,\delta)$. Furthermore, the spectral bound of the generator, $s(\delta):=\sup\{\Re\lambda|\;\lambda\in\sigma(\delta)\}$, satisfies
$$
-\infty\leq s(\delta)\leq w_0^\alpha<\infty,
$$
where $w_0^\alpha:=\inf\big\{w\in\R|\;\exists M_w\geq 1:\|\alpha^t\|\leq M_we^{wt},\;\forall t\geq 0\big\}$ is the growth bound.
\end{theo}
The next task is to take the opposite direction, i.e., to find all the operators which generate a strongly continuous one-parameter group. To demand the properties of the infinitesimal generators mentioned above, linearity, closedness, dense domain and their spectrum lying in some proper left half-plane, is not sufficient, since one can construct groups fulfilling all these conditions but still not mapping $\X$ into itself. An additional assumption on the resolvent will extract the correct candidates for real generators.

\begin{theo}
Let $\delta$ be a linear operator on a Banach space $\X$ and $M\geq 1$, $N\in\R$ arbitrary constants, then the following statements are equivalent:

\begin{itemize}
\item[(i)] $\delta$ generates a strongly continuous group $(\alpha^t)_{t\in\R}$ satisfying the growth estimate
$$
\|\alpha^t\|\leq Me^{Nt},\qquad t\in\R.
$$
\item[(ii)] $\delta$ is closed, densely defined and for every $\lambda\in\R$, $|\lambda|>N$, one has $\lambda\in\rho(\delta)$ and 
$$
\big\|\big[(|\lambda|-N)R(\lambda,\delta)\big]^n\big\|\leq M,\qquad\forall n\in\N.
$$
\end{itemize}

\end{theo}

\begin{theo}[Stone]

Let $\delta$ be a densely defined operator on a Hilbert space $\H$. Then $\delta$ is the infinitesimal generator of a unitary group $(\alpha^t)_{t\in\R}$ on $\H$ if and only if it is skew-adjoint, i.e., $\delta^*=-\delta$.

\end{theo}
Let us consider the abelian Banach algebra $\X:=\mathcal{C}^0(\R^n)$ equipped with the supremum norm $\|f\|:=\sup\{|f(x)||\;x\in\R^n\}$ and a continuously differentiable vector field $F:\R^n\longrightarrow\R^n$ which satisfies the estimate $\sup_x\|DF(x)\|<\infty$, where $DF(x)$ denotes the derivative of $F$ at $x\in\R^n$. $F$ induces a continuous flow 
\begin{align*}
\beta_t:\R\times\R^n&\longrightarrow\R^n\\
(t,x)&\mapsto\beta_t(x),
\end{align*}
i.e., $\beta_{t+s}(x)=\beta_t\beta_s(x)$ and $\beta_0(x)=x$. Additionally, the flow satisfies the following differential equation,
$$
\frac{\partial}{\partial_t}\beta_t(x)=F\big(\beta_t(x)\big),
$$
and we can formulate

\begin{prop}\label{generator}
Let $\X:=\mathcal{C}^0(\R^n)$ then
$$
\alpha^t(f)(x):=f\big(\beta_t(x)\big),\qquad f\in\X,x\in\R^n,
$$
defines a strongly continuous group $(\alpha^t)_{t\in\R}$ on $\X$, and its infinitesimal generator is given by the closure of the differential operator
\begin{align*}
\delta f(x):=\langle\grad f(x),F(x)\rangle&=\sum_{i=1}^n F_i(x)\frac{\partial f}{\partial x_i}(x)\\
&=\sum_{i=1}^n F_i\big(\beta_0(x)\big)\frac{\partial f}{\partial x_i}(x),\\
\mathbf{D}(\delta):&=\mathcal{C}_c^1(\R^n).
\end{align*}

\end{prop}
From now on we focus on the one-parameter groups of $\*$-automorphisms of $C\*$-algebras or von Neumann algebras and use the algebraic setting for this purpose. The fundamental algebraic tool for the investigation of infinitesimal generators is the symmetric derivation. The defining characteristics of derivations are naturally motivated by the main algebraic properties of the groups:
\begin{gather*}
\alpha^t(A)^*=\alpha^t(A^*)\quad\text{and}\\
\alpha^t(AB)=\alpha^t(A)\alpha^t(B).
\end{gather*}  

\begin{defi}
A symmetric derivation $\delta$ of a $C\*$-algebra $\Aa$ is a linear operator from a $\*$-subalgebra $\mathbf{D}(\delta)$, the domain of $\delta$, into $\Aa$ satisfying, for all $A,B\in\mathbf{D}(\delta)$, the conditions:
\begin{itemize}
\item[(i)] $\delta(A)^*=\delta(A^*)$.
\item[(ii)] $\delta(AB)=\delta(A)B+A\delta(B)$.
\end{itemize}

\end{defi}
For the discussion of automorphism groups the notion of spatial derivations is of great importance, due to the fact that they occur as their infinitesimal generators.

\begin{defi}
A symmetric derivation $\delta$ of a $C\*$-algebra $\Aa$ of bounded operators on a Hilbert space $\H$ is called spatial or inner if there exists a symmetric operator $H\in\Aa$ with the properties
\begin{gather*}
\delta(A)=i[H,A],\quad A\in\mathbf{D}(\delta),\\
\text{and}\quad\mathbf{D}(\delta)\mathbf{D}(H)\subseteq\mathbf{D}(H).
\end{gather*}
$H$ is said to implement $\delta$. 

\end{defi}

In the case of a $C\*$-algebra $\Aa$ a linear operator $\delta$ on $\Aa$ is the generator of a uniformly continuous one-parameter group of $\*$-automorphisms $\alpha^t$ if and only if it is a symmetric derivation of $\Aa$ with $\mathbf{D}(\delta)=\Aa$. Then the existence of a self-adjoint operator $H\in\pi(\Aa)''$, where $\pi$ is an arbitrary representation of $\Aa$, is  ensured, and the group can be described as 
$$
\pi\big(\alpha^t(A)\big)=e^{itH}\pi(A)e^{-itH}
$$
for all $A\in\Aa$ and $t\in\R$. If the group $\alpha^t$ is to be strongly continuous then more information is needed. We only give one of many possible combinations.

\begin{theo}
The densely defined closed linear operator $\delta$ on a $C\*$-algebra $\Aa$ is the infinitesimal generator of a strongly continuous one-parameter group of $\*$-automorphisms if and only if the following conditions hold:

\begin{itemize}
\item[(i)] $\delta$ is a symmetric derivation, and its domain is a $\*$-algebra. 
\item[(ii)] $\delta$ has a dense set of analytic elements in $\Aa$.
\item[(iii)] $\|(\1+M\delta)(A)\|\geq\|A\|$ for all $A\in\mathbf{D}(\delta)$ and $M\in\R$.
\end{itemize}

\end{theo}
If within the framework of von Neumann algebras the second condition is replaced by

\begin{itemize}
\item[$(ii')$] $(\1+M\delta)\big(\mathbf{D}(\delta)\big)=\Ma\;$ for all $M\in\R\backslash\{0\}$,
\end{itemize}
then the above theorem is also valid for arbitrary von Neumann algebras $\Ma$. In this case we know more about the derivations.

\begin{theo}\label{implementation}
Every derivation of a von Neumann algebra is inner.
\end{theo}
If two separable $C\*$-algebras $\Aa$ and $\Ba$ are connected via a surjective morphism $\pi:\Aa\longrightarrow\Ba$, then for every derivation $\delta_\Ba$ on $\Ba$ one can always find a derivation $\delta_\Aa$ on $\Aa$ such that 
$$
\pi\circ\delta_\Aa=\delta_\Ba\circ\pi\quad\text{and}\quad\|\delta_\Aa\|=\|\delta_\Ba\|.
$$

\begin{defi}\label{innerautomorphism}
An automorphism $\alpha$ acting on the C$\*$-algebra $\Aa$ is said to be inner, $\alpha\in\mathbf{Int}(\Aa)$, if there exists a unitary operator $U\in\Aa$ such that
$$
\alpha(A):=UAU^{-1}
$$
holds for all elements $A\in\Aa$. If the automorphism is not inner then we call it outer, $\alpha\in\mathbf{Out}(\Aa)$. The automorphism $\alpha$ is called approximately inner if $\alpha\in\overline{\mathbf{Int}(\Aa)}$. 
\end{defi}
For one-parameter groups of automorphisms $\alpha^t$ the set of all $t\in\R$ such that $\alpha^t$ is inner establish a subgroup of $\R$, which can be investigated more extensively through spectral theory. \\
Each uniformly continuous one-parameter group $(\alpha^t)_{t\in\R}$ of a separable and unital $C\*$-algebra $\Aa$ can be approximated by inner automorphisms, i.e., there exists a sequence  $(\alpha_n^t)_{n\in\N}$ of inner automorphism groups such that 
$$
\|\alpha^t(A)-\alpha_n^t(A)\|\overset{n\rightarrow\infty}{\longrightarrow} 0 
$$
uniformly in $t$ on any compact subset of $\R$ for all $A\in\Aa$.

\section{Conformal Transformations}

The metric $g_{\mu\nu}:=\text{diag}(1,-1,-1,-1)$ first describes distances between events and second defines the causal structure of spacetime, i.e., for each point $x$ it divides the spacetime in timelike, lightlike and spacelike regions. If one drops the first property and requires only conservation of the causal structure, then one obtains the highest spacetime symmetry possible, the so-called conformal symmetry. The conformal group is also the largest group which preserves the light cone. The fundamental physical laws are expected to be invariant under the conformal group, which contains the Poincar\'e group and the transformations preserving angles between world lines. The following summary is extracted from \cite{Schottenloher:1997pw} and \cite{Fradkin:1996is}.

Let the pair $(\M,g)$ be a semi-Riemannian manifold consisting of a smooth manifold $\M$ and a differentiable tensor field $g$ which maps every point $a\in\M$ into a non-degenerate, symmetric, bilinear form on the tangent space $T_a\M$,
$$
g_a:T_a\M\times T_a\M\longrightarrow\R.
$$
Equivalently, one can describe the bilinear form by means of local coordinates $(x^1,x^2,\cdots,x^n)$ of $\M$ as
\begin{gather*}
g_a(X,Y)=g_{\mu\nu}X^\mu Y^\nu,\\
\text{where}\quad X:=X^\mu\frac{\partial}{\partial x^\mu},\quad Y:=Y^\nu\frac{\partial}{\partial x^\nu}\in T_a(\M),
\end{gather*}
and with the properties
$$
\det\big(g_{\mu\nu}(a)\big)\neq 0\quad\text{and}\quad\big(g_{\mu\nu}(a)\big)^t=\big(g_{\mu\nu}(a)\big).
$$

\begin{defi}
Let $\U\subset\M$ and $\U'\subset\M'$ be two open subsets of the semi-Riemannian manifolds  $(\M,g)$ and $(\M',g')$, respectively, then a differentiable map $\phi:\U\longrightarrow\U'$ is called a conformal transformation if there exists a differentiable function $\Omega:\U\longrightarrow\R_+$, the so-called conformal factor for $\phi$, such that 
$$
\phi^*g'(X,Y):=g'\big(T\phi(X),T\phi(Y)\big)=\Omega^2g(X,Y),
$$
where $T\phi:T\U\longrightarrow T\U'$ is the derivation of $\phi$.
\end{defi}
The relation given above can equivalently be described by means of local coordinates as
$$
(\phi^*g')_{\mu\nu}(a)=g'_{\alpha\beta}\big(\phi(a)\big)\partial_\mu\phi^\alpha\partial_\nu\phi^\beta=\Omega^2(a)g_{\mu\nu}(a).
$$
From now on we restrict ourselves to local one-parameter groups of isometries $\varphi^t=:e^{tX}$ satisfying the differential equation
$$
\frac{d}{dt}\varphi^t(a)=X\varphi^t(a)\quad\text{with}\quad\varphi^0(a)=a,\quad a\in\M,
$$
where $X$ is the infinitesimal generator of the group.

\begin{defi}
  The vector field $X:\R^{p,q}\supset\M\longrightarrow\R^n$ on $\M\subset\R^{p,q}$ is said to be a conformal Killing field if $\varphi^t=e^{tX}$ is conformal in a neighbourhood of $a=0$ for all $t\in\R$.
\end{defi}
For the tensor field 
$$g(X,Y)=g^{p,q}(X,Y):=\sum_{i=1}^pX^iY^i-\sum_{i=p+1}^{p+q}X^iY^i
$$
 and the conformal Killing field $X=(X^1,...,X^n)=X^\nu\partial_\nu$ one can always find a twice differentiable function $\kappa:\M\longrightarrow\R$, which satisfies the so-called conformal Killing equation,
$$
\partial_\nu g_{\mu\lambda}X^\lambda+\partial_\mu g_{\nu\lambda}X^\lambda=\kappa g_{\nu\mu}.
$$
This fact prompts the next 

\begin{defi}
A differentiable function $\kappa:\M\longrightarrow\R$ is called conformal Killing factor if there is a conformal Killing field $X$ fulfilling the equation

\begin{equation}\label{Killing field}
\partial_\nu X_\mu+\partial_\mu X_\nu=\kappa g_{\nu\mu}.
\end{equation}

\end{defi}

\begin{cor}
Let $\Delta_g:=g^{\alpha\beta}\partial_\alpha\partial_\beta$ be a Laplace-Beltrami operator, then the function $\kappa:\M\supset\U\longrightarrow\R$ is a  conformal Killing factor if and only if 
$$
(n-2)\partial_\mu\partial_\nu\kappa+g_{\mu\nu}\Delta_g\kappa=0
$$
holds.

\end{cor}
This means that in the case $n=2$ the condition for $\kappa$ being a conformal Killing factor reduces to $\Delta_g\kappa=0$. In the case of $p=2$ and $q=0$, the Euclidean plane, $\phi=(u,v):\M\longrightarrow\R^{2,0}\cong\C$ is a holomorphic function on an open subset $\M\subset\R^{2,0}$ with non vanishing derivation $D\phi$ if and only if $\phi$ is a conformal and orientation-preserving transformation. Here, the conformal factor is determined by
$$
\Omega^2=u_x^2+u_y^2=\det(D\phi).
$$
The other possibility, $p=q=1$, represents the two-dimensional Minkowski space, and the differentiable function $\phi=(u,v):\M\longrightarrow\R^{1,1}$ on an open, connected subset $\M\subset\R^{1,1}$ is conformal if and only if 

\begin{itemize}
\item $u_x^2>u_y^2\;$ and
\item $u_x=v_y$, $u_y=v_x\;$ or $\;u_x=-v_y$, $u_y=-v_x$. 
\end{itemize}
The extant case, i.e., $p+q>2$, which will be considered in this thesis, allows for the conditions
\begin{gather*}
\partial_\mu\partial_\nu\kappa=0\quad\text{ for}\quad \mu\neq\nu\quad\text{and}\\
\partial_\mu\partial_\mu\kappa=\pm(n-2)^{-1}\Delta_g\kappa\quad\text{ for}\quad \mu=\nu.
\end{gather*}
The second equation leads to $\Delta_g\kappa=0$ and therefore to $\partial_\mu\partial_\mu\kappa=0$. Thus we conclude $\partial_\mu\partial_\nu\kappa=0$ for all $\mu,\nu\leq n=p+q$ and 
$$
\partial_\mu\kappa(x)=\partial_\mu\kappa(x^1,\cdots,x^n)=a_\mu,
$$
where $x\in\M$ and $a_\mu\in\R$ is an arbitrary constant. The solutions of the differential equation for a conformal Killing factor $\kappa$ are linear functions of the form
\begin{equation*}
\kappa(x)=\alpha_\nu x^\nu+\lambda\qquad\forall x\in\M,
\end{equation*}
where $\lambda,\alpha_\nu\in\R$ are arbitrary constants.

We are now interested in the conformal Killing fields $X$ with reference to the Killing factor $\kappa$. First of all we can choose $\kappa=0$ with the consequence that, due to the relations
\begin{align*}
\partial_\mu X_{\mu}+\partial_\mu X_{\mu}=0&\quad\Longrightarrow\quad X^\mu\text{ is independent of }x^\mu,\\
\partial_\nu X_{\mu}+\partial_\mu X_{\nu}=0&\quad\Longrightarrow\quad \partial_\nu X^\mu=0,
\end{align*}
the Killing fields have the following structure,
\begin{equation*}
X^\mu(x)=\o_\nu^\mu x^\nu+c^\mu\qquad\text{with}\quad\o_\nu^\mu,c^\mu\in\R.
\end{equation*}
Therefore we obtain three different possibilities:

\begin{itemize}
\item[(i)] $\o_\nu^\mu=0,c^\mu\neq 0$ determines the conformal transformation as the translation $\phi_c(x)=x+c$.
\item[(ii)] $\o_\nu^\mu\neq0,c^\mu=0$ leads to $\phi_\Lambda(x)=\Lambda x$ with\\ $\Lambda\in\mathbf{O}(p,q):=\big\{\tilde{\Lambda}\in\R^{n\times n}|\;\tilde{\Lambda}^t g^{p,q}\tilde{\Lambda}=g^{p,q}\big\}$.
\item[(iii)] $\o_\nu^\mu\neq 0,c^\mu\neq 0$ is a linear combination of the first two items.
\end{itemize}
The choice $\kappa=\lambda=\const\neq 0$ determines the conformal Killing field $X(x)=\lambda x$ and consequently the conformal transformation as
\begin{itemize}
\item[(iv)] the dilation $\phi(x)=e^\lambda x$.
\end{itemize} 
For a non-constant conformal Killing factor $\kappa$ one can verify by straightforward calculation that the conformal Killing field, defined as 
\begin{gather*}
X^\mu(x)=2(x,b)x^\mu-(x,x)b^\mu
\end{gather*}
with
\begin{gather*}
(x,b):=g_{\mu\nu}x^\mu b^\nu\quad\text{and}\quad b\in\R^n\backslash\{0\},
\end{gather*}
solves equation \eqref{Killing field}. The associated conformal transformation is the
\begin{itemize}
\item[(v)] proper conformal transformation or conformal translation, 
$$
\phi(x)=\frac{x^\mu-(x,x)c^\mu}{1-2(x,c)+(x,x)(c,c)}.
$$

\end{itemize}
Contrary to the first four possibilities, the proper conformal transformation has no extension from $\M\subset\R^{p,q}$ onto the whole space $\R^{p,q}$. These transformations are singular for any point $x'^\mu$ on the hypersurface $1-2(x,c)+(x,x)(c,c)=0$, and for this reason they cannot constitute a global symmetry.

To put it in a nutshell the following statement holds.

\begin{cor}
The composition of two conformal transformations is again conformal. Let $\M\subset\R^{p,q}$, $p+q=d\geq3$, be open, then each conformal transformation $\rho:\M\longrightarrow\R^{p,q}$ can be described as a composition of

\begin{itemize}
\item[(i)] a translation $x'^\mu=x^\mu+a^\mu$, $a\in\R^d$, 
\item[(ii)] an orthogonal transformation $x'^\mu=\Lambda x$, $\Lambda\in\mathbf{O}(p,q)$,
\item[(iii)] a dilation $x'^\mu=\lambda x^\mu$, $\lambda\in\R$, and
\item[(iv)] a special conformal transformation $x'^\mu=\frac{x^\mu-(x,x)c^\mu}{1-2(x,c)+(x,x)(c,c)}$, $c\in\R^d$. 
\end{itemize}

\end{cor} 
If we consider an infinitesimal conformal transformation of a point $x$,
$$
x^\mu\mapsto x^\mu+\epsilon u^\mu(x),
$$
with $\epsilon\rightarrow 0$, then  $u$ has to be of the following form:
\begin{gather}\label{conformaltrafo}
u^\mu(x)=a^\mu+g^{\mu\lambda}\o_{\lambda\nu}x^\nu+b\,x^\mu+2x^\mu c_\lambda x^\lambda-x_\lambda x^\lambda c^\mu. 
\end{gather}
In this equation the quantities $a^\mu$, $b$ and $c^\mu$ are arbitrary constants. In the case of a $d$-dimensional spacetime $u^\mu(x)\partial_\mu$ is generating a $\frac{1}{2}(d+1)(d+2)$- dimensional Lie group, the so-called conformal group, more specifically one obtains the following infinitesimal generators:\\
\begin{table}[h]\label{generatortable}
\begin{tabular}{lcl}
             &  \\
$P:=a^\mu\partial_\mu$ &:& translations ($d$ generators),\\
$M:=g^{\mu\lambda}\o_{\lambda\nu}x^\nu\partial_\mu$ &:& Lorentz transformations \big($\frac{d(d-1)}{2}$ generators\big),\\
$D:=b\,x^\mu\partial_\mu$ &:& dilations (1 generator),\\
$K:=\big(x^\mu c_\lambda x^\lambda-x_\lambda x^\lambda c^\mu\big)\partial_\mu$ &:&special (proper) conformal transformations\\ 
 &&($4$ generators). 

\end{tabular}
\end{table}
\\ \\
The generators of the Lorentz transformation can be decomposed in $d-1$ generators for the Lorentz boosts and $\frac{1}{2}(d-1)(d-2)$ generators for the rotations. The commutation relations of these generators read: 
\begin{gather}\label{commutation-relations}
[P_\mu,P_\nu]=0,\quad[P_\alpha,M_{\mu\nu}]=i(g_{\alpha\mu}P_\nu-g_{\alpha\nu}P_\mu),\notag\\
[P_\mu,D]=iP_\mu,\quad[K_\nu,P_\mu]=2i(g_{\mu\nu}D+M_{\mu\nu}),\notag\\
[M_{\mu\nu},M_{\alpha\beta}]=i(g_{\mu\beta}M_{\nu\alpha}+g_{\nu\alpha}M_{\mu\beta}-g_{\mu\alpha}M_{\nu\beta}-g_{\nu\beta}M_{\mu\alpha}),\\
[M_{\mu\nu},D]=0,\quad[K_\alpha,M_{\mu\nu}]=i(g_{\alpha\mu}K_\nu-g_{\alpha\nu}K_\mu),\notag\\
[D,D]=0,\quad[D,K_\mu]=iK_\mu,\quad[K_\mu,K_\nu]=0.\notag
\end{gather}
Because of the local isomorphism between the conformal group $\mathbf{Conf}(\R^{p,q})$ and the group of pseudo-orthogonal transformations,
$$
\mathbf{Conf}(\R^{p,q})\cong\mathbf{SO}(p+q,q+1),
$$
which in our case reads
$$
\mathbf{Conf}(\R^{3,1})\cong\mathbf{SO}(4,2),
$$
we will, for the sake of technical simplicity, also make use of the so-called pseudo-orthogonal transformations, which are described in terms of coordinates $\xi^\alpha$, $\alpha=0,1,\cdots,d+2$, of a real space equipped with the metric
$$
g_{\alpha\beta}:=\begin{cases}\delta_{\alpha\beta} & ,\alpha=0,d+2,\\ -\delta_{\alpha\beta} & ,\alpha=1,2,\cdots,d+1.\end{cases}
$$
The group transformations in the new coordinates have the form
$$
\xi^\alpha\mapsto\xi'^\alpha:=M_\beta^\alpha\xi^\beta,
$$
where $M$ is the matrix determined by the equations
$$
g_{\alpha\beta}M_\gamma^\alpha M_\delta^\beta=g_{\gamma\delta}\quad\text{and}\quad \det M=1.
$$
Their relation to the Minkowski coordinates are
\begin{gather*}
x^\mu=\xi_+^{-1}\xi^\mu\quad\text{and}\quad (x,x)=g_{\mu\nu}x^\mu x^\nu=\frac{\xi_-}{\xi_+}\\
\text{with}\quad \xi_\pm:=\xi^{d+2}\pm\xi^{d+1}.
\end{gather*}
The conformal transformations in Minkowski space have their analogue in the pseudo-orthogonal coordinates:

\begin{itemize}
\item[(i)] Translations correspond to transformations: $\xi^\mu \mapsto\xi'^\mu:=\xi^\mu+\xi_+c^\mu$.
\item[(ii)] Homogeneous Lorentz transformations correspond to  pseudo-rotations in a $d$-dimensional subspace:
$$
\xi^\mu \mapsto\xi'^\mu:=\Lambda_\nu^\mu\xi^\nu,\qquad\mu,\nu=0,1,\cdots,d-1.
$$
\item[(iii)] Dilations correspond to pseudo-rotations in a $2$-dimensional subspace:
\begin{gather*}
\xi^{d+1}\mapsto\xi'^{d+1}:=\xi^{d+2}\sinh\tau+\xi^{d+1}\cosh\tau,\\
\xi^{d+2}\mapsto\xi'^{d+2}:=\xi^{d+2}\cosh\tau+\xi^{d+1}\sinh\tau.
\end{gather*}
\item[(iv)] Special conformal transformations correspond to the transformations of the type: $\xi^\mu\mapsto\xi'^\mu:=\xi^\mu+\xi_-c^\mu$.
\end{itemize}
The correspondence between the infinitesimal generators $J_{\alpha\beta}=-J_{\beta\alpha}$ of the pseudo-orthogonal group  and the ones of transformations in Minkowski space are given by
\begin{gather*}
P_\mu=J_{\mu\; d+2}-J_{\mu\;d+1},\qquad M_{\mu\;\nu}=J_{\mu\;\nu},\\
D=J_{d+2\;d+1},\qquad K_\mu=J_{\mu\;d+2}+J_{\mu\;d+1},
\end{gather*}
and the commutation relations of $J_{\alpha\beta}$ can be summarized as
$$
[J_{\alpha\beta},J_{\gamma\delta}]=-i\big(g_{\alpha\gamma}J_{\beta\delta}+g_{\beta\delta}J_{\alpha\gamma}-g_{\beta\gamma}J_{\alpha\delta}-g_{\alpha\delta}J_{\beta\gamma}\big).
$$
For our purposes, i.e., the special case of a four-dimensional Minkowski space, the following quantities will be of greater interest,
$$
g_{\alpha\beta}:=\begin{cases}\delta_{\alpha\beta} & ,\alpha=0,5\\ -\delta_{\alpha\beta} & ,\alpha=1,2,3,4.\end{cases}
$$
The two different coordinate systems are linked  to each other by the relation
$$
x^\mu=(\xi^4+\xi^5)^{-1}\xi^\mu.
$$
The variation of the distance between two events under a conformal transformation is given by a conformal factor $N$ as 
$$
(x_1'-x_2')^2=N(x_1)^{-1}N(x_2)^{-1}(x_1-x_2)^2.
$$
It can be calculated for each type of transformations as
\begin{gather*}
N=\begin{cases} 1 & \text{for translations},\\ 1 & \text{for Lorentz transformations},\\ \lambda^{-1} & \text{for dilations},\\ \eta^{-1}\eta' & \text{for special conformal transformations},\end{cases} 
\end{gather*}
where $\eta_+':=\eta_+-2(\xi,c)+(\xi^4-\xi^5)(c,c)$ and $\eta_+:=\xi^4+\xi^5$.

For the analysis of modular automorphism groups in quantum field theory the so-called conformal inversion map will be of great significance, more precisely, if one wishes to transfer the modular action on the Rindler wedge $\W_R:=\{x|\;x^3>|x^0|\}$ to that on the forward light cone $\V_+:=\{x\in\R^4|\;(x,x)>0\text{ and }x^0>0 \}$ or the double cone $\D:=(\V_+-e_0)\cap(\V_-+e_0)$, where $\V_-:=\{x\in\R^4|\;(x,x)>0\text{ and }x^0<0 \}$ is the backward light cone. The inversion map is defined as
\begin{align*}
\rho:\M&\longrightarrow\M\\
x^\mu&\mapsto\rho(x^\mu):=\frac{-x^\mu}{(x,x)}.
\end{align*}
First, while the inversion map and the Lorentz transformation commute and the consecutive application of $\rho$ and the dilation leads to a change of the parameter, i.e.,
\begin{gather*}
x^\mu\;\overset{\Lambda}{\mapsto}\;\Lambda_\nu^\mu\,x^\nu\overset{\rho}{\mapsto}\;(\Lambda_\nu^\mu\,x^\nu)^{-2}\Lambda_\nu^\mu\,x^\nu=\frac{1}{(x,x)}\Lambda_\nu^\mu\,x^\nu,\\
x^\mu\overset{\rho}{\mapsto}\;\frac{x^\mu}{(x,x)}\;\overset{\Lambda}{\mapsto}\;\Lambda_\nu^\mu\,\frac{x^\nu}{(x,x)},%\;\overset{\rho}{\mapsto}\Lambda_\nu^\mu x^\nu\quad\text{and}\\
\\x^\mu\overset{\rho}{\mapsto}\;\frac{x^\mu}{(x,x)}\;\overset{\text{dil.}}{\mapsto}\;\lambda\frac{x^\mu}{(x,x)}\;\overset{\rho}{\mapsto}\frac{1}{\lambda}x^\mu,
\end{gather*}
the successive application of $\rho$ and the translation has the proper conformal transformation as a consequence:
\begin{gather*}
x^\mu\overset{\rho}{\mapsto}\;\frac{x^\mu}{(x,x)}\;\overset{\text{trans}}{\mapsto}\;\frac{x^\mu}{(x,x)}+c^\mu\;\overset{\rho}{\mapsto}\;\frac{x^\mu+(x,x)c^\mu}{1+2(x,c)+(x,x)(c,c)}.
\end{gather*}
The second interesting feature of the inversion map is that it maps the right Rindler wedge onto the forward light cone and onto the double cone with radius one $\D_1$ via the equations 
\begin{gather*}
\rho(\D_1-e_0)=\V_++\frac{e_0}{2}\quad\text{and}\\
\rho(\D_1+e_3)=\W_R+\frac{e_3}{2}.
\end{gather*}
Here the canonical orthonormal basis of $\R^4$ is denoted by $\big\{e_0,e_1,e_2,e_3\big\}$ .\\ 
These two properties are of interest insofar as in a free massless scalar theory the inversion map can be represented by a vacuum-preserving, unitary operator,
\begin{gather}\label{ray-inversion-operator}
U_\rho\varphi(x)U_\rho^{-1}=\frac{1}{(x,x)^3}\varphi\big(\rho(x)\big)\\
\text{and}\quad U_\rho\Omega=\Omega,\notag
\end{gather}
on the one-particle Hilbert space, and $U_\rho$ itself through its second quantisation operator $\Gamma(U_\rho)$, a vacuum-preserving, unitary operator on the Fock space, such that
\begin{gather*}
\Gamma(U_\rho)\varphi(x)\Gamma(U_\rho)^{-1}=\frac{1}{(x,x)^3}\varphi\big(\rho(x)\big),\quad\text{and}\\
\Gamma(U_\rho)\Aa(\O)\Gamma(U_\rho)^{-1}=\Aa\big(\rho(\O)\big),
\end{gather*}
where $\O$ is an arbitrary double cone.

It is noteworthy that the requirement of vacuum invariance is stronger than the assumption of conformal symmetry.

\begin{defi}
An equation for a field $\varphi$ is called conformally invariant if there exists a constant $s\in\R$, the so-called conformal weight of the field $\varphi$, such that $\varphi$ is a solution with respect to the metric $g_{\mu\nu}$ if and only if $\tilde{\varphi}:=\Omega^s\varphi$ is a solution with respect to the metric $\tilde{g}_{\mu\nu}:=\Omega^s g_{\mu\nu}$, where $\Omega$ is the conformal factor. 
\end{defi}
For the conformal symmetry to hold vacuum invariance under the group of translations, Lorentz transformations and dilations is sufficient.

Let $U_P$, $U_D$ and $U_K$ be the unitary operators induced by translations, dilations and special conformal transformations, respectively, then 
\begin{gather*}
U_P(a)\varphi(x)U_P^{-1}(a)=\varphi(x+a),\\
U_D(\lambda)\varphi(x)U_D^{-1}(\lambda)=\varphi(\lambda x),\\
U_K(c)\varphi(x)U_K^{-1}(c)=\big(1-2cx+c^2(x,x)\big)^{-3}\varphi\left(\frac{x-c(x,x)}{1-2cx+c^2(x,x)}\right),\\
U_\rho U_D(\lambda)U_\rho=U_D(\lambda^{-1}),\\
\text{and}\quad U_K(c)=U_\rho U_P(c)U_\rho.
\end{gather*}
The latter two relations hold under the assumption that the action of $U_\rho$ is
$$
U_\rho\varphi(x)U_\rho=(x,x)^{-3}\varphi\big(\rho(x)\big).
$$
These relations can be transferred to the infinitesimal generators of translations, Lorentz transformations, dilations and proper conformal transformations,
\begin{gather*}
U_\rho P_\mu U_\rho=K_\mu,\\
U_\rho M_{\mu\nu}U_\rho=M_{\mu\nu},\quad\text{and}\\
U_\rho DU_\rho=-D,
\end{gather*}
which means that the generator of the Lorentz transformations commutes and that of the dilations anti-commutes with the operator $U_\rho$, respectively. These relations in particular show that the algebra of the conformal group is invariant under $U_\rho$.

%%%%%%%%%%%%%%%%%%%%%%%%%%%%%%%%%%%%%%%%%%%%%%%%%%%%%%%%%%%%%%%%%%%%%%%%%%%%

\chapter{Modular Theory and Quantum Field Theory}

\begin{flushright}
  \emph{Tu as voulu de l'alg\`ebre,\\
et tu en auras jusqu'au menton.}\\
\vspace{0,5cm}
J. Verne\\
Autour de la Lune
\end{flushright}
\vspace{0,5cm}

In this chapter we give a short introduction to modular theory, which has been formulated by Tomita but first published by Takesaki \cite{Takesaki:1970}, and a general survey of its most innovative applications in mathematics and physics. 

While the assumptions for the formulation of modular theory are only the underlying von Neumann algebra $\Ma$ and a cyclic and separating vector $\Omega$ in the representing Hilbert space $\H$, the consequences of its main statement, namely Tomita's theorem which relates $\Ma$ with its commutant $\Ma'$ and ensures the existence of a group of automorphisms $\sigma^t:\Ma\longrightarrow\Ma$, $t\in\R$, is immense.

In mathematics, the classification of von Neumann algebras would be hardly imaginable, in particular Connes' classification of type $III$ factors \cite{Connes:1973}, Jones' classification of type $II_1$ subfactors \cite{Jones:1983kv} and it generalisation by Kosaki to arbitrary subfactors \cite{Kosaki:1986}, which will be addressed in the first section. These achievements again have made possible many developments in mathematics, e.g., the index theory of Jones yields groundbreaking insights in algebraic topology with a new polynomial invariant for knots and links in $\R^3$.

Since local quantum physics, due to the Reeh-Schlieder Theorem \ref{Reeh-Schlieder}, from the outset brings along the conditions for modular theory, both theories are fitted perfectly to each other as reflected in many results. The very first fruit has been the generalisation of Gibbs' equilibrium states to KMS states formulated by Haag, Hugenholtz and Winnink \cite{Haag:1967sg}. Another branch is the determination of local algebras as hyperfinite factors of type $III_1$ by the collaboration of many authors through a long process, and the classification of conformal local subfactors by Kawahigashi and Longo \cite{Kawahigashi:2002px}. This thesis is mainly concerned with the third branch, the modular action. The investigation of Bisognano and Wichmann yield the surprising insight that for wedge regions the modular action coincides with Lorentz boosts \cite{Bisognano:1975ih}, \cite{Bisognano:1976za}. An immediate consequence of this result is the discovery of black hole evaporation, the so-called Hawking radiation. Buchholz can identify the modular action on forward light cones as dilations \cite{Buchholz:1977ze}, and Hislop and Longo identify it as conformal transformations in the case of double cones \cite{Hislop-Longo:1981uh}. The consequence of the geometric modular action for the development of the algebraic quantum field theory is revolutionary, because it serves to prove some main pillars and indispensable concepts of quantum field theory. We will mention only the most important ones, as the investigations on the PCT theorem by Borchers \cite{Borchers:1991xk}, the construction of the Poincar\'e group by Brunetti, Guido and Longo \cite{Brunetti:1992zf}, the algebraic spin-statistics theorem by Guido and Longo \cite{Guido:1995fy} and the modular nuclearity condition by Buchholz, D'Antoni and Longo \cite{Buchholz:1989bj}.

\section{Modular Theory in Mathematics}\label{modulartheory}

 In this section we give a short and straightforward introduction to the Tomita-Takesaki theory, also called modular theory \cite{Takesaki:1970}. Although one can formulate it in a more general setting, namely in terms of left or right Hilbert algebras which are identical for isometrical involutions, we will restrict ourselves to what is absolutely necessary for our aims. We will follow the standard literature, e.g., \cite{Takesaki:2001}, \cite{Stratila:1981}, \cite{Bratteli:1979tw} and \cite{Kadison:1983}. Confer also Appendix A for definitions and notations which are used but are not mentioned here.

The modular theory has also been investigated in the framework of O$\*$-algebras, i.e., $\*$-algebras of closable operators, see e.g. \cite{Inoue:1998}.

 Our starting point is a von Neumann algebra $\mathfrak{M}$ acting on a Hilbert space $\H$  and $\Omega\in\H$ with a cyclic, i.e., $\mathfrak{M}\Omega$ is dense in $\H$, and separating vector, i.e., $A\Omega=0$ implies $A=0$ for $A\in\mathfrak{M}$. Because cyclicity of $\Omega$ for $\Ma$ is equivalent to it being separable for the commutant $\mathfrak{M}'$, the vector $\Omega$ transports these two properties from the algebra onto its commutant. Thus the following two anti-linear operators are well defined:
\begin{align*}
S_0:\D(S_0)=\mathfrak{M}\Omega\subset\H&\longrightarrow\H\\
A\Omega&\mapsto S_0 A\Omega:=A^*\Omega,\\
F_0:\D(F_0)=\mathfrak{M}'\Omega\subset\H&\longrightarrow\H\\
A\Omega&\mapsto F_0 A\Omega:=A^*\Omega.\\
\end{align*}
Both operators are closable, and one defines $F:=\overline{F_0}=S_0^*$ and $S:=\overline{S_0}=F_0^*$. Therefore, the Tomita operator $S$ allows for a unique polar decomposition into the positive, selfadjoint operator $\Delta$ and the anti-unitary operator $J$, the so-called modular operator and modular conjugation with respect to the pair $(\mathfrak{M},\Omega)$, respectively:
\begin{equation*}
S=J\Delta^{1/2}.
\end{equation*}
One may easily verify the following relations:
\begin{align*}
\Delta&=FS,& J&=J^*,\\
\Delta^{-1}&=SF,&J^{2}&=\1,\\
F&=J\Delta^{-1/2},&\Delta^{-1/2}&=J\Delta^{1/2}J.
\end{align*}
We are now already in the position to formulate the core of the modular theory.
\begin{theo}
For the von Neumann algebra $\mathfrak{M}$ and the associated modular operator and modular conjugation the relations
\begin{equation*}
J\mathfrak{M} J=\mathfrak{M}'\quad\text{and}\quad\Delta^{it}\Ma\Delta^{-it}=\mathfrak{M}
\end{equation*}
hold for all $t\in\R$.
\end{theo}

\begin{defi}
A von Neumann algebra is said to be $\sigma$-finite if it contains (at most) countably many pairwise orthogonal projections. 
\end{defi}
In statistical quantum mechanics and quantum field theory only $\sigma$-finite  von Neumann algebras appear, which therefore can be represented in a separable Hilbert space, while von Neumann algeras, which can be represented in a separable Hilbert space, need not be $\sigma$-finite in general.

\begin{lem}
For the von Neumann algebra $\Ma$ acting on the Hilbert space $\H$ the following statements are equivalent:

\begin{itemize}
\item[(i)] $\Ma$ is $\sigma$-finite.
\item[(ii)] There exists a countable subset of $\H$, which is separating for $\Ma$.
\item[(iii)] There exists a faithful and normal weight on $\Ma$.
\item[(iv)] $\Ma$ is isomorphic to a von Neumann algebra $\pi(\Ma)$ which admits a cyclic and separating vector.
\end{itemize}

\end{lem}
Given a faithful and normal weight $\o$, one can derive the associated cyclic representation $(\H_{\o},\pi_{\o},\Omega_{\o})$ through the GNS construction and the modular operator for the pair $(\pi_{\o}(\mathfrak{M}),\Omega_{\o})$. The above theorem ensures the existence of a  $\sigma$-weakly continuous one-parameter group of $\*$-automorphisms, $\big(\sigma_{\o}^t\big)_{t\in\R}$,

\begin{align*}
\sigma_{\o}^t:\mathfrak{M}&\longrightarrow\mathfrak{M}\\
A&\mapsto\sigma_{\o}^t(A):=\pi_{\o}^{-1}\big(\Delta^{it}\pi_{\o}(A)\Delta^{-it}\big),
\end{align*}
the so-called modular automorphism group associated with $\big(\pi_{\o}(\mathfrak{M}),\Omega_{\o}\big)$. Since we are concerned with von Neumann algebras, due to Theorem \ref{weakstrongtopology}, the modular group is continuous with respect to the strong topology, too. The modular group is a powerful and constructive tool for the investigation of von Neumann algebras and has made possible many applications in mathematics and theoretical physics. The main linkage between the modular theory and physics is the following property, 
\begin{align*}
\big(\Delta^{1/2}\pi_\o(A)\Omega_\o,\Delta^{1/2}\pi_\o(B)\Omega_\o\big)&=\big(J\pi_\o(A^*)\Omega_\o,J\pi_\o(B^*)\Omega_\o\big)\\
 &=\big(\pi_\o(B^*)\Omega_\o,\pi_\o(A^*)\Omega_\o\big),
\end{align*}
the so-called modular condition, which can equivalently be described by means of the modular group itself,
$$
\o\big(\sigma_\o^{i/2}(A)\sigma_\o^{-i/2}(B)\big)=\o(BA)
$$ 
for all $A,B\in\Ma$.

As previously mentioned, one-parameter groups of inner automorphisms can be analysed more elaborately with the help of spectral theory, but since in this thesis we deal with von Neumann algebras of type $III_1$, the next statement rules out this possibility in the present context.
\begin{prop}
The modular group associated with a faithful normal state on a von Neumann algebra $\Ma$ is inner if and only if $\Ma$ is semifinite, i.e., if the decomposition of $\Ma$ has no part of type III.
\end{prop}
Because, due to Theorem \ref{implementation}, every derivation on a von Neumann algebra is inner, for the derivation of the infinitesimal generator of the modular group it is sufficient to know the generator of the modular group.

\begin{theo}
Let $\Ma$ be a von Neumann algebra on a Hilbert space $\H$, $\Delta_\Omega$ the modular operator with respect to the cyclic and separating vector $\Omega\in\H$, and, with the help of the self-adjoint operator $H\in\Ma$, $H\Omega=0$, define the domain
$$
\mathbf{D}(\delta):=\big\{A\in\Ma|\;i[H,A]\in\Ma\big\}.
$$ 
Then the following two conditions are equivalent:

\begin{itemize}
\item[(i)] $e^{itH}\Ma e^{-itH}=\Ma$ for all $t\in\R$. 
\item[(ii)] $\mathbf{D}(\delta)\Omega$ is a core for $H$, and $H$ and $\Delta$ commute strongly, i.e., $\Delta^{it}H\Delta^{-it}=H$ for all $t\in\R$.
\end{itemize}

\end{theo}

One may also ask for the inverse problem in the modular theory, namely investigate the question, how to characterise all von Neumann algebras $\mathfrak{N}$ which are isomorphic to a given von Neumann algebra $\Ma$ and share with $\Ma$ the same cyclic and separating vector $\Omega_\Ma$ and modular objects $\Delta_\Ma$ and $J_\Ma$. Boller and Wollenberg investigate the case of factors \cite{Wollenberg:1998fp}, \cite{Boller:2000}. Let us write for the set of these factors $NF(\Ma,\Omega_\Ma,\Delta_\Ma,J_\Ma)$.
Their result is as follows:

\begin{theo}
Let $\mathcal{F}_{\infty}:=\L\big(L^2(\R,dx)\big)$, $\rho$ a faithful, normal and semifinite state on $\mathcal{F}_{\infty}$, $\H_{\rho}$ the GNS space for $(\mathcal{F}_{\infty},\rho)$, $\pi_{\rho}$ the isomorphism from $\mathcal{F}_{\infty}$ onto its GNS representation, $\mathcal{F}_{\rho}:=\pi_{\rho}(\mathcal{F}_{\infty})$ and $\o_\Ma(\cdot):=(\Omega_\Ma,\cdot\;\Omega_\Ma)$. \\
Then a factor $\mathfrak{N}$ belongs to $NF(\mathfrak{M},\Omega_\mathfrak{M},\Delta_\mathfrak{M},J_\mathfrak{M})$ if there exists a unique operator $U$ with the properties:
\begin{itemize}
\item[(i)] $\Na=U\mathfrak{M} U^*$.
\item[(ii)] There are unitaries $K\in\{\Delta_\mathfrak{M}\bigotimes\Delta_{\rho}\}',Y_1\in\mathfrak{M}\overline{\otimes}\mathcal{F}_{\rho},Y_2\in\mathfrak{M}'\overline{\otimes}\mathcal{F'}_{\rho}$ on $\H\overline{\otimes}\H_{\rho}$ such that
\begin{equation*}
U\otimes\1_{\rho}=KY_1Y_2.
\end{equation*}
\item[(iii)] $\o_\Ma\otimes\rho_1(\cdot)=c(\o_\Ma\otimes\rho_1)(K\cdot K^*)$ with $c>0$.

\end{itemize}

\end{theo}

One of our approaches for the derivation of the modular group with respect to the massive algebra of local observables will be to formulate, in a first step, the modular group on the massive algebra with respect to the `wrong' massless vacuum state, and then, in a final step, bridge the gap to the modular group with respect to the `right' massive vacuum state. The modular theory makes some tools available for this purpose.

\begin{theo}
For the modular groups $\sigma_{\o_1}^t$ and $\sigma_{\o_2}^t$ on the von Neumann algebra $\Ma$ such that respect to the faithful normal states $\o_1$ and $\o_2$ the following statements are equivalent:

\begin{itemize}
\item[(i)] $\o_1$ is $\sigma_{\o_2}^t$-invariant. 
\item[(ii)] $\o_2$ is $\sigma_{\o_1}^t$-invariant. 
\item[(iii)] $\sigma_{\o_1}^t$ and $\sigma_{\o_2}^t$ commute. 
\item[(iv)] There exists a unique positive injective operator $B$ affiliated with $\Ma^{\sigma_{\o_1}}\cap\Ma^{\sigma_{\o_2}}$, i.e., $B$ commutes with all unitaries from $\Ma^{\sigma_{\o_1}}\cap\Ma^{\sigma_{\o_2}}$, with $\o_2(A)=\o_1(BA)$ for all $A\in\Ma$.
\end{itemize}
\end{theo}

\begin{lem}\label{quasi-equivalence}
Let $\mathfrak{M}$ and $\mathfrak{N}$ be two von Neumann algebras and $\alpha:\mathfrak{M}\longrightarrow\mathfrak{N}$ an isomorphism. If $\o$ is a faithful, normal and semifinite state on $\mathfrak{N}$, then
\begin{equation*}
\sigma_{\o\circ\alpha}^t=\alpha^{-1}\circ\sigma_{\o}^t\circ\alpha
\end{equation*}
holds.
\end{lem}
If there exists more than one faithful, normal and semifinite state on $\Ma$, it can be shown that the modular automorphism is unique up to unitaries. Let $\o_1$ and $\o_2$ be two faithful, normal and semifinite states on $\Ma$ and $\sigma^t_{\o_1}$ and $\sigma^t_{\o_2}$ the corresponding modular groups. Let us consider the faithful and normal weight 
\begin{equation*}
\rho\left[\begin{pmatrix}A_{11} & A_{12}\\A_{21} & A_{22}\end{pmatrix}\right]:=\frac{1}{2}\big(\o_1(A_{11})+\o_2(A_{22})\big)
\end{equation*}
on $\Ma\otimes M_2$ and the corresponding modular group $\sigma^t_{\rho}$, then the unitaries $\Gamma_t$, defined as 
\begin{equation*}
\begin{pmatrix}0 & \Gamma_{t}\\ 0 & 0\end{pmatrix}:=\sigma_{\rho}^t\left[\begin{pmatrix} 0 & \1\\ 0 & 0\end{pmatrix}\right],
\end{equation*}
connect the two original modular groups. The next theorem is dealing exactly with this situation, but first we have to introduce the following notions. We denote by $\A(\mathbf{D})$ the set of bounded and holomorphic functions on the domain $\mathbf{D}$ and define
$$
\mathrm{n}_\o:=\big\{A\in\Ma|\;\o(A^*A)<\infty\big\}.
$$

\begin{theo}[Connes' Cocycle Derivative]\label{Connes-theorem}
If  $\o_1$, $\o_2$ are two faithful, normal and semifinite states on $\mathfrak{M}$, then there exists a $\sigma$-strongly continuous one-parameter family of unitaries $\big(\Gamma_{t}\big)_{t\in\R}$ in $\mathfrak{M}$ with the following properties:

\begin{itemize}
\item [(i)] $\Gamma_{s+t}=\Gamma_{s}\sigma_{\o_1}^s(\Gamma_{t}),\quad s,t\in\R$.
\item [(ii)] $\Gamma_{s}\sigma_{\o_1}^s(\mathrm{n}_{\o_1}\cap \mathrm{n}^*_{\o_2})=\mathrm{n}_{\o_1}\cap \mathrm{n}^*_{\o_2}$.
\item [(iii)] For every $x\in\mathrm{n}^*_{\o_1}\cap \mathrm{n}_{\o_2}$ and $y\in\mathrm{n}_{\o_1}\cap \mathrm{n}^*_{\o_2}$ there exists an element $F\in\A(\mathbf{D})$ such that
\begin{equation*}
F(t)=\o_2\big(\Gamma_{t}\sigma_{\o_1}^t(y)x\big)\quad\text{and}\quad F(t+i)=\o_1\big(x\Gamma_{t}\sigma_{\o_1}^t(y)\big).
\end{equation*}
\item [(iv)]$\sigma_{\o_2}^t(A)=\Gamma_{t}\sigma_{\o_1}^t(A)\Gamma_{t}^*,\quad A\in\mathfrak{M},\;t\in\R$.
\end{itemize} 
The third condition determines the family of unitaries uniquely.

\end{theo}
The main statement of this theorem is that two arbitrary modular automorphism groups are equivalent up to inner automorphisms. One of our approaches will be based mainly on the cocycle theorem. The specific problem with its statement is that it assures the existence of the unitaries only, but fails to give a method for their construction.

The above theorem still holds true with few adjustments if the faithfulness is given only for one of the weights. The family of unitaries is called cocycle derivative of $\o_1$ with respect to $\o_2$ and denoted by $(D_{\o_2}:D_{\o_1})_t:=\Gamma_t$. For two faithful, normal and semifinite states we have 
$$
(D_{\o_2}:D_{\o_1})_t=(D_{\o_2}:D_{\o_2})_t^{-1},\quad t\in\R.
$$
If an additional third faithful, normal and semifinite state on $\Ma$ is given, then one can establish the chain rule,
\begin{equation*}
(D_{\o_1}:D_{\o_2})_t=(D_{\o_1}:D_{\o_3})_t(D_{\o_3}:D_{\o_2})_t, \quad t\in\R.
\end{equation*}
Consequently, the equivalence of two cocycle derivatives of $\o_1$ with respect to $\o_2$ and $\o_3$ uniquely determines the identity of  $\o_2$ and $\o_3$:
$$
(D_{\o_1}:D_{\o_2})_t=(D_{\o_1}:D_{\o_3})_t\quad\forall t\in\R\quad\Longleftrightarrow\quad\o_2=\o_3.
$$
On the other hand, the knowledge of a modular group $\sigma_{\o_1}^t$ and a family of unitaries is sufficient for the existence of another faithful, normal and semifinite state, namely:

\begin{theo}
For a faithful, normal and semifinite state $\o_1$ on $\mathfrak{M}$ and a $\sigma$-strongly continuous one parameter family of unitaries $\Gamma_{t}$ in $\mathfrak{M}$, satisfying the cocycle identity 
\begin{equation*}
\Gamma_{s+t}=\Gamma_{s}\sigma_{\o_1}^t(\Gamma_{t}),
\end{equation*}
there exists a second faithful, normal and semifinite state $\o_2$ on $\mathfrak{M}$ such that
\begin{equation*}
(D_{\o_2}:D_{\o_1})_t:=\Gamma_t,\quad t\in\R.
\end{equation*}
\end{theo}
In some special cases more information about the cocycle derivative is known which could lead to its explicit derivation. For the next theorem we will need the notion of the centralizer of the state $\o$ on the von Neumann algebra $\Ma$ defined as
$$
\Ma^\o:=\big\{A\in\Ma|\;\sigma_\o^t(A)=A\text{ for all } t\in\R\big\}
$$

\begin{theo}[Pedersen, Takesaki]
Let $\o_1$ and $\o_2$ be two faithful, normal and semifinite states on the von Neumann algebra $\Ma$, then the following statements are equivalent:

\begin{itemize}
\item[(i)] $\o_2\circ\sigma_{\o_1}^t=\o_2\quad\forall t\in\R$.
\item[(ii)] $(D_{\o_2}:D_{\o_1})_t\in\Ma^{\o_2}\quad\forall t\in\R$.
\item[(iii)] $(D_{\o_2}:D_{\o_1})_t\in\Ma^{\o_1}\quad\forall t\in\R$.
\item[(iv)] $\big\{(D_{\o_2}:D_{\o_1})_t\big\}_{t\in\R}$ is a strongly continuous group of unitary elements of $\Ma$.
\item[(v)] There exists a positive and self-adjoint operator $A$ affiliated with $\Ma^{\o_1}$, i.e., $A$ commutes with all unitary elements of $\Ma^{\o_i}$, such that $\o_2(B)=\o_1(A^{1/2}BA^{1/2})$ holds for all $B\in\Ma_+$.
\end{itemize}
\end{theo}

One of the most important mathematical applications of the Tomita-Takesaki theory is the classification of factors. If for a fixed $t_0\in\R$ and a particular faithful, normal and semifinite state $\o$ its modular group $\sigma^{t_0}_{\o}$ is inner, then, because of the cocycle theorem, $\sigma_{\o'}^{t_0}$ is inner for any faithful, normal and semifinite weight $\o'$. Thus the modular period group 
\begin{equation*}
T(\mathfrak{M}):=\big\{t\in\R|\;\sigma^{t}_{\o} \text{ is inner}\big\}
\end{equation*}
characterises the von Neumann algebra. Let $\o$ be an arbitrary faithful and semifinite state, then $T(\Ma)$ is related to the so-called modular spectrum of $\Ma$, 
\begin{equation}\label{modularspectrum}
S(\mathfrak{M}):=\underset{\o}{\bigcap}\;{\bf Spec}\Delta_{\o},
\end{equation}
via the inclusion
$$
\ln\big(S(\Ma)\backslash\{0\}\big)\subset\big\{s\in\R|\;e^{ist}=1\quad\forall t\in T(\Ma)\big\}.
$$
This modular spectrum can be used for the classification.

\begin{theo}\label{type}
Let $\Ma$ be a factor. Then the following statements hold:

\begin{itemize}
\item[(i)] $\mathfrak{M}$ is of type $I$ or type $II$, if $S(\mathfrak{M})=\{1\}$;
\item[(ii)] $\mathfrak{M}$ is of type $III_0$, if $S(\mathfrak{M})=\{0,1\}$;
\item[(iii)] $\mathfrak{M}$ is of type $III_\lambda$, if $S(\mathfrak{M})=\{0\}\cup\{\lambda^{n}|\;0<\lambda<1,n\in \mathbb{Z}\}$;
\item[(iv)] $\mathfrak{M}$ is of type $III_1$, if $S(\mathfrak{M})=\R_+$.
\end{itemize}

\end{theo}
We want to have a closer look at the factors of type $III$, in particular at those of  type $III_1$. While the structure of factors of type $I_n$ is well understood, as isomorphic to the algebra $M_n(\C)$ for finite $n$ and to $\L(l^2)$ for infinite $n$, the detailed classification for general factors is not known. There is one special class of factors for which one can extract more details about their structure.

\begin{defi}\label{AFD}
A separable von Neumann algebra $\Ma$ is said to be hyperfinite or approximately finite-dimensional, abbreviated AFD, if for any $A_1, A_2,\cdots,A_n\in\mathfrak{M}$ and any $\sigma$-strong neighborhood $\mathfrak{U}$ of $0$ in $\mathfrak{M}$ there exists a finite-dimensional *-subalgebra $\mathfrak{N}$ with
\begin{equation*}
A_i\in\mathfrak{N}+\mathfrak{U},\quad i=1,2,\cdots,n.
\end{equation*}

\end{defi} 
In the algebraic setting of quantum field theory factors of type $III_1$ play a significant role, because the algebras of local observables are of this form. Haagerup has even shown that they are AFD \cite{Haagerup:1987}.

It should be mentioned, that subalgebras of AFD von Neumann algebras are not automatically AFD, whereas the $W^*$-inductive limit of increasing sequences of AFD von Neumann algebras is again AFD.

Because every factor of type $I$ is approximately finite-dimensional, its structure could be already analysed. Factors of type $II_1$ are AFD if and only if they are isomorphic to the infinite tensor product of $M_2(\C)$, defined as the inductive limit
\begin{align*}
\prod_{n=1}^{\infty}{}^{\otimes} M_2(\C)_n:&=\lim_{\longrightarrow}\big\{\overline{M_2(\C)},\pi_n\big\}\\
&=\lim_{\longrightarrow}\big\{M_2(\C)_1\otimes_\text{min} M_2(\C)_2\otimes_\text{min}\cdots\otimes_\text{min} M_2(\C)_n,\pi_n\big\},
\end{align*}
where $\pi_n$ is the isomorphism $\overline{M_2(\C)}_n\ni X\mapsto X\otimes\1\in\overline{M_2(\C)}_{n+1}$. For $0<\lambda<1$ let
\begin{align*}
\o_n\left[\begin{pmatrix}A_{11} & A_{12}\\A_{21} & A_{22}\end{pmatrix}\right]:=\frac{\lambda^{-1/2}}{\lambda^{-1/2}+\lambda^{1/2}}A_{11}+\frac{\lambda^{1/2}}{\lambda^{-1/2}+\lambda^{1/2}}A_{22}
\end{align*} 
be a state on $M_2(\C)$. Then the von Neumann algebra
\begin{align*}
\mathcal{R}_{\lambda}:=\prod_{n=1}^{\infty}{}^{\otimes}\{M_2(\C)_n,\o_n\},\quad\text{where}\quad \o_{\lambda}:=\prod_{n=1}^{\infty}{}^{\otimes}\o_n
\end{align*} 
is the product state on $\prod_{n=1}^{\infty}{}^{\otimes} M_2(\C)_n$, 
\begin{gather*}
\o\big(A_1\otimes A_2\cdots\otimes\1\otimes\1\otimes\cdots\cdots\big)=\prod_{n=1}^\infty\o_n(A_n),
\end{gather*}
is an AFD factor of type $III_{\lambda}$, and one can prove even more.

\begin{theo}
Every AFD factor of type $III_{\lambda}$ is isomorphic to $\mathcal{R}_{\lambda}$.
\end{theo}

\begin{defi}
Let $\pi_\alpha$ and $\lambda$ be the representations of the von Neumann algebra $\Ma$ and the locally compact group $\Gr$, respectively. Then the von Neumann algebra generated by $\pi_\alpha(\Ma)$ and $\lambda(\Gr)$ with
\begin{gather*}
\big(\pi_\alpha(A)f\big)(s):=\alpha_s^{-1}(A)f(s)\quad\text{and}\\
\big(\lambda(t)f\big)(s):=f(t^{-1}s),\qquad A\in\Ma,\;f\in L^2(\Gr,\H),\;s,t\in\Gr,
\end{gather*}
is called the crossed product of $\Ma$ and $\alpha$ and denoted by $\Ma\rtimes_{\alpha}\Gr$.
\end{defi}
For a factor of type $III_0$ the following statement holds.

\begin{theo}
Every separable AFD factor of type $III_0$ is of the form $\Ma\rtimes_{\alpha}\Z$, where $\alpha$ is an arbitrary ergodic automorphism of an abelian and separable von Neumann algebra $\Ma$.
\end{theo}

\begin{theo}\label{aut-int}The following statements hold:
\begin{itemize}
\item[(i)] All AFD factors of type $III_1$ are mutually isomorphic.
\item[(ii)] If $\mathfrak{M}_{\lambda}$ and $\mathfrak{M}_{\mu}$ are AFD factors of type $III_{\lambda}$ and $III_{\mu}$, $0<\lambda,\mu<1 $ respectively, then $\mathfrak{M}_{\lambda}\overline{\bigotimes}\mathfrak{M}_{\mu}$ is an AFD factor of type $III_1$.
\item[(iii)] If $\Ma$ is an AFD factor of type $III_1$, then $\mathbf{Aut}(\mathfrak{M})=\overline{\mathbf{Int}(\mathfrak{M})}$. 
\end{itemize}
\end{theo}

For AFD factors of type $III_1$ we may drop the condition of the state being semifinite in order to obtain at least an approximately inner modular group, see Definition \ref{innerautomorphism}.

\begin{theo}
If $\o$ is an arbitrary faithful normal state on $\Ma$ then, due to $(iii)$ of Theorem \ref{aut-int}, the modular group of automorphisms $\sigma_\o^t$, $t\in\R$, is approximately inner.
\end{theo}

We close this section with the groundbreaking result of Jones on the structure of subfactors of type $II_1$ \cite{Jones:1983kv}, which has many innovative applications in both disciplines, mathematics and physics. Jones' result is useful particularly in algebraic quantum field theory as this formulation is based on von Neumann algebras. Two prominent examples are the classification of conformal local nets for $c<1$ by Kawahigashi and Longo \cite{Kawahigashi:2002px} and the investigation of Longo on spin and statistics \cite{Longo:1989tt}. Both subjects will be discussed later, see Theorem \ref{Kawahigashi-Longo} and Theorem \ref{spin-statistics}, respectively.

Let us consider a factor $\Ma$ of type $II_1$ and a subfactor $\Na\subset\Ma$, then the (global) index of $\Na$ in $\Ma$, the so-called Jones index, is defined by means of the dimension of $\H$ relative to $\Ma$ and relative to $\Na$, $\dim_\Ma(\H)$ and $\dim_\Na(\H)$, respectively, as
\begin{gather*}
[\Ma:\Na]:=\frac{\dim_\Na(\H)}{\dim_\Ma(\H)}=\dim_\Na L^2(\Ma),
\end{gather*}
whenever $\Ma$ is represented on $\H$ with finite commutant $\Ma'$. 

\begin{theo}[Jones]\label{Jones}
Let $\Ma$ be a factor of type $II_1$ and $\Na$ a subfactor of $\Ma$, then for the index of $\Na$ in $\Ma$ one has:
\begin{gather*}
[\Ma:\Na]\in\Big\{4\cos^2\frac{\pi}{n}|\;n\in\N,n\geq3\Big\}\cup\big[4,\infty\big].
\end{gather*}
Moreover, each number in this union of sets is realised as an index for some subfactor.
\end{theo}
This result has been a complete surprise. Since one deals here with continuous-dimensional objects (von Neumann algebras), everyone had expected the index to be allowed to take each value from the interval $[1,\infty]$.
 
Kosaki extends Theorem \ref{Jones} to arbitrary factors \cite{Kosaki:1986}. He formulates a generalised version of the index as
\begin{gather*}
\text{Index}\,E:=E^{-1}(\1),
\end{gather*}
where $E$ is an operator-valued weight from $\Ma$ to $\Na$, i.e., a mapping 
\begin{gather*}
E:\:\Ma_+\longrightarrow\widehat{\Na_+},
\end{gather*}
where $\widehat{\Na_+}$ denotes the extended positive cone of $\Na$, the set of all lower semi-continuous weights $m:\Na_{+,*}\longrightarrow[0,\infty]$. In the case of type $II_1$ factors $\Ma$, $[\Ma:\Na]$ can be shown to coincide with $\text{Index}\,E$, for which Kosaki proves the statement of Theorem \ref{Jones}.

\section{The Algebraic Approach to Quantum Field Theory}

The usual formulation of quantum field theory is based on the representation of states as unit rays in a Hilbert space, with observables as operators acting on them. The algebraic ansatz proceeds in the opposite direction. There, the starting point are observables as elements of an abstract $\*$-algebra, on which the states are introduced as normalised, positive and linear functionals. For more details we refer the reader to the main source \cite{Haag:1992hx}, but also to \cite{Araki:1999ar} and \cite{Horuzhy:1986}.

First one constructs a net of $C^*$-algebras $\{\Aa(\O)\}_{\O\subset\M}$, the so-called local algebras, i.e., to each open subset $\O$ of the spacetime $\M$ a $C^*$-algebra $\Aa(\O)$ is assigned that represents physical quantities to be measured in $\O$,
\begin{equation*}
\O\mapsto\Aa(\O).
\end{equation*}
Since all physical information is assumed to be encoded in this mapping, its knowledge allows one in principle to extract all kinds of physical data. The $C^{*}$-Algebra $\Aa:=\overline{\bigcup_{\O}\Aa(\O)}$, i.e., the $C^*$-inductive limit of the net $\{\Aa(\O)\}_{\O\subset\M}$, is called the quasi-local algebra of observables and the bicommutant $\mathfrak{A}''$ of $\Aa$ the global algebra of observables. Two nets of local observables, $\Aa(\O)$ and $\tilde{\Aa}(\O)$, respectively, are said to be mutually isomorphic if there is an isomorphism $i:\Aa\rightarrow\tilde{\Aa}$  with  $i[\Aa(\O)]=\tilde{\Aa}(\O)$. In addition the net is required to satisfy the following conditions:

\begin{itemize}
\item[(i)] Isotony: $O_1\subset\mathcal{O}_{2}\;\Longrightarrow\;\Aa(\O_1)\subset\Aa(\O_2)$.    
\item[(ii)] Locality: $[\Aa(\O_1),\Aa(\O_2)]=\{0\}$ for spacelike separated regions $\O_1$ and $\O_2$.
\item[(iii)] Additivity: $\O=\underset{i}{\bigcup}\O_i\;\Longrightarrow\;\Aa(\O)=\left(\underset{i}{\bigcup}\Aa(\O_i)\right)''$.
%\item[(iv)] Causality: $\O_1\subset \O_2^c\;\Longrightarrow\;\Aa(\O_1)\subseteq\Aa(\O_2)'$, where $\O_2^c$ denotes the causal complement of $\O_2$.
\item[(iv)] Covariance: There is a strongly continuous unitary representation $U(\P)$ of the Poinc\'are group $\P$ with
\begin{equation}
\Aa(g\O)=U(g)\Aa(\O)U(g)^{-1},\quad g\in\P.
\end{equation}
\item[(v)] Spectrum condition: $\mathbf{Spec}\,U(g)\subseteq\V_+$.
\item[(vi)] Vacuum sector: There exists a vector $\Omega\in\H,\;\|\Omega\|=1$, such that $U(g)\Omega=\Omega,\;g\in\P$, and $\left(\underset{\O\subset\M}{\bigcup}\Aa(\O)\right)\Omega$ is dense in $\H$.
\end{itemize}
If  $\M=\mathbb{M}$ is the Minkowski space, then the isometries turn out to be the Poincar\'e transformations, and the last condition becomes the Poincar\'e covariance and cyclicity of the vacuum vector $\Omega$. The former four axioms have been proposed by Haag and Kastler, who were the first to formulate quantum field theory on Minkowski space in the algebraic framework \cite{Haag:1963dh}.

A state on the observable algebra $\Aa$ is represented by a linear functional $\o :\Aa\longrightarrow\C$, which is normalised, i.e., $\omega(\1)=1$, and positive, i.e. $\o(A^*A)\geq 0$ for all $A\in\Aa$. Thus the state takes on the role of the expectation value functional. Unfortunately, the set of states introduced in this way is too large, and by far not all of them make sense physically. Even the restriction to quasifree states, see \eqref{quasifreestate} below, is from a physical point of view not sufficient, because there still exist quasifree states with an unbounded expectation value for the energy-momentum tensor. One of the main procedures for the extraction of physically relevant states is the requirement of  the Hadamard form proposed by Brehme and de Witt \cite{DeWitt:1960fc}. A more fruitful equivalent characterisation of Hadamard states is given by Radzikowski in the context of microlocal analysis \cite{Radzikowski:1996pa}, namely in terms of the wave front set of the two-point function, see Proposition \ref{wavefrontset} and \eqref{npointfunction}.

The `usual' and the algebraic formulation of quantum field theory can be connected via the GNS representation.

The Hilbert space $\H$ occurring in the last axiom (vii) can be written as a direct sum of subspaces $\H_i$, the so-called coherent subspaces or superselection sectors. These subspaces are defined as subsets, wherein the superposition principle is valid without restriction. The local net containing observable fields determined by the causality assumption is operating only inside these sectors, whereas the unobservable fields `communicate' between different sectors. They may be interpreted as charge-carrying fields transporting some kind of quantum quantities from one sector to another. The quasi-equivalent nets of local observables $\Aa_i|_{\H_i}$ restricted to the superselection sectors $\H_i$ still do include all physically relevant information, and superselection rules are needed only if the observable algebra $\Aa$ has inequivalent representations on a Hilbert space. For some investigations of inner symmetries it is more convenient to extend the local nets to the so-called field algebra, which in addition to the observable fields also contains the unobservable ones. 

The net of von Neumann algebras $\{\Fa(\O)\}_{\O\subset\M}$ on a Hilbert space $\H$, where $\O$ is a double cone, consists of the so-called local field algebras if they satisfy the following assumptions:

\begin{itemize}
\item[(i)] Irreducibility, i.e., $\bigcap_\O F(\O)'=\C\1\;$ for all $F\in\Fa(\O)$.
\item[(ii)] There exists a strongly continuous and unitary representation $U$ of the covering group of $\P_+^\uparrow$ in $\H$ such that 
$$
U(g)\Fa(\O)U(g)^{-1}=\Fa(g\O)\quad\text{for all}\quad g\in\P_+^\uparrow.
$$
The infinitesimal generators $P_\mu$ of the translations fulfill the spectrum condition, i.e., $\spec P_\mu\subset\V_+$. Furthermore, there exists a Poincar\'e-invariant vector $\Omega\in\H$, the vacuum vector, which is uniquely determined up to a phase factor.
\item[(iii)] There is a strongly continuous, faithful and unitary representation $U$ of a compact group $\mathcal{G}$, the so-called global gauge group, in $\H$ such that
\begin{gather*}
U(g)\Fa(\O)U(g)^{-1}=\Fa(\O),\quad U(g)\Omega=\Omega,\quad\text{and}\\
U(g)U(g')=U(g')U(g)\quad\text{for all }\quad g\in\mathcal{G},\,g'\in\P_+^\uparrow.
\end{gather*} 
\item[(iv)] There exists an element $g\in\mathcal{G}$ with $g^2=1$ such that for spacelike separated regions $\O_1$ and $\O_2$ and for elements $F_\pm^{(i)}:=F_\pm(\O_i)\in\Fa(\O)$, $i=1,2$, the relations
\begin{gather}\label{gradedlocality}
\alpha_g\Big(F_\pm^{(i)}\Big):=U(g)F_\pm^{(i)}U(g)^{-1}=\pm F_\pm^{(i)}\quad\text{and}\notag\\
\Big[F_+^{(1)},F_+^{(2)}\Big]=\Big[F_+^{(1)},F_-^{(2)}\Big]=\Big[F_-^{(1)},F_-^{(2)}\Big]_+=0\quad\text{(graded locality)}
\end{gather}
hold.
\item[(v)] Additivity:  $\O=\underset{i}{\bigcup}\O_i\;\Longrightarrow\;\Fa(\O)=\left(\underset{i}{\bigcup}\Fa(\O_i)\right)''$.
\item[(vi)] Haag duality ($\mathcal{G}$-invariant): $\left(\underset{\O\subset\O^c}{\bigcup}\Fa(\O)\cap U(\mathcal{G})'\right)''=\Fa(\O)'\cap U(\mathcal{G})'$.
\end{itemize}
The closure of the union of the local field algebras is called the field algebra,
$$
\Fa:=\overline{\bigcup_\O \Fa(\O)}.
$$
The observable fields can now be introduced as gauge-invariant elements of the field algebra $\Fa$:
\begin{align*}
\Aa(\O)&=\Big\{A\in\Fa(\O)|\;\alpha_g(A)=A\text{ for all }g\in\mathcal{G}\Big\}\\
&=\Fa(\O)\cap U(\mathcal{G})'.
\end{align*}
The opposite direction is also ensured. Via the Doplicher-Roberts reconstruction procedure one is able to show that the field algebra $\Fa$ together with the gauge group $\mathcal{G}$ are uniquely characterized by the local algebra of observables $\Aa$ under physically well-motivated assumptions .

\subsection{The Free Klein-Gordon Field}

In this thesis we are dealing only with the simplest case of a quantum field theory, i.e., the free scalar field $\varphi[f]$ satisfying the Klein-Gordon equation
\begin{equation*}
\varphi\Big(\left(\Box_g+m^{2}\right)f\Big)=0,
\end{equation*}
where $\Box_g:=|g|^{-1/2}\partial_\mu g^{\mu\nu}|g|^{1/2}\partial_\nu$ is the d'Alembert operator for which the Cauchy problem is well-posed \cite{Dimock:1980}.

\begin{defi}
The submanifold $\Sigma$ of a Lorentz manifold $\M$ is said to be a Cauchy surface, if it owns only spacelike tangent spaces and if each non-extendible causal curve meets $\Sigma$ exactly once.  
\end{defi}
If $\M$ is the Minkowski spacetime $\mathbb{M}$, then every subspace with $t=\const$ is a Cauchy surface. Lorentz manifolds possessing a Cauchy surface $\Sigma$ are diffeomorphic to the manifold $\R\times\Sigma$. 
\begin{theo}[Cauchy Problem]
Let $\Sigma$ be a Cauchy surface and $u_{0},u_{1}\in\mathcal{C}_{0}^{\infty}(\Sigma)$, then there exists a uniquely defined function $u\in\mathcal{C}^{\infty}(\mathcal{M})$ such that 
\begin{gather*}
\left(\Box_g+m^{2}\right)u=0,\quad\rho_{0}(u)=u_{0},\quad\rho_{1}(u)=u_{1},\\
\text{and}\qquad\supp(u)\subset\bigcup_{i}\bigcup_{\pm}J^{\pm}\big(\supp(u_{i})\big), 
\end{gather*}
where $\rho_{0}(u):\mathcal{C}^{\infty}(\mathcal{M})\rightarrow\mathcal{C}^{\infty}(\Sigma)$ is the restriction operator and $\rho_{1}(u):\mathcal{C}^{\infty}(\mathcal{M})\rightarrow\mathcal{C}^{\infty}(\Sigma)$ the normal derivation on $\Sigma$.
\end{theo}
Therefore, for each test function $f\in\mathcal{D}(\mathcal{M})$ there exist the advanced and retarded fundamental solutions 
\begin{gather*}
E_{av/ret}:\mathcal{D}(\mathcal{M})\longrightarrow\mathcal{E}(\mathcal{M})\notag
%E_{av/ret}:(f,g)\mapsto E_{av/ret}(f,g):=\int f(x)(E_{av/ret}g)(x)g^{\frac{1}{2}}d^{4}x
\end{gather*}
with the properties
\begin{gather*}
\left(\Box_g+m^{2}\right)E_{av/ret}f=E_{av/ret}\left(\Box_g+m^{2}\right)f=\text{id},\\
\supp(E_{av}f)\subset J^{+}\big(\supp(f)\big)\quad\text{and}\quad \supp(E_{ret}f)\subset J^{-}\big(\supp(f)\big).
\end{gather*}
Their kernels are distributions on the set of test functions $\D(\M)\times\D(\M)=\D(\M\times\M)$, defined as
\begin{gather*}
E_{av/ret}(f,g):=\int f(x)(E_{av/ret}g)(x)g^{1/2}d^{4}x.
\end{gather*}
The difference $E:=E_{ret}-E_{av}$ is an antisymmetric distribution, called the fundamental solution or the propagator of the Klein-Gordon equation. It satisfies the conditions :
\begin{gather*}
\left(\Box_g+m^{2}\right)Ef=E\left(\Box_g+m^{2}\right)f=0,\quad\text{and}\\
\supp(Ef)\subset J^{+}\big(\supp(f)\big)\cup J^{-}\big(\supp(f)\big),\quad f\in\mathcal{D}(\mathcal{M}).
\end{gather*}
By means of the field equation and the commutation relations on an arbitrary Cauchy surface $\Sigma$, 
\begin{equation*}
\begin{split}
\big[\varphi(x),\varphi(y)\big]&=0,  \\
\big[\varphi(x),\partial_{\Sigma}\phi(y)\big]&=i\delta_{\Sigma}(x,y),\quad\text{and} \\
\big[\partial_{\Sigma}\varphi(x),\partial_{\Sigma}\varphi(y)\big]&=0\qquad\text{for all } x,y\in\Sigma, \\
\end{split}
\end{equation*}
one obtains the commutation relations on the entire manifold $\mathcal{M}\times\mathcal{M}$:
\begin{equation*}
\big[\varphi(x),\varphi(y)\big]=iE(x,y)\qquad\text{for all } x,y\in\M.
\end{equation*}
For the sake of proper definition of the quantum field as an operator, we have to continue our formulation of the theory with the smeared-out fields, 
\begin{equation*}
\varphi[f]=\int\varphi(x)f(x)d^{4}x\quad\text{with}\quad f\in\mathcal{D}(\mathcal{M}).
\end{equation*}
The quantum field will be considered as an operator-valued distribution, i.e., as a linear map $\varphi:\mathcal{D}(\mathcal{M})\longrightarrow\mathcal{L}(\mathcal{H})$. To summarise the above discussion, we are dealing with a unital ${}^{*}$-algebra $\mathfrak{A}$ which is generated by the quantum fields $\varphi[f]$, $f\in\mathcal{D}(\mathcal{M})$, and fulfills the following requirements:

\begin{enumerate}
\item[(i)] The map $\varphi:\mathcal{D}(\mathcal{M})\longrightarrow\mathcal{L}(\mathcal{H})$ is linear.
\item[(ii)] $\varphi[f]^{*}=\varphi[\bar{f}]$.
\item[(iii)] $\varphi\big[\left(\Box_{g}+m^{2}\right)f\big]=0$.
\item[(iv)] $\big[\varphi[f],\varphi[g]\big]=iE(f,g)\quad\forall f,g,\in\mathcal{D}(\mathcal{M})$.
\end{enumerate}  
One obtains the desired algebra by dividing the Borchers algebra over $\mathcal{D}(\mathcal{M})$, i.e., the tensor algebra with the particular ${}^{*}$-operation $f^{*}:=\bar{f}$, by the ideal which is determined by the field equation and the commutation relations. Then the local algebras $\mathfrak{A}(\mathcal{O})$ are generated by the elements $\varphi[f]$, $f\in\mathcal{D}(\mathcal{O})$. Since the commutation relations do not allow any representation by bounded operators, one has to obtain a $C^*$-norm via the Weyl algebra, that is the algebra generated by the elements $e^{i\varphi[f]}$.\\
A state on this $C^*$-algebra $\mathfrak{A}$ is determined uniquely by its $n$-point functions $\omega_{n}$, $n\in\N$,
\begin{equation}\label{npointfunction} 
\omega_{n}(f_{1},\cdots,f_{n}):=\omega\big(\varphi[f_{1}]\cdots\varphi[f_{n}]\big),\quad f_{i}\in\mathcal{D}(\mathcal{M}),
\end{equation}
where $\omega_{n}$ is a distribution with respect to each component. Due to the Schwartz kernel theorem, the  $n$-point functions are well-defined distributions on the entire manifold $\mathcal{M}^{n}$. The relations within the algebra are enforcing the following properties for the $n$-point functions:
\begin{itemize}
\item[(i)] In each component they are solutions of the Klein-Gordon equation. 
\item[(ii)] Because of the commutation relations they satisfy
\begin{gather*}
\omega_{n}(x^{1},\cdots,x^{k},x^{k+1},\cdots,x^{n})-\omega_{n}(x^{1},\cdots,x^{k+1},x^{k},\cdots,x^{n})=\\iE(x^{k},x^{k+1})\omega_{n-2}(x^{1},\cdots,x^{k-1},x^{k+2},\cdots,x^{n}).
\end{gather*}
\item[(iii)] Normalization: $\omega_{0}\equiv1$, due to $\omega(\1)=1$ for general states $\o$.
\item[(iv)] Positivity: 
\begin{gather*}
\sum_{i,j}\omega_{i+j}(f_{i}^{*}\otimes f_{j})\geq 0,\quad\text{with}\quad f_{k}\in\mathcal{D}(\mathcal{M}^{k}),\;k=1,\cdots,n,\\
\text{and}\qquad f_{k}^{*}(x^{1},\cdots,x^{k})=\overline{f_{k}(x^{1},\cdots,x^{k})}.
\end{gather*}
\end{itemize}
We will be concerned only with quasifree states, whose $n$-point functions with odd $n$ vanish, while those with even $n$ are characterised uniquely by their two-point function $\Lambda\equiv\omega_{2}$:
\begin{equation}\label{quasifreestate}
\omega_{2n}(f_{1},\cdots,f_{2n})=\sum_{\sigma}\prod_{i=1}^{n}\Lambda\big(f_{\sigma(i)},f_{\sigma(i+n)}\big),\quad n\in\N.
\end{equation} 
Here the sum extends over all permutations $\sigma$ of $\{1,\cdots,2n\}$ with $\sigma(1)<\sigma(2)<\cdots<\sigma(n)$ and $\sigma(i)<\sigma(i+n)$.

We state the properties of the 2-point function in a nutshell:

\begin{itemize}
\item[(i)] $\Lambda\Big(\left(\Box_g+m^{2}\right)f,h\Big)=\Lambda\Big(f,\left(\Box_g+m^{2}\right)h\Big)=0\quad\forall f,h\in\D(\mathcal{M})$.
\item[(ii)] $\Lambda(f,h)=\overline{\Lambda(h,f)}$.
\item[(iii)] $\text{Im}\Lambda(f,h)=\frac{1}{2}E(f,h)$.
\item[(iv)]$ \Lambda(\bar{f},f)\geq0$.
\end{itemize}

\section{Type of Local Algebras and KMS States}\label{localgKMS}

The starting point for the application of modular theory to the algebraic formulation of quantum field theory, i.e., the existence of a cyclic and separating vector for a von Neumann algebra of local observables, has been established by Reeh and Schlieder \cite{Reeh-Schlieder:1961}. Their result is a direct consequence of the axioms for the nets of local observables and reads as follows.

\begin{theo}[Reeh-Schlieder]\label{Reeh-Schlieder}
The vacuum vector $\Omega\in\H$ is cyclic and separating for the polynomial algebra $\mathfrak{P}(\O)$ generated by the operators 
$$
A:=\sum_i\int f_i(x_1,x_2,\cdots,x_i)\varphi(x_1)\varphi(x_2)\cdots\varphi(x_i)\prod d^4x_i
$$
with $\supp f_i\subset\O$, provided the causal complement $\O^c$ of $\O$ contains an open region.
\end{theo}

\sproof First, one shows the cyclic property of the vector $\Omega$. Let us consider for this purpose the matrix elements,
\begin{align*}
F(x_1,x_2,\cdots,x_n)&=\big(\psi,\varphi(x_1)\varphi(x_2)\cdots\varphi(x_n)\Omega\big)\\
&=\Big(\psi,e^{iPx_1}\varphi(0)e^{iP(x_2-x_1)}\varphi(0)\cdots e^{iP(x_n-x_{n-1})}\varphi(0)\Omega\Big),
\end{align*}
where $P$ is an arbitrary polynomial of field operators, which are boundary values of analytic functions $F(z)$ defined by
\begin{gather*}
F(z_1,z_2,\cdots,z_n):=\Big(\psi,e^{iPz_1}\varphi(0)e^{iPz_2}\varphi(0)\cdots e^{iPz_n}\varphi(0)\Omega\Big).
\end{gather*}
The analyticity domain of $F$ is the tube,
\begin{gather*}
 z_1:=x_1+i\eta_1,z_2:=x_2-x_1+i\eta_2\;\cdots\;, z_n:=x_n-x_{n-1}+i\eta_n,
\end{gather*}
where the real parts are arbitrary, but all imaginary parts $\eta_j$ are contained in the forward light cone $\V_+$. Now, choose a vector $\psi$ from the orthogonal complement of $\mathfrak{P}(\O)\Omega$, i.e., $\psi$ is orthogonal to all vectors $\varphi(x_1)\varphi(x_2)\cdots\varphi(x_n)\Omega$ with $x_j$ lying in $\O$. One has $F(z_1,z_2,\cdots,z_n)=0$ inside $\O$ and therefore, due to the edge of the wedge Theorem \ref{Edge of the wedge}, $F$ vanishes for all $x$ in the Minkowski space. Consequently, $\psi$ has to be orthogonal to the whole Wightman domain. But since this domain is dense one gets $\psi=0$, i.e., the orthogonal complement of $\mathfrak{P}(\O)\Omega$ is trivial, and therefore $\mathfrak{P}(\O)\Omega$ is a dense subspace of the Hilbert space $\H$ and $\Omega$ a cyclic vector. 

In order to prove the separability property for $\Omega$, let us consider an operator $P$ such that $P\Omega=0$. For an arbitrary polynomial $\tilde{P}\in\mathfrak{P}(\O)$ we obtain $\tilde{P}P\Omega=0$, and, because of locality, $P\tilde{P}\Omega=0$. But since $\mathfrak{P}(\O)\Omega$ is a dense subset of $\H$, $P=0$ identically and hence $\Omega$ is also a separable vector. For more details of the proof confer \cite{Haag:1992hx}.

\qed

\hspace*{-0.65cm}It is worth mentioning that the vacuum vector in the Reeh-Schlieder theorem can be replaced by an arbitrary element of the Hilbert space with bounded energy.

\subsection{Type of Local Algebras}

The quest for the type of the local algebra of observables has its root in the early sixties. Since in quantum theory a factor of type $I$ can be associated to each system, from the very beginning of algebraic quantum theory the question arose, how far the Haag-Kastler axioms determine the type of the underlying von Neumann algebra.

The first result is due to Kadison \cite{Kadison:1963} and Guenin and Misra \cite{Guenin-Misra:1963} who prove that the local algebras cannot be of finite type. Under general assumptions Borchers shows in \cite{Borchers:1967} that if $\O_1$ is contained in $\O$ then every non-zero projection of the local algebra $\mathfrak{M}(\O_1)$ is equivalent to $\1$ in any $\mathfrak{M}(\O)$.

\begin{theo}[Borchers]
For the von Neumann algebras of local observables $\Ma(\O_i)$, $i=1,2$, the following statements hold:
\begin{itemize}
\item[(i)] Let $\O_1\subset\O_2$ such that there is $\O_3\subset(\O_2\bigcap\O_1')$. If $E$ is a projection in $\Ma(\O_1)$, then $E$ is equivalent to its central support in $\mathfrak{M}(\O_2)$, mod $\mathfrak{M}(\O_2)$.
\item[(ii)] If $\O_1+x\subset\O_2$ for $x$ in some open neighborhood of $\R^d$, then the central support of $E$ in $\mathfrak{M}(\O_2)$ belongs to the center of the global algebra.
\end{itemize}
\end{theo}  
To get a richer structure one has to add additional assumptions. Araki gives a more precise characterisation in showing, by explicit calculation, that the von Neumann algebras associated with free fields are factors of type $III$ \cite{Araki:1964}. Araki an Woods conjecture in \cite{Araki-Woods:1963} that for local subspaces $\spec\big(4\delta(\delta-\1)^{-2}\big)$, where $\delta$ is the infinitesimal generator of the modular operator $\Delta$, contains continuous parts so that the local algebra is therefore of type $III_1$. For special nonlocal algebras Sto\hspace{-0.17cm}{\scriptsize/}rmer proves the type $III$ structure in \cite{Stormer:1973es}, and Driessler establishes the type $III_1$ form in \cite{Driessler:1975cm}. In an other publication he succeeds in proving the following sufficient condition for von Neumann algebras to be of type $III$ \cite{Driessler:1976ky}.

\begin{theo}
Let $\H$ be a separable Hilbert space and $\Ma\subset\L(\H)$ a von Neumann algebra with separating vector $\Omega\in\H$. Let furthermore $\Na\subset\Ma$ be a subalgebra of infinite type and $(\alpha_n)_{n\in\mathbb{N}}\subset\mathbf{Aut}\big(\L(\H)\big)$ a sequence of transformations such that

\begin{itemize}
\item[(i)] $\alpha_n(\Na)\subset\Na\quad\forall n\in\N$,
\item[(ii)] $\underset{n\rightarrow\infty}{\text{w-lim}}\,\alpha_n(A)=\o(A)\1\quad\forall A\in\Na\text{ with }\o\in\Na^*,\o\neq 0$, and
\item[(iii)] $\underset{n\rightarrow\infty}{\text{w-lim}}\big[\alpha_n(A),B\big]=0\quad\forall A\in\Na\;\forall B\in\Ma$.
\end{itemize} 
Then $\Ma$ is of type $III$. 
\end{theo}
As an application of this theorem he shows that for dilation-invariant local systems the local algebra is of type $III$. This conclusion is more precise than Roberts' statement which rules out only the type $I$ structure for such algebras \cite{Roberts:1974}.

 Hislop and Longo are the first to show the type $III_1$ structure for the massless case \cite{Hislop-Longo:1981uh} by using unitary equivalence of free local algebras, a result of Eckmann and Fr\"ohlich \cite{Eckmann:1974}. In the same year Longo generalises their achievement to the massive case. Haagerup proves the hyperfinite (AFD) property of the local von Neumann algebras, see Definition \ref{AFD}, and the uniqueness of the hyperfinite factor of type $III_1$ up to $W^*$-isomorphisms \cite{Haagerup:1987}.

Fredenhagen's analysis on the type of double cone algebras is based on the standard algebraic postulates, the Bisognano-Wichmann property, see Theorem \ref{Bisognano}, and the compliance with a scaling property introduced by Haag, Narnhofer and Stein in \cite{Haag:1984xa}, which is expected to hold in all renormalizable field theories with an ultraviolet fixed point \cite{Fredenhagen:1984dc}.

\begin{defi}\label{scaling}
 A Wightman field $\varphi(x)$ is said to have a Haag-Narnhofer-Stein scaling limit if there exists a positive and monotone scaling function $N(\lambda)$, $\lambda>0$, such that for all $n\in\mathbb{N}$ 
\begin{gather*}
N(\lambda)^n\big(\Omega,\varphi(\lambda x^1)\cdots\varphi(\lambda x^n)\Omega\big)
\end{gather*}
converges for $\lambda\longrightarrow 0$ to a non vanishing Wightman field.
\end{defi}
Fredenhagen chooses a double cone in the corner of the Rindler wedge and then derives the Connes invariant for the double cone via the known modular group of automorphisms for the Rindler wedge, which has been calculated by Bisognano and Wichmann in \cite{Bisognano:1975ih} and \cite{Bisognano:1976za}. We will give more details of this investigation later.

Instead of the Bisognano-Wichmann property, see Theorem \ref{Bisognano}, Buchholz, D'Antoni and Fredenhagen propose in \cite{Buchholz:1986bg} a nuclearity condition introduced by Buchholz and Wichmann \cite{Buchholz:1986dy}, which will also be explained more extensively later, see \eqref{Buchholz-Wichmann}, and obtain the following result for the double cone.

\begin{theo}
Let the net of local algebras $\Ma(\O)$ satisfy the nuclearity condition and consider two double cones $\D_1\subset\D_2$ such that $\overline{\D_1}\subset\overset{\circ}{\D_2}$, where $\overset{\circ}{\D}_2$ denotes the interior of $\D_2$. Then there exists a factor $\tilde{\Ma}$ of type I with
$$
\Ma(\D_1)\subset\tilde{\Ma}\subset\Ma(\D_2).
$$
\end{theo} 
Borchers combines this with Fredenhagen's result and discovers further structural properties \cite{Borchers:2000pv}.

\begin{theo}
Let us assume the net of local algebras $\Ma(\O)$ in the vacuum sector to satisfy the (Buchholz-Wichmann) nuclearity condition, see \eqref{Buchholz-Wichmann}, and the (Haag-Narnhofer-Stein) scaling property, Definition \ref{scaling}. Furthermore, let $\Ma(\D)$ be continuous from the inside, i.e., 
$$
\Ma(\D)=\left(\bigcup_i\Ma(\D_i)\right)'' \quad\text{with}\quad\D=\bigcup_i\D_i\quad\text{and}\quad\overline{\D_i}\subset\overset{\circ}{\D}_{i+1}
$$  
(or equivalently from the outside). Then there exists a unique hyperfinite factor $\tilde{\Ma}$ of type $III_1$ such that
$$
\Ma(\D)\cong\tilde{\Ma}\;\overline{\otimes}\Za(\Ma),
$$
where $\Za(\Ma)$ is the center of $\Ma(\D)$.
\end{theo} 

As already mentioned, in algebraic quantum field theory all physical information is encoded in the net structure,
\begin{gather*}
\O\mapsto\Ma(\O).
\end{gather*}
Therefore, the ultimate challenge in the analysis of algebras of local observables is the determination of their subnet structure. A first and important step has been taken by Kawahigashi and Longo who give a complete classification of the discrete series of local conformal nets, i.e., for the case where the central charge is $c<1$ \cite{Kawahigashi:2002px}. 

\begin{theo}\label{Kawahigashi-Longo}
An irreducible local conformal net $\Ma$ with the central charge $c<1$ is completely classified by the pair $(m,s)$, where the number of finite-index conformal subnets can take the values $s=1,2,3$ and $c=1-\frac{6}{m(m+1)}$. For any $m\in\N$ the following values for $s$ can be realised:
\begin{itemize}
\item[(i)] $s=1$ for all $m\in\N$,
\item[(ii)] $s=2$ if $m=1,2$ mod $4$, and if $m=11,12$, and
\item[(iii)] $s=3$ for $m=29,30$.
\end{itemize}
\end{theo}

\subsection{KMS States}

Modular theory can also be connected with algebraic quantum field theory by KMS states, the generalisation of Gibbs states. For a survey we refer the reader to \cite{Bratteli:1996xq} (confer also \cite{Saffary2:2001}).

One of the possible formulations of equilibrium states is the grand canonical ensemble, where the energy as well as the particle number are variable. Here, an equilibrium state, the Gibbs state, is represented by a state over $\L(\H)$ defined as
\begin{gather*}
\o_{\beta,\mu}(A):=\frac{\text{Tr}_\H\big(e^{-\beta(H-\mu N)}A\big)}{\text{Tr}_\H\big(e^{-\beta(H-\mu N)}\big)},
\end{gather*}
where $H$ is the Hamiltonian, $N$ the particle number operator and $\beta,\mu\in\R$. But the assumption in this approach that $e^{-\beta(H-\mu N)}$ be a trace-class operator is unfortunately not fulfilled in infinite-dimensional systems. However, this assumption can be circumvented by introducing the time evolution
\begin{align*}
\sigma^t:\L(\H)&\longrightarrow\L(\H)\\
A&\mapsto\sigma^t(A):=e^{it(H-\mu N)}Ae^{-it(H-\mu N)},
\end{align*}
which, as long as $H$ is self-adjoint, always exists and can be used to introduce the so-called KMS states.\\
The KMS condition has first been formulated by Kubo \cite{Kubo:1957mj}, Martin and Schwinger \cite{Martin:1959jp}, but Haag, Hugenholtz and Winnink \cite{Haag:1967sg} are the first to impose it as a criterion for equilibrium states.

\begin{defi}\label{KMS}
Let $(\Ma,\sigma)$ be a $W^{*}$-dynamical system. A state $\omega:\Ma\rightarrow\C$ is called $\sigma$-KMS state with reference to $\beta\in\R$, or $(\sigma,\beta)$-KMS state if it satisfies the KMS condition,
\begin{equation*}
\omega\big(A\sigma^{i\beta}(B)\big)=\omega(BA),
\end{equation*}
for all $A,B$ in a norm-dense, $\sigma$-invariant $\*$-subalgebra of $\Ma$.
\end{defi}

This definition is equivalent to the requirement that for all $A,B\in\Ma$ there exists a function $F_{A,B}$ which is analytic in the strip 
$$
\mathbf{D}_\beta:=\begin{cases}\big\{z\in\C|\;0<\Im z<\beta\big\}& \text{if}\quad \beta\geq 0, \\ \big\{z\in\C|\;0>\Im z>\beta\big\}& \text{if}\quad \beta\leq 0,\end{cases}
$$
and continuous in the closure $\overline{\mathbf{D}}_\beta$ with
\begin{gather*}
F_{A,B}(t)=\o\big(A\sigma^t(B)\big)\quad\text{and}\\
F_{A,B}(t+i\beta)=\o\big(\sigma^t(B)A\big)
\end{gather*}
for all $t\in\R$.

KMS states have received a lot of attention since their introduction and physical interpretation, in particular due to their formulation through modular theory. Tomita associates to an arbitrary normal and faithful state $\o$ on a von Neumann algebra an automorphism group $\sigma_\o^t$, and Takesaki succeeds in showing that this state is a KMS state with respect to $\sigma_\o^t$. 

\begin{theo}[Takesaki]\label{Takesaki-KMS}
Let $\o$ be a normal state on the von Neumann algebra $\Ma$. Then the following statements are equivalent:

\begin{itemize}
\item[(i)] $\o$ is a faithful state on $\pi_{\o}(\Ma)$, i.e., there exists a projection $E\in\Ma\cap\Ma'$ with $\o(\1-E)=0$, and $\o\arrowvert_{\Ma E}$ is faithful.
\item[(ii)] There exists a $\sigma$-weak and continuous one-parameter ${}^{*}$-automorphism group $\sigma_\o^t$, $t\in\R$, of $\Ma$ such that $\o$ is a \t.
\end{itemize}
If these conditions are satisfied, then $\sigma_{t}^\o(E)=E$ holds for all $t\in\R$, and the restriction of $\sigma_\o^t$ onto $\Ma E$ is the modular automorphism group on $\mathfrak{M} E$, uniquely determined by $\o$.
\end{theo}

%\mar{S,4.11}
\begin{cor}
Let $\o_1$ be a normal, faithful and semifinite state on the von Neumann algebra $\Ma$ with center $\Za(\Ma)$, and let $\o_2$ be a normal semifinite state on $\Ma$. Then the following statements are equivalent:

\begin{itemize}
\item[(i)] $\o_2$ satisfies the KMS condition with respect to $\sigma_{\o_1}^t$.
\item[(ii)] Let $E,F\in\Ma$ be two projections defined as 
\begin{align*}
\Ma E:&=\overline{\{A\in\Ma|\;\o_2(A^*A)<\infty\}}\qquad\text{and}\\
 \Ma F:&=\{A\in\Ma|\;\o_2(A^*A)=0\},
\end{align*}
where the closure is meant with respect to the $\sigma$-strong topology, and $\mathbf{s}(\o_2):=E-F$, the so-called support of $\o_2$. Then one has $\mathbf{s}(\o_2)\in\Za(\Ma)$ and $\sigma_{\o_2}^t=\sigma_{\o_1}^t\big\arrowvert_{\Ma\mathbf{s}(\o_2)}$ for all $t\in\R$.
\item[(iii)] There exists a positive self-adjoint operator $A\in\Ma$ affiliated with the center $\Za(\Ma)$ such that $\o_2(B)=\o_1\big(A^{1/2}BA^{1/2}\big)$ for all $B\in\Ma_+$. 
\end{itemize}
\end{cor}

%\mar{S,4.13}
\begin{cor}
Let us assume that $\o$ is a normal, faithful and semifinite state on the von Neumann algebra $\Ma$ of type III and $\beta\in\R$, $\beta\neq1$, then there is no normal, faithful, semifinite state on $\Ma$ satisfying the KMS condition with respect to $\sigma_\o^t$.  
\end{cor}
Over the years the interpretation of KMS states as equilibrium states has gained additional support. On the one hand one can show that KMS states satisfy certain stability conditions, and on the other hand one is able to derive them from some realistic stability  criteria. A  prominent example from quantum field theory on curved spacetime, where the KMS property has played an essential role, is the Hawking effect \cite{Hawking:1974sw}. The proof in \cite{Fredenhagen:1990kr} is based  on an idealised detector which is simulated with the help of the KMS condition (see also \cite{Saffary:2001ru}).

In the investigation of stability properties of KMS states one compares a general $C^{*}$-dynamical system $(\Aa,\tau)$ with a disturbed system $(\Aa,\tau^{P})$, where $P=P^{*}\in\Aa$ denotes a small perturbation. Here, the disturbed automorphism group  $\tau^{P}$ is generated by the infinitesimal generator,
$$
\delta+\delta_{P}\quad\text{with}\quad \delta_{P}(A):=i[P,A]\quad\forall A\in\Aa,
$$
where $\delta$ is the infinitesimal generator of the undisturbed automorphism group $\tau$. One distinguishes between two different approaches, the ``time-independent'' one of Connes and Araki, which is of advantage for the comparison of $\tau$-KMS states with $\tau^{P}$-KMS states, and the ``time-dependent'' one of Robinson, which is easier to use in the case  of general states on systems having ergodicity properties, for example $\L_1(\Aa_0)$-asymptotic commutativity, i.e., satisfying the requirement
$$
\int_{-\infty}^\infty\big\|\big[A,\sigma^t(B)\big]\big\|dt<\infty,\quad A,B\in\Aa_0,
$$
where $\Aa_0$ is a norm-dense subalgebra of $\Aa$. We give here only one example of many stability properties.

\begin{theo}
Let $\omega$ be a \t  over the $C^{*}$-dynamical or $W^{*}$-dynamical system $(\Aa,\sigma)$ with the strong cluster-property

\begin{equation*}
\underset{t\rightarrow\pm\infty}{\lim}\omega\big(A\sigma^{t}(B)\big)=\omega(A)\omega(B)\qquad\forall A,B\in\Aa.
\end{equation*} 
Then 
\begin{equation*}
\begin{split}
\underset{T_{1}\rightarrow\pm\infty}{\lim}\cdots&\underset{T_{n}\rightarrow\pm\infty}{\lim}i^{n}\int^{T_{1}}_{0}dt_{1}\int^{T_{2}}_{t_{1}}dt_{2}\cdots\int^{T_{n}}_{t_{n-1}}dt_{n}\omega\left(\Big[\sigma^{t_{n}}(B)\big[\cdots[\sigma^{t_{1}}(B),A]\big]\Big]\right)\\
&=\int_{0}^{\beta}ds_{1}\cdots\int_{0}^{s_{n-1}}ds_{n}\omega_{T}\big(A;\sigma^{is_{n}}(B);\cdots;\sigma^{is_{1}}(B)\big),
\end{split}
\end{equation*}
where $\o_T$ denotes the truncated state. In particular,
\begin{equation*}
\underset{T\rightarrow\infty}{\lim}\int^{T}_{-T}\omega\big([A,\sigma^{t}(B)]\big)dt=0
\end{equation*}
for all $A,B\in\Aa$.
\end{theo}
One expects an equilibrium state not only to be stable under small perturbations, but also to be derivable from physically motivated stability assumptions. One postulates for KMS states the following physically reasonable conditions:

\begin{itemize}
\item[(i)] $\sigma^t$-invariance, i.e., stationarity in time;
\item[(ii)] ergodicity of $(\Aa,\sigma^t)$, e.g., asymptotic commutativity;
\item[(iii)] relative ``purity'' of $\o$, e.g., $\o$ should be extremal under the $\sigma^t$-invariant states;
\item[(iv)] stability under perturbations.
\end{itemize}
These postulates can be shown to lead to the following stability criterion, 
\begin{equation*}
\int_{-\infty}^{\infty}\o\big([A,\sigma^{t}(B)]\big)dt=0,\quad\forall A,B\in\Aa,
\end{equation*}
which is a strong indication that the KMS property could be a consequence of the postulates given above. One may lend weight to these postulates differently. We give here one version which demands weak cluster properties instead of $\L_1(\Aa_0)$-asymptotic commutativity for the state \cite{Bratteli:1978pe}.

\begin{theo}[Bratteli, Kishimoto, Robinson]
Let $(\Aa,\sigma)$ be a n $\L_1(\Aa_0)$-asymptotic commutative $C^{*}$-dynamical system and $\o$ a $\sigma$-invariant state on $\Aa$. Let us further assume the validity of the following two conditions:

\begin{itemize}
\item[(i)]  Either
\begin{equation*}
\underset{\underset{i\neq j}{\inf}|t_{i}-t_{j}|\rightarrow\infty}{\lim}\,\omega\Big(\sigma^{t_{1}}(A_{1})\sigma^{t_{2}}(A_{2})\sigma^{t_{3}}(A_{3})\Big)=\omega(A_{1})\omega(A_{2})\omega(A_{3})\quad\forall A_{k}\in\Aa,
\end{equation*}
or $\o$ is a factor state.
\item[(ii)]  $\o$ satisfies the stability condition,
\begin{equation*}
\int_{-\infty}^{\infty}\omega\big([A,\sigma^{t}(B)]\big)dt=\infty\qquad\forall A,B\in\Aa_{0}.
\end{equation*}
\end{itemize}
Then $\o$ is an extremal \t with $\beta\in\overline{\R}$.
\end{theo}

The aforementioned theorems bound up the notion of stability with KMS states for $\L_1(\Aa_0)$-asymptotic commutative $C^{*}$-dynamical systems, and one can therefore justify the interpretation of KMS states as equilibrium states of physical systems.

Despite of all these properties of the KMS condition for thermal states, one last desire remains to be fulfilled, namely its relativistic formulation. Ojima shows that a KMS state cannot be Lorentz-invariant \cite{Ojima:1985}, because the KMS condition implies a distinguished time axis. To be more precise, the reason for the breakdown of the Lorentz invariance is due to the fact that every automorphism group of the local algebra $\Aa(\O)$ leaving the KMS state invariant can be implemented by a unitary operator which commutes with the modular operator and the modular conjugation. Therefore, since the time translation commutes with the modular group of automorphisms but the Lorentz boosts do not commute with the time translations, see \eqref{commutation-relations}, the KMS state cannot be Lorentz-invariant.

This problem is addressed by Bros and Buchholz \cite{Bros:1998} who propose the following first version of the KMS condition.

\begin{defi}
A state $\o:\Aa\longrightarrow\C$ satisfies the so-called relativistic KMS condition with respect to the automorphism group $\sigma$ at inverse temperature $\beta>0$ if and only if for all $A,B\in\Aa$ there exist a positive timelike vector $e\in\V_+$, $e^2=1$, and a function $F$, which is analytic in the tube $\mathbf{T}:=\big\{z\in\C^d|\;\Im z\in\V_+\cap(\beta e+\V_-)\big\}$ and continuous at the boundary sets $\Im z=0$ and $\Im z=\beta e$ with
\begin{gather*}
F_{A,B}(x)=\o\big(A\sigma^x(B)\big)\quad\text{and}\\
F_{A,B}(x+i\beta e)=\o\big(\sigma^x(B)A\big)
\end{gather*}
for all $x\in\R^d$.
\end{defi}

Contrary to Definition \ref{KMS}, this definition treats all spacetime coordinates equally, and therefore any observer moving with constant velocity experiences a relativistic KMS state as an equilibrium state with a distinguished rest frame and time axis along $e$.

\section{Modular Action}

We start with the introduction of the CPT operator which will appear throughout the following investigation. The CPT operator is defined uniquely as an anti-unitary operator via the relation
\begin{gather*}
\Theta\varphi(x)\Theta^{-1}=(-1)^m(-i)^n\varphi^*(-x),\\
\Theta\Omega=\Omega,\\
n:=\begin{cases} 0, & \varphi\text{  is a Bose field},\\ 1, & \varphi\text{  is a Fermi field},\end{cases}
\end{gather*}
where $\Omega\in\H$ represents the vacuum vector and $m$ is the number of dotted spinor indices of $\varphi$, i.e., these spinors transform under the complex conjugation representation $\alpha\longrightarrow\bar{\alpha}$, where $\alpha$ is the representation of $SL(2,\C)$ onto the complex vector space of covariant spinors, confer \cite{Haag:1992hx} for more details. For our analysis the commutation relation with the representation of the Poincar\'e group,
\begin{equation*}
\Theta U(a,\alpha)=U(-a,\alpha)\Theta,
\end{equation*} 
will be of interest.

\begin{defi}\label{wedgeduality}
The net of von Neumann algebra $\Ma(\O)$ is said to satisfy wedge duality, if
\begin{gather*}
\big(\Ma(\W_R)\big)'=\Ma(\W_R'),
\end{gather*}
where $\W_R'$ denotes the interior of the spacelike complement of $\W_R$.
\end{defi}

The ground-breaking result, proved by Bisognano and Wichmann in \cite{Bisognano:1975ih}, states that, since, due to the Reeh-Schlieder Theorem \ref{Reeh-Schlieder}, the von Neumann algebra of (finite-component) Wightman fields on the right Rindler wedge $\W_R$ satisfies the conditions for the modular theory, the resulting modular conjugation implements a combination of charge conjugation, time and spatial reflection, while the modular operator implements the Lorentz boosts.

\begin{theo}[Bisognano-Wichmann]\label{Bisognano}
In the Wightman framework the modular objects for the von Neumann algebra of observable fields $\mathfrak{M}(\W_R)$ with reference to the vacuum state are
\begin{align*}
J_{\W_R}=\Gamma\big(\Theta U\big(R_1(\pi)\big)\big),\qquad \Delta_{\W_R}^{it}=\Gamma\big[U(\Lambda_s)\big],
\end{align*}
where $R_1$ denotes the spatial rotation around the $x^1$-axis, and the modular automorphism group acts geometrically as pure Lorentz transformations:
\begin{gather*}
\sigma_{\W_R}^t\big(\varphi[f]\big)=\Delta_{\W_R}^{it}\varphi[f]\Delta_{\W_R}^{-it}=\varphi[f_s],\\
\text{where}\quad f_s(x):=f\big(\Lambda_{s}(x)\big),\quad s:=2\pi t,\\
\Lambda_{s}:=\begin{pmatrix} \cosh s & \sinh s & 0 & 0\\  \sinh s & \cosh s & 0 & 0\\  0 & 0  & 1 & 0\\ 0 & 0 & 0 & 1 \end{pmatrix}.
\end{gather*}
Furthermore, the theory fulfills wedge duality.% i.e., $\Ma(\W)'=\Ma(\W')$ for an arbitrary wedge, where $\W'$ denotes the interior of the spacelike complement of $\W$. 
\end{theo}

\sproof First of all the Tomita operator is established as
$$
S:=\Theta U\big(R_1(\pi)\big)U\big(\Lambda(i\pi)\big),
$$
by making use of the spectrum condition for the energy-momentum. Since $R_1$ is the spatial rotation around the $x^1$-axis, $R_1(\pi)$ causes the inversion of the signs of $x^i$, $i=2,3$. $S$ really being the Tomita operator, is proved by verification of its properties. To this end Bisognano and Wichmann show that the transformation law of covariant fields under the action of $S$ is
$$
S\varphi(x^1)\varphi(x^2)\cdots\varphi(x^n)\Omega=\varphi^*(x^1)\varphi^*(x^2)\cdots\varphi^*(x^n)\Omega. 
$$
This property is generalised for all elements $A$ in the polynomial algebra $\mathfrak{P}(\W_R)$ generated by operator-valued distributions with support in the Rindler wedge $\W_R$ to yield
$$
SA\Omega=A^*\Omega.
$$ 
Finally, it is shown that the domain $\mathfrak{P}(\W_R)\Omega$ constitutes a core for the operator $S$ which is thus uniquely defined. The conditions of the modular theory are therefore satisfied and one concludes,
$$
J_{\W_R}=\Theta U\big(R_1(\pi)\big),\quad\text{and}\quad\Delta_{\W_R}^{1/2}=U\big(\Lambda(i\pi)\big).
$$

%\QED
\begin{flushright}$\Box$\end{flushright}

The geometric r\^ole of the modular conjugation is to transform the right wedge onto the left one, i.e., $J_{\W_R}\Ma(\W_R)J_{\W_R}=\Ma(\W_L)$.

The infinitesimal generator $\delta_{\W_R}$ of the modular group $\sigma_{\W_R}^t$ can directly be computed as in Proposition \ref{generator}:
\begin{gather}\label{BW-generator}
\delta_{\W_R}\varphi[f]:=\partial_s\sigma_{\W_R}^t\big(\varphi[f]\big)\big\arrowvert_{s=0}=\varphi\big[\partial_s f_s\big]\big\arrowvert_{s=0}
\end{gather}
with
\begin{align*}
\partial_s f_s(x)\big\arrowvert_{s=0}&=x^1\partial_{x^0}f(x)+x^0\partial_{x^1}f(x)
\end{align*}

It is worth mentioning that the modular group and the modular conjugation of wedge regions act locally and map local algebras into local ones. This property will be of importance in proving the PCT theorem.
 
In a subsequent paper, Bisognano and Wichmann verify these statements in a more general setting, namely for charged Bose and Fermi fields fulfilling various physically reasonable assumptions, e.g., covariance with respect to finite-dimensional representations of the proper orthochronous Lorentz group \cite{Bisognano:1976za}.

Although this theorem is based on a fairly general formalism, the finiteness of the components of the fields poses a restriction which can be dropped. This is done by Kuckert \cite{Kuckert:1998pb} using a result of Borchers \cite{Borchers:1991xk}. Borchers proves that the spectrum condition only implies the commutation relations between the modular objects and the translations $U(a)$,
\begin{gather*}
J_{\W_R}U(a)J_{\W_R}=\Theta U(R_1(\pi)),\\
\Delta_{\W_R}^{is}U(a)\Delta_{\W_R}^{-is}=U(\Lambda_{-s}(a)).
\end{gather*}
Wiesbrock shows that these commutation relations are not only necessary but even sufficient for the spectrum condition \cite{Wiesbrock:1992mg}.

\label{Wiesbrock-erratum} Unfortunately, a mistake enters the proof of Wiesbrock's main Theorem 3, Corollary 6 and Corollary 7 in \cite{Wiesbrock:1992mg}, mentioned in \cite{Wiesbrock-erratum:1997}, which has recently been remedied by Araki and Zsid\'o \cite{Araki-Zsido:2005}. They even generalise Wiesbrock's result, formulating their statement for normal, semifinite, faithful weights $\varphi$, whereas Wiesbrock demands $\varphi$ to be bounded.

If additionally some symmetry condition is given, i.e., the associated modular group should map local nets into local nets again, Borchers' commutation relations can be shown to lead to the Bisognano-Wichmann property.

\begin{theo}[Kuckert]
Let $\O\subset\R^n$ be an arbitrary double cone, then the following statements hold:

\begin{itemize}
\item[(i)] If for every $\O$ there exists an open set $\M_\O$ with
$$
J_{\W_R}\mathfrak{M}(\O)J_{\W_R}=\mathfrak{M}(\M_\O),
$$
then the modular conjugation acts as given in the Bisognano-Wichmann Theorem \ref{Bisognano}. 
\item[(ii)] If for every $\O$ and all $t\in\R$ there exists an open set $\M_\O^t$ such that
$$
\Delta_{\W_R}^{it}\mathfrak{M}(\O)\Delta_{\W_R}^{-it}=\mathfrak{M}(\M_\O^t),
$$
then the modular operator acts geometrically as Lorentz boosts given in the Bisognano-Wichmann Theorem \ref{Bisognano}.
\end{itemize}
\end{theo}

Bisognano and Wichmann deal with local algebras which are generated by a finite number of Poincar\'e-covariant Wightman fields and therefore not with the general case, since there exists an infinite number of quantum fields satisfying Poincar\'e covariance. Borchers gives a derivation of the Bisognano and Wichmann property in a purely algebraic setting including these fields, too \cite{Borchers:1998nf}. Let us consider the so-called characteristic two-plane of the wedge
$$
\W(l_1,l_2):=\big\{\lambda l_1+\mu l_2+l_\perp,|\;\mu<0<\lambda,(l_1,l_\perp)=(l_2,l_\perp)=0\big\},
$$
where $l_1$ and $l_2$ are two linearly independent lightlike vectors and the vector $l_\perp$ is perpendicular to $l_1$ and $l_2$, and let $\Lambda_{l_1,l_2}$ be the Lorentz boost which leaves $\W(l_1,l_2)$ invariant. Then the elements $A\in\mathfrak{M}\big(\W(l_1,l_2)\big)$ such that $U\big(\Lambda_{l_1,l_2}(2\pi t)\big)A^*\Omega$ can be analytically continued into the strip $\mathbf{S}(-\frac{1}{2},0):=\{a+ib\in\C|\;-\frac{1}{2}<b<0\}$. They form a dense set in $\mathfrak{M}(\W(l_1,l_2))$, and there exist elements $\hat{A}$ and $\tilde{A}$ affiliated with $\mathfrak{M}\big(\W(l_1,l_2)\big)'\equiv\mathfrak{M}\big(\W(l_2,l_1)\big)$ that satisfy
\begin{equation*}
\hat{A}\Omega:=U\big(\Lambda_{l_1,l_2}(-i\pi)\big)A\Omega \quad\text{and}\quad \tilde{A}\Omega:=U\big(\Lambda_{l_1,l_2}(-i\pi)\big)A^*\Omega. 
\end{equation*}

%\begin{defi}\label{wedgeduality}
%The net of von Neumann algebra $\Ma(\O)$ is said to satisfy wedge duality, if
%\begin{gather*}
%\big(\Ma(\W_R)\big)'=\Ma(\W_R'),
%\end{gather*}
%%where $\W_R'$ denotes the interior of the spacelike complement of $\W_R$.
%\end{defi}

\begin{defi}\label{reality}
A Poincar\'e-covariant theory of local observables in the vacuum sector satisfying wedge duality is said to fulfill the reality condition, if for elements $A\in\mathfrak{M}\big(\W(l_1,l_2)\big)$ with the aforementioned property one has $\hat{A}^*=\tilde{A}$, i.e., $U\big(\Lambda_{l_1,l_2}(-i\pi)\big)A^*\Omega=\hat{A}^*\Omega$.
\end{defi}
Borchers then shows the following

\begin{theo}[Borchers]
In a representation of a Poincar\'e-covariant theory of local observables with reference to the vacuum state, the Bisognano-Wichmann property holds for the net if and only if the theory fulfills wedge duality and the reality condition with respect to the Lorentz transformation.
\end{theo}
Borchers' reality condition implies the assumption of Schroer and Wiesbrock \cite{Schroer:1998ax} that the mapping
\begin{equation*}
A\Omega\mapsto U\left(\Lambda_{l_1,l_2}\left(-\frac{i}{2}\right)\right)A^*\Omega,\quad A\in\mathfrak{M}\big(\W(l_1,l_2)\big),
\end{equation*}
is uniformly bounded. This is the basis of their algebraic proof of the Bisognano-Wichmann property.

The first algebraic investigation of the Bisognano-Wichmann theorem has been carried out by Brunetti, Guido and Longo in the case of conformal quantum field theory \cite{Brunetti:1992zf}. Their proof is based on the assumption that the modular group already acts geometrically on an arbitrary wedge region. In addition to that, the split property is assumed to be satisfied by the local net, i.e., for all standard diamonds $\D_1\subset\D_2$ there exists a factor $\mathfrak{N}$ of type $I$ satisfying
$$
\mathfrak{M}(\D_1)\subset\mathfrak{N}\subset\mathfrak{M}(\D_2).
$$
Furthermore, they show that conformal theories automatically satisfy essential duality and PCT symmetry .

In \cite{Mund:2001sv} Mund investigates the Bisognano-Wichmann theorem in an algebraic context for massive theories. He assumes that

\begin{itemize}
\item there exist massive particles with their scattering states spanning the whole Hilbert space,
\item within each charge sector the occurring particle masses have to be isolated eigenvalues of the mass operator,
\item the representation of the covering group of the Poincar\'e group should not have `accidental' degeneracies, and
\item localisation in spacelike cones should be given.
\end{itemize}

All investigations mentioned so far are concerned with one preferred spacetime region, namely the right Rindler wedge $\W_R$. The wish to transport the result of Bisognano and Wichmann to other interesting spacetime regions, like the light cone $\V_\pm$ and the double cone $\D$, seems obvious. The action of the modular group was first computed for the forward light cone by Buchholz \cite{Buchholz:1977ze} and then some years later for the double cone by Hislop and Longo \cite{Hislop-Longo:1981uh}. Unfortunately, these results deal with the massless case only, whereas the Bisognano-Wichmann theorem is valid for all masses $m\geq0$. In the sequel we first present the derivation of Hislop and Longo for $\D_1$ and then proceed to Buchholz' result on $\V_\pm$.

Via the rotations in the $\xi^1-\xi^4$-plane, 
\begin{gather*}
T_{41}(\varphi)\xi:=\begin{cases} \xi^1(\varphi)=-\xi^4\sin\varphi+\xi^1\cos\varphi, &\\  \xi^4(\varphi)=-\xi^4\cos\varphi+\xi^1\sin\varphi, & \varphi\in[0,\pi],\\ \xi^i(\varphi)=\xi^i, & i=0,2,3,5,\end{cases}
\end{gather*}
the wedge $\W_R$ can be mapped conformally onto the double cone $\D_1$:
\begin{equation*}
T_{41}\left(\frac{\pi}{2}\right)\W_R=\D_1.
\end{equation*}
In $x$-coordinates the transformation reads 
\begin{equation*}
x^\mu(\varphi):=\xi^\mu(\varphi)\big(\xi^4(\varphi)+\xi^5(\varphi)\big)^{-1},\quad \mu=0,1,2,3.
\end{equation*}
The idea now is to transport the known modular action from $\W_R$ onto $\D_1$ through the relation 
\begin{align*}
T_{04}(s)\xi:&=T_{41}\left(\frac{\pi}{2}\right)T_{10}(s)\left(T_{41}\left(\frac{\pi}{2}\right)\right)^{-1}\xi\\
&=\begin{cases} \xi^0(s)=\xi^0\cosh s+\xi^4\sinh s, &\\  \xi^4(s)=\xi^0\sinh s+\xi^4\cosh s, & s\in[0,\pi],\\ \xi^i(s)=\xi^i, & i=1,2,3,5,\end{cases}
\end{align*}
where $T_{10}$ are the Lorentz boosts written as pseudo-rotations in the $(\xi^1-\xi^0)$-plane:
\begin{gather*}
T_{10}(s)\xi:=\begin{cases} \xi^0(s)=-\xi^0\cosh s+\xi^4\sinh s, &\\  \xi^1(s)=-\xi^0\sinh s+\xi^1\cosh s, & s\in[0,\pi],\\ \xi^i(s)=\xi^i, & i=2,3,4,5.\end{cases}
\end{gather*}
The new conformal transformation in the $x$-space then is:
\begin{gather}\label{Hislopaction}
x(s)=\begin{cases}x^0(s)=N(s)^{-1}\big(x^0\cosh s+\frac{1}{2}(1+(x,x))\sinh s\big), &\\ x^i(s)=N(s)^{-1}x^i, & i=1,2,3, \end{cases}
\end{gather}
where
\begin{gather*}
N(s):=x^0\sinh s+\frac{1}{2}\big(1+(x,x)\big)\cosh s+\frac{1}{2}\big(1-(x,x)\big).\notag
\end{gather*}
This can be written in a more compact form as:
\begin{gather*}
x_\pm(s)=\frac{1+x_\pm-e^{-s}(1-x_\pm)}{1+x_\pm+e^{-s}(1-x_\pm)}\\
\text{with}\quad x_+:=x^0+|\mathbf{x}|\quad\text{and}\quad x_-:=x^0-|\mathbf{x}|.
\end{gather*}
If the net structure is respected by the conformal group, i.e., for each $g$ in a neighborhood $\mathcal{N}_\1$ of the identity in the conformal group one has an algebraic isomorphism $\alpha_g:\mathfrak{M}(\O)\longrightarrow\mathfrak{M}(g\O)$, where $\O$ is a bounded region, with
\begin{equation*}
\alpha_g:\mathfrak{M}(\O_1)\longrightarrow\mathfrak{M}(g\O_1),\quad\O_1\subset\O,g\in\mathcal{N}_\1,
\end{equation*}
and if the vacuum state $\o_0$ is invariant under the conformal group, i.e.,
\begin{equation*}
\o_0\big(\alpha_g\varphi[f]\big)=\o_0\big(\varphi[f]\big),
\end{equation*}
then the isomorphism $\alpha_g$ can be implemented unitarily:
\begin{equation*}
U_g\varphi[f]\Omega=\big(\alpha_g\varphi[f]\big)\Omega.
\end{equation*}
Both conditions defining conformal invariance of a theory are given in a free massless field theory, only. Thus we get for the modular objects on the double cone $\D_1$,
\begin{gather*}
J_{\D_1}=U_{T_{41}}J_{\W_R}U_{T_{41}}^{-1},\quad\text{and}\\
\Delta_{\D_1}^{it}=U_{T_{41}}\Delta_{\W_R}^{it}U_{T_{41}}^{-1}.
\end{gather*}
Finally, one arrives at

\begin{theo}[Hislop-Longo]\label{Hislop-Longo}
In a free massless quantum field theory the modular objects for the von Neumann algebra of observable fields $\mathfrak{M}(\D_1)$ with reference to the vacuum state are
\begin{align*}
J_{\D_1}=\Gamma\big(I_tU_\rho\big)\quad\text{and}\quad \Delta_{\D_1}^{-it}=\Gamma\big[U(s)\big],
\end{align*}
where $I_t$ is the time reversal operator, $U_\rho$ the conformal ray inversion operator, defined in \eqref{ray-inversion-operator}, and $s:=2\pi t$. The modular automorphism group acts geometrically as special conformal transformations:
\begin{gather*}
\sigma_{\D_1}^t\big(\varphi[f]\big)=\Delta_{\D_1}^{it}\varphi[f]\Delta_{\D_1}^{-it}=\varphi[f_s],\\
\text{where}\quad f_s(x):=\gamma\big(x^0,x^3,s\big)f\big(x_\pm(-s)\big),\\
\gamma(x^0,x^3,s):=2^6\big(1+z_++e^{-s}(1-z_+)\big)^{-3}\big(1-z_-+e^{s}(1+z_-)\big)^{-3},\\
z_+:=x^0+ x^3,\quad z_-:=x^0-x^3.
\end{gather*}
\end{theo}

The geometric action of the modular conjugation is illustrated in Figure 4.1, namely $J_{\D_1}$ transforms the double cone $\D_1$ onto the shaded region.

\begin{figure}[here]\label{grafik1}
\begin{center}
 \includegraphics[height=7cm]{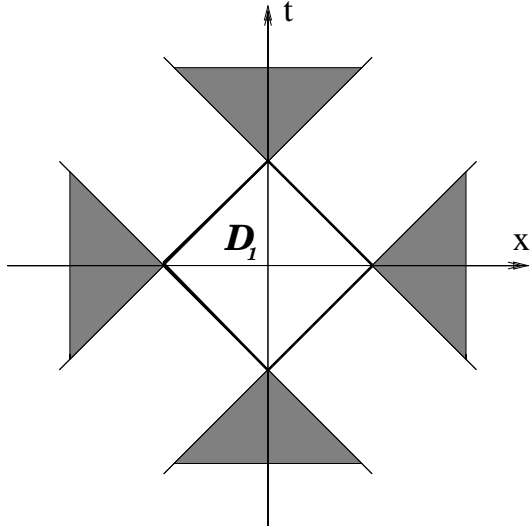}
    \caption[]{ \footnotesize \em Geometric action of $J_{\D_1}$}
\end{center}
\end{figure}

% Fuer PDF-datei

%\begin{figure}[here]\label{grafik1}
%\begin{center}
%\includegraphics[height=7cm]{doublecone.png} 
%\caption[]{ \footnotesize \em Geometric action of $J_{\D_1}$}
%\end{center}
%\end{figure}

%\begin{center}
%\begin{picture}(140,140)
%\thinlines{
%\put(0,70){\vector(1,0){140}}
%\put(75,140){\scriptsize{$t$}}
%\put(70,0){\vector(0,1){140}}
%\put(140,75){\scriptsize{$\x$}}}
%\thicklines{
%\put(20,50){\line(1,1){70}}
%\put(50,20){\line(1,1){70}}
%\put(20,90){\line(1,-1){70}}
%\put(20,90){\line(1,-1){70}}
%\put(50,120){\line(1,-1){70}}
%\put(20,90){\line(1,-1){70}}}
%\put(57,57){\scriptsize{$\mathcal{D}_1$}}
%\end{picture}
%\end{center}

The infinitesimal generator $\delta_{\D_1}$ of the modular group $\sigma_{\D_1}^t$ can directly be computed as in Proposition \ref{generator}. It seems to be a linear combination of the generator of time translations and that of proper conformal transformations:
\begin{gather*}
\delta_{\D_1}\varphi[f]:=\partial_s\sigma_{\D_1}^t\big(\varphi[f]\big)\big\arrowvert_{s=0}=\varphi\big[\partial_s f_s\big]\big\arrowvert_{s=0}
\end{gather*}
with
\begin{align*}
\partial_s f_s(x)\big\arrowvert_{s=0}&=\partial_s\gamma(x^0,x^3,s)\big\arrowvert_{s=0}f(x)+\partial_s f\big(x^\mu(-s)\big)\big\arrowvert_{s=0}\\
&=\Big(3x^0+\partial_s x^\mu(-s)\big\arrowvert_{s=0}\partial_{x^\mu}\Big)f(x)\\
&=\Big(3x^0+\frac{1}{2}\big(-1+(x^0)^2+\x^2\big)\partial_{x^0}+x^0\sum_ix^i\partial_{x^i}\Big) f(x),
\end{align*}
where we have made use of \eqref{Hislopaction}.\\

The forward light cone is another important example for which the modular objects can be computed. Analogous to the previous case one has the continuous transformation 
\begin{gather*}
T_{05}(\varphi)\xi:=\begin{cases} \xi^0(\varphi)=-\xi^0\cos\varphi+\xi^4\sin\varphi, &\\  \xi^5(\varphi)=-\xi^0\sin\varphi+\xi^1\cos\varphi, & \varphi\in[0,\pi],\\ \xi^i(\varphi)=\xi^i, & i=1,2,3,4,\end{cases}
\end{gather*}
of the double cone $\D_1$ onto the forward light cone $\V_+$:
\begin{equation*}
T_{05}\left(\frac{\pi}{2}\right)\D_1=\V_+,
\end{equation*}
\begin{align*}
T_{54}(s)\xi:&=T_{05}\left(\frac{\pi}{2}\right)T_{04}(s)\left(T_{05}\left(\frac{\pi}{2}\right)\right)^{-1}\xi\\
&=\begin{cases} \xi^4(s)+\xi^5(s)=e^{-s}(\xi^4+\xi^5), &\\ \xi^4(s)-\xi^5(s)=e^{s}(\xi^4-\xi^5), & s\in[0,\pi],\\ \xi^i(s)=\xi^i, & i=0,1,2,3.\end{cases}
\end{align*}
For conformally invariant theories this mapping is also unitarily implementable, and one directly obtains the modular objects for $\V_+$ as
\begin{gather*}
J_{\V_+}=U_{T_{05}}J_{\D_1}U_{T_{05}}^{-1}\;,\quad\text{and}\\
\Delta_{\V_+}^{it}=U_{T_{05}}\Delta_{\D_1}^{it}U_{T_{05}}^{-1}.
\end{gather*}

\begin{theo}[Buchholz]\label{Buchholz}
In a free massless quantum field theory the modular objects for the von Neumann algebra of observable fields $\Ma(\V_+)$ with reference to the vacuum state are
\begin{align*}
J_{\V_+}=\Gamma\big(I_tI_s\big)\quad\text{and}\quad\Delta_{\V_+}^{it}=\Gamma\big[U(e^{-s})\big],
\end{align*}
where $I_s$ is the space reflection operator and $s:=2\pi t$. The modular automorphism group acts geometrically as dilations:
\begin{gather*}
\sigma_{\V_+}^t\big(\varphi[f]\big)=\Delta_{\V_+}^{it}\varphi[f]\Delta_{\V_+}^{-it}=\varphi[f_{-s}]\\
\text{with}\quad f_s(x):=e^{3s}f(e^sx).
\end{gather*}
\end{theo}
Thus, the modular conjugation maps the forward light cone onto the backward light cone, i.e., $J_{\V_+}\Ma(\V_+)J_{\V_+}=\Ma(\V_-)$.

In this case the infinitesimal generator is given by
$$
\delta_{\V_+}\varphi[f]:=\partial_s\sigma_{\V_+}^t\big(\varphi[f]\big)\big\arrowvert_{s=0}=\varphi\big[\partial_s f_s\big\arrowvert_{s=0}\big]
$$
with
$$
\partial_sf_s(x)\big\arrowvert_{s=0}=\big(-3+x^\mu\partial_{x^\mu}\big)f(x).
$$
\\

Trebels examines the opposite direction \cite{Trebels:1997}. He starts from a physically well-defined one-parameter group of unitaries and compares their action with the known modular action on $\W_R$, $\V_+$ and $\D_1$. For double cones he can show that the assumption of local action implies the Hislop-Longo transformation \eqref{Hislopaction} up to a scaling factor.

\begin{defi}
A unitary transformation $V:\mathfrak{M}(\mathcal{K})\longrightarrow\mathfrak{M}(\mathcal{K})$, where $\mathcal{K}\subset\M$ is open, with $V\Omega=\Omega$ is called geometric, causal and order-preserving if there exists an isomorphism $g:\mathcal{K}\longrightarrow\mathcal{K}$ with the following properties:

\begin{itemize}
\item[(i)] $x\in\mathcal{K}\Longrightarrow g(x),g^{-1}(x)\in\mathcal{K}$.
\item[(ii)] If $x,\,y\in\mathcal{K}$ and $x-y$ is spacelike, then so are $g(x)-g(y)$ and $g^{-1}(x)-g^{-1}(y)$.
\item[(iii)] $(x-y)\in\V_+\Longrightarrow g(x)-g(y),\;g^{-1}(x)-g^{-1}(y)\in\V_+$.
\item[(iv)] $V\mathfrak{M}(\mathcal{K}_1)V^{-1}=\mathfrak{M}\big(g(\mathcal{K}_1)\big)$  for all subsets $\mathcal{K}_1\subset\mathcal{K}$.
\end{itemize}
\end{defi}
Geometric and causal one-parameter groups are automatically order preserving.

\begin{theo}\label{Trebels}
In the context of a free massless quantum field theory in the vacuum sector let a subset $\mathcal{K}$ of Minkowski space be given. Let us further assume that the unitary group $U_t$ has a geometric and causal action $g_t\neq\1_\mathcal{K}$ and that  for all $A\in\mathfrak{M}(\mathcal{K})$ the mapping $t\mapsto U_tA\Omega$ has  an analytic continuation on the strip $\mathbf{S}(0,\gamma)$ for some $\gamma>0$, then the following statements hold:

\begin{itemize}
\item[(i)] For $\mathcal{K}=\W_R\;$ $g_t$ is identical  to the action of Bisognano and Wichmann up to a scaling parameter. 
\item[(ii)]  For $\mathcal{K}=\V_+\;$ $g_t$ coincides with the dilation up to a scaling parameter.
\item[(iii)] For $\mathcal{K}=\D_1\;$ $g_t$ is identical to the Hislop-Longo transformation $g(t)=e^{mt\delta_{\D_1}}$ up to a scaling parameter.
\end{itemize}
\end{theo}
One can fix the scaling parameter to $1$ for wedges, forward and backward light cones with the help of translations which leave it invariant. Due to the absence of such translations for double cones, the scaling parameter $m$ is only known to be positive in general. For special cases of quantum field theories it can be shown that $0<m\leq 1$.

Fredenhagen \cite{Fredenhagen:1984dc} compares the result of Bisognano and Wichmann with that of Hislop and Longo. One expects a relativistic quantum field theory to be conformally invariant in the short distance limit (because of 1. the importance of chirality in the standard model and 2. the existence of an `ultraviolet fixed point', of a scaling limit, is of great significance for quantum field theories in curved spacetime). If one chooses a double cone $\D\subset\W_R$ which contains the origin in its closure, one gets
\begin{gather*}
|x|^{-1}\big(x_\pm(s)-\Lambda_s\big)x\longrightarrow 0\quad\text{as}\quad x\longrightarrow 0
\end{gather*}  
for $x\in\D$. The natural question is, if this coincidence of the two actions in the neighbourhood of the origin is of general validity. 

\begin{prop}
For each $f\in\mathbf{L}_1(\R)$ and $\lambda\in(0,1)$ there exists a constant $c_f(\lambda)>0$ such that the following statements hold:

\begin{itemize}
\item[(i)] $\|\int dt f(t)(\Delta_\D^{-it}-\Delta_{\W_R}^{-it})A\Omega\|^2\leq c_f(\lambda)\big(\|A\Omega\|^2+\|A^*\Omega\|^2\big)$ for all $A\in\mathfrak{M}(\lambda\D)$, and
\item[(ii)] $c_f(\lambda)\longrightarrow 0$ for $\lambda\longrightarrow 0$.
\end{itemize}
The constant $c_f(\lambda)>0$ depends neither on the details of the theory nor on the size of $\D$.
\end{prop}
Thus the action of the modular group for the double cone $\sigma_\D^t$ on $\mathfrak{M}(\lambda\D)$ becomes approximately geometrical, and it converges to that of the modular group for the wedge $\sigma_{\W_R}^t$. Therefore the spectra of the modular operators $\Delta_{\W_R}$ and $\Delta_\D$ have also to coincide, provided that the theory is asymptotically scaling-invariant. This information leads to the identification of the type of local von Neumann algebras.
\begin{prop}
If $\o$ is a faithful, normal state on a von Neumann algebra $\mathfrak{M}$ then a necessary and sufficient condition for $\lambda\in\R$ to be in $\mathbf{Spec}\Delta_\o$ is that for each $\epsilon>0$ there exists an $A\in\mathfrak{M}$ with $\o(A^*A)=1$ such that 
\begin{equation*}
|\o(AB)-\lambda\o(BA)|\leq\epsilon\big(\o(B^*B)+\lambda\o(BB^*)\big)^{\frac{1}{2}}
\end{equation*}
for all $B\in\mathfrak{M}$.
\end{prop}  
Fredenhagen then proves that the above condition is satisfied for all positive numbers $\lambda$ in a scaling-invariant theory.

\begin{theo}
Let $\mathfrak{M}$ be a von Neumann algebra with $\mathfrak{M}(\D)\subset\mathfrak{M}\subset\mathfrak{M}(\W_R)$. If a Wightman field theory with a scale-invariant limit is associated to the local net then the modular spectrum, see \eqref{modularspectrum}, is $S(\mathfrak{M})=\R_+$.
\end{theo}
Consequently, due to Theorem \ref{type}, he obtains the following result.

\begin{cor}
The von Neumann algebras of local observables are of type $III_1$.
\end{cor}

One might get the wrong impression of the modular group always acting locally, but Yngvason \cite{Yngvason:1994nk} dashes this hope with the construction of a simple counter-example. In the case of the left and right wedges he constructs a group of automorphisms which he proves to be the modular group by verifying the KMS condition. 

In the case of the right wedge the group is defined by the unitary operator $V_{\W_R}$ which acts first on the one-particle Hilbert space $\H_1:=\L^2\big(\R^n,M(p)d\mu(p)\big)$, as
\begin{gather*}
V_{\W_R}(\lambda)\varphi(p):=\frac{F(-\lambda p_+,-\lambda^{-1} p_-,-\hat{p})}{F(-p_+,-p_-,-\hat{p})}\varphi(\lambda p_+,\lambda^{-1} p_-,\hat{p}),
\end{gather*}
and then is lifted on to the Fock space by the second quantisation procedure. Here the function $F$ is assumed to have some kind of analytic properties by means of $V_{\W_R}$. The group reads  
\begin{gather*}
\sigma_{\W_R}^t\big(W(f)\big):=V_{\W_R}\big(e^{-2\pi t}\big)W(f)V_{\W_R}\big(e^{2\pi t}\big),
\end{gather*}
with the coordinates $\hat{p}:=(p^2,\cdots,p^n)$ and $p_\pm:=p^0\pm p^1$ and can explicitely be shown to act non locally in the directions parallel to the edge of the wedges. Since this pathological behaviour is of interest for our investigation in the sense that the non local action could give us a hint for the form of the infinitesimal generator of the modular group acting on the massive algebra, we will analyse this group in detail in the next chapter.

\section{More Applications of the Modular Action}

In Section \ref{localgKMS} we have already seen two important applications of modular theory in the algebraic approach to quantum field theory, namely the discussion on the type of local algebras and the mathematically rigorous formulation of thermal equilibrium states, the KMS states. Fortunately, these are not the only fields where modular theory enters the physical arena leading to great breakthroughs. In this section we present the main and prominent consequences of modular theory in quantum field theory, the Hawking effect, the PCT theorem, the construction of the Poincar\'e group, the algebraic spin-statistics theorem and the modular nuclearity condition. There are many other applications, some of them still under investigation, which we are not able to incorporate for the sake of brevity. For more information we refer the reader to \cite{Borchers:2000pv}.

\subsection*{Hawking Radiation and Unruh Effect}

Sewell shows in \cite{Sewell:1980} that the Bisognano-Wichmann theorem is closely related to the Unruh effect and Hawking radiation. In a classical treatment black holes can only absorb, and consequently their mass increases permanently. But if one takes into account also quantum field theoretical arguments, then this picture changes significantly. Hawking proves in \cite{Hawking:1974sw} that, due to vacuum fluctuations near the Schwarzschild radius, static and uncharged black holes are not so black, but rather radiate constantly like black bodies with the temperature 
\begin{gather}\label{Haking-temperatur}
T=\frac{a}{2\pi},
\end{gather}
where $a$ is the acceleration. Motivated by this achievement, Unruh finds that a uniformly accelerated observer feels the Minkowski vacuum like a thermal bath at the so-called Unruh temperature given above in \eqref{Haking-temperatur} \cite{Unruh:1976db}. 

Now let us imagine, for the sake of simplicity, an observer uniformly accelerated in the $x^1$-direction, i.e., his acceleration vector is given as $(0,a,0,0)^t$, where $a$ is the magnitude of the absolute value of the acceleration. The observer's worldline is then given by the trajectory of the point
$$
x=\left(0,a^{-1},0,0\right)^t
$$
under the boosts in $x^1$-direction, the modular action in the right Rindler wedge of Theorem \ref{Bisognano},
$$
\Lambda_sx=\begin{pmatrix} \cosh s & \sinh s & 0 & 0\\  \sinh s & \cosh s & 0 & 0\\  0 & 0  & 1 & 0\\ 0 & 0 & 0 & 1 \end{pmatrix}\begin{pmatrix}0 \\a^{-1} \\ 0 \\0\end{pmatrix}=\begin{pmatrix}a^{-1}\sinh s \\a^{-1}\cosh s \\ 0 \\0\end{pmatrix}.
$$
Replacing the boost parameter by the proper time $\tau=a^{-1}s$, we find that the operator $aK$, where $K$ is the infinitesimal generator of the Lorentz boosts in $x^1$-direction, generates time translation in the observer's rest frame, i.e.,
$$
\Lambda_{a\tau}x(\tau_0)=x(\tau_0+\tau).
$$ 
Applying the Bisognano-Wichmann Theorem \ref{Bisognano}, we can compare this operator with the modular operator of the modular group for the Rindler wedge, and see that
$$
U(\Lambda_{a\tau})=e^{ia\tau K}=e^{2\pi itK}=\Delta_{\W_R}^{it}
$$
holds if and only if $a\tau=2\pi t$. Hence we get the Unruh temperature,
$$
T_{\W_R}=\frac{1}{\beta}=\frac{t}{\tau}=\frac{a}{2\pi}, 
$$
where $\beta$ is the inverse of the temperature defined through the KMS condition. 

The Unruh effect does not explain the Hawking effect  in a mathematically rigorous manner, and therefore we should not expect the Bisognano-Wichmann theorem to do so. Nevertheless, their close relationship may be justified easily insofar as the boundary of the Rindler wedge can be interpreted as a simple case of a horizon, because a signal sent across this boundary will never be responded. 

Since the Bisognano-Wichmann theorem is only built upon the general framework of axiomatic quantum field theory, it ensures the validity of the Unruh effect for all field theories, in particular for the interacting ones. Unruh investigates only free field theories. On the other hand, the Bisognano-Wichmann Theorem can hardly be generalised to curved spacetimes, because it is based on a self-adjoint and semi-bounded Hamiltonian operator whose existence is not assured in non-stationary spacetimes.

In the same manner, i.e., with the help of the modular operators for double cones and forward light cones, one can calculate the Unruh effect for the diamond and the forward light cone, as done recently by Martinetti and Rovelli \cite{Martinetti:2002sz}. Their analysis is mainly based on the known modular actions on the wedges, Theorem \ref{Bisognano}, light cones, Theorem \ref{Buchholz}, and double cones, Theorem \ref{Hislop-Longo}, and the so-called thermal time hypothesis, which has first been introduced by Connes and Rovelli \cite{Connes:1994hv}. This hypothesis says that the physical time is state-dependent, and when the system is in a state $\o$ then time is defined by the modular group with respect to $\o$. One consequence is that if there exist to independent definitions of time flow, e.g., thermal and geometrical ones, then the temperature is defined by their ratio.

Martinetti and Rovelli investigate the case of an observer with finite lifetime and calculate the diamond's temperature to be
$$
T_\D(\tau)=\frac{a^2L}{2\pi\big(\sqrt{1+a^2L^2}-\cosh a\tau\big)},
$$
where $L$ is the radius of the diamond's basis, $|x_0|+|\x|<L$. First of all, for large acceleration $a$, representing an observer travelling near the boundary of the diamond and for large $L$ we obtain the temperature
$$
T_\D(\tau)=\frac{a}{2\pi\big(1-\frac{\cosh a\tau}{aL}\big)}.
$$
This function rapidly approaches infinity at the boundary of the observer's proper time, but nearly stays constant for most of his lifetime at the Unruh temperature. Thus, the observer with finite lifetime $\frac{L}{2}$ experiences the Unruh effect for most of his lifetime; however, the situations shortly after his birth and before his death remain to be discussed.

This investigation is even more striking for an unaccelerated observer, as the assumption $a=0$ leads to a non vanishing temperature which has its maximum
$$
\beta_\D=\frac{1}{\pi L}
$$
at $x^0=0$. In the case of the forward light cone the temperature is given by 
$$
T_{\V_+}(\tau)=\frac{1}{2\pi}e^{-a\tau},
$$
and consequently the temperature at the birth of a uniformly accelerated observer with $a\neq 0$ is positive and converges rapidly to zero. An unaccelerated observer would still feel a non vanishing temperature.

\subsection*{PCT Theorem}

Jost gives the first proof of the PCT theorem in 1957 \cite{Jost:1957}. It took more than three decades to give an algebraic version of this theorem, i.e., within the framework of the algebra of local observables $\Ma(\O)$. This is mainly due to Borchers \cite{Borchers:1991xk}, \cite{Borchers:2000fw}. 

The PCT theorem states that the product of space reflection, charge conjugation and time reversal is a symmetry. The algebraic proof of this statement in principle reduces to establishing the existence of the PCT operator $\Theta$, which is determined by the following properties:

\begin{itemize}
\item[(i)] The operator $\Theta$ is anti-unitary, and for all bounded regions $\O\subset\M$
$$
\Theta\pi_0\big(\Ma(\O)\big)\Theta=\pi_0\big(\Ma(-\O)\big),
$$
where $\pi_0$ is the cyclic representation with respect to the vacuum state.
\item[(ii)] $\Theta$ fulfills the following commutation relation with Poincar\'e transformations,
$$
\Theta U(\Lambda,a)\Theta=U(\Lambda,-a).
$$
\item[(iii)] $\Theta$ maps a charge sector into its conjugate sector.
\end{itemize}
Up to the formulation of the modular theory and its application in algebraic quantum field theory, the Bisognano-Wichmann Theorem, it seems to have been impossible to verify the PCT theorem in terms of local algebras. But the Bisognano-Wichmann Theorem supplies Borchers with a possible candidate for the PCT operator, 
$$
\Theta=J_{\W_+}U\big(R_1(\pi)\big).
$$
He shows the defining properties of $\Theta$ by clarifying the commutation relation between the modular automorphism group and the translations in a first step.

\begin{theo}[Borchers]
Let $\Ma$ be a von Neumann algebra with a cyclic and separating vector $\Omega$  and $U(t)$, $t\in\R$, a one-parameter group fulfilling the requirements:

\begin{itemize}
\item[(i)] $U(t)\Omega=\Omega$ for all $t\in\R$.
\item[(ii)] $\spec U(t)\subset\R_+$.
\item[(iii)] $U(t)\Ma U(-t)\subset\Ma$ for all $t\in\R$.
\end{itemize}
Then the following statements hold:
\begin{gather*}
\Delta^{it}U(s)\Delta^{-it}=U\left(e^{-2\pi t}s\right),\\
\text{and}\quad JU(s)J=U(-s).
\end{gather*}
\end{theo}  

A converse of this theorem has been proved by Wiesbrock \cite{Wiesbrock:1992mg}, \cite{Wiesbrock:1996rq}, \cite{Wiesbrock:1996rp}. He starts from a von Neumann subalgebra $\Na$ of $\Ma$ satisfying the condition of $+$-half sided modular inclusion (or $-$-half sided modular inclusion), i.e., $\Na$ has the same cyclic and separating vector $\Omega$ and 
\begin{equation}\label{half-sided}
\Delta^{it}\Na\Delta^{-it}\subset\Na 
\end{equation}
for all $t\geq 0$ ($t\leq 0$). Then he shows that there always exists a one-parameter group $U(t)=e^{iHt}$ with positive infinitesimal generator $H$, which leaves $\Omega$ invariant and satisfies
$$
U(1)\Ma U(-1)=\Na.
$$
Apart from yielding the PCT operator, the Bisognano-Wichmann Theorem ensures the validity of wedge duality, see Definition \ref{wedgeduality}, which is required for the next statement.

\begin{theo}[PCT, Borchers]
Every Poincar\'e-covariant quantum field theory of local observables satisfying wedge duality and the reality condition, see Definition \ref{reality}, is invariant under the PCT operations.
\end{theo}

\subsection*{Construction of the Poincar\'e Group}

We have seen that the modular group for wedge regions acts as Lorentz boosts, and therefore one can pose the question if this fact is sufficient for the construction of a representation of the Poincar\'e group. If the answer is positive, then this fact can be used to discriminate the real relativistic, i.e., Poincar\'e-covariant, vacuum from all the other cyclic and separating states fulfilling the Reeh-Schlieder property of Theorem \ref{Reeh-Schlieder}. This would also mean that all spacetime symmetries are already encoded intrinsically in the net of von Neumann algebras of local observables.

 Borchers shows in \cite{Borchers:1991xk} that the two-dimensional case, where all wedges are described by translations of the left or right wedge, can be covered with the help of Wiesbrock's construction of the translations \cite{Wiesbrock:1992mg}\footnote{Concerning the results of this paper confer the remark on p. \pageref{Wiesbrock-erratum}.}. The higher-dimensional problem has been solved by Brunetti, Guido and Longo whose proof is built up on Moore's theory of central extension of locally compact groups by Polish groups \cite{Brunetti:1992zf}. They construct a representation only of the covering $\tilde{\mathcal{P}}_+^\uparrow$ of the proper orthochronous Poincar\'e group $\mathcal{P}_+^\uparrow$. 

\begin{theo}[Brunetti, Guido, Longo]
Let the modular group with respect to the cyclic and separating vector $\Omega$ act on the von Neumann algebra of local observables $\Ma(\O)$ associated with an open and bounded region $\O$. If the Bisognano-Wichmann property holds, then there exists a unitary representation $U$ of $\tilde{\mathcal{P}}_+^\uparrow$, determined by the equation
$$
U\big(\tilde{\Lambda}_{\W_R}(t)\big)=\Delta_{\W_R}^{it},\quad t\in\R,
$$
which preserves the vacuum, $U\Omega=\Omega$, satisfies the spectrum condition and is $\Ma(\O)$-covariant, i.e., the local algebras transform covariantly under $U$. Furthermore, wedge duality holds.
\end{theo}
The invariance of the vector $\Omega$ with respect to the representation $U$ of $\tilde{\mathcal{P}}_+^\uparrow$ justifies treating $\Omega$ as the vacuum vector. Consequently, if one restricts the discussion to this case, then it should be possible to solve the problem for $\mathcal{P}_+^\uparrow$ itself. In a later publication Guido and Longo prove the existence of the unitary $U$ for the Poincar\'e group $\mathcal{P}_+^\uparrow$ as a byproduct of their general algebraic spin-statistics theorem \cite{Guido:1995fy}.

Borchers succeeds in showing directly from the Bisognano-Wichmann property the existence of such a representation for $\mathcal{P}_+^\uparrow$ by making use of the modular intersection property, a concept of Wiesbrock \cite{Wiesbrock:1996rq}, \cite{Wiesbrock:1996rp}.

\begin{defi}
The triple $(\Ma,\Na,\Omega)$, where $\Ma$ and $\Na$ are von Neumann algebras with the same cyclic and separating vector $\Omega$, is said to fulfill the $\pm$-modular intersection property if the following two conditions hold:

\begin{itemize}
\item[(i)] $\Ma\cap\Na$ satisfies the condition of $\pm$-half sided modular inclusion \eqref{half-sided} with respect to $\Ma$ and $\Na$.
\item[(ii)] $J_\Na\big(s-lim_{t\rightarrow\pm\infty}\Delta_\Na^{it}\Delta_\Ma^{-it}\big)J_\Na=s-lim_{t\rightarrow\pm\infty}\Delta_\Na^{it}\Delta_\Ma^{-it}$.
\end{itemize}
\end{defi} 
Borchers continues his presentation with von Neumann algebras $\Ma_{i,k}$,where $i\neq k$ and $\Ma_{i,k}'=\Ma_{k,i}$, which satisfy the subsequent requirements (apart from some additional technical ones):

\begin{itemize}
\item[(i)] All algebras $\Ma_{i,k}$ have a common cyclic and separating vector $\Omega$, and the triples $(\Ma_{i,k},\Ma_{i,j},\Omega)$ fulfill the conditions of $+$-modular intersection and of $-$-modular intersection.
\item[(ii)] The modular operators $\Delta_{j,k}^{it_{j,k}}$ generate a six-dimensional Lie group. 
\item[(iii)] There exists an algebra $\Ma_{1,2,\lambda}$ such that the triple $(\Ma_{1,2,\lambda},\Ma_{1,2},\Omega)$ satisfies the condition of $+$-half sided inclusion.
\end{itemize}
Using all these assumptions he then shows in \cite{Borchers:1998nf} that, if $\Ma_{i,j}$, $1\leq i\leq j\leq 4$, are six von Neumann algebras satisfying the conditions $(i)$ and $(ii)$ and $\Ma_{1,2,\lambda}$ fulfills the condition $(iii)$, then the corresponding modular groups generate a continuous representation of the Poincar\'e group $\mathcal{P}_+^\uparrow$ (in four dimensions) which satisfies the spectrum condition.

\subsection*{Algebraic Spin-Statistics Theorem}

Another main supporting pillar of quantum field theory, which can be verified algebraically with the help of modular theory, is the connection between spin and statistics, namely that it is not possible to quantise consistently an integer spin system by Fermi statistics and a half-integer system by Bose statistics. This has been established by Guido and Longo in \cite{Guido:1995fy}, starting from the local field algebra $\Fa(\O)$ and assuming the following conditions to hold:

\begin{itemize}
\item[(1)] The Reeh-Schlieder property, Theorem \ref{Reeh-Schlieder}, for spacelike cones.
\item[(2)] Normal commutation relations, i.e., there exists a vacuum-preserving, selfadjoint and unitary operator $\Gamma$, the so-called statistics operator, which implements an automorphism on every local field algebra $\Fa(\O)$, and the field algebras
$$
\Fa_\pm(\O):=\big\{F\in\Fa(\O)|\;\Gamma F\Gamma=\pm F\big\}
$$
satisfy graded locality \eqref{gradedlocality}.
\item[(3)] Modular covariance, i.e., the field algebra fulfills the relation
$$
\Delta_\W^{it}\Fa(\O)\Delta_\W^{-it}=\Fa\big(\Lambda_\W(t)\O\big),\quad t\in\R,
$$
for all wedges $\W$ and all regions $\O$.
\end{itemize}
Their proof is based on a former result of Longo \cite{Longo:1989tt}, \cite{Longo:1990zp}, who connects the statistical dimension with the Jones index, see Theorem \ref{Jones}. Let us consider for this purpose an irreducible endomorphism $\rho:\Ma\longrightarrow\Ma$ localised in a double cone $\O$ and a unitary element $U\in\Ma$ such that the endomorphism, 
$$
\rho'(\cdot):=U\rho(\cdot)U^{-1},
$$
is localised in the spacelike complement of $\O$. Then the statistical operator, defined as 
\begin{gather*}
\epsilon:=U^{-1}\rho(U),
\end{gather*}
can be shown to be a unitary element of $\rho^2(\Ma)\cap\Ma$ and to fulfill
\begin{gather*}
\rho(\epsilon)\epsilon\rho(\epsilon)=\epsilon\rho(\epsilon)\epsilon.
\end{gather*}
In the context of Doplicher-Haag-Roberts (DHR) theory, where $\rho$ represents superselection sectors, a representation of the permutation group $\mathbb{S}_\infty$ can be introduced for spacetime dimension four via the equation
\begin{gather*}
\epsilon_i:=\rho^{i-1}(\epsilon), 
\end{gather*}
which can be classified by the value of the so-called statistical parameter,
\begin{gather*}
\lambda_\rho:=\phi(\epsilon)=0,\pm1,\pm\frac{1}{2},\pm\frac{1}{3},\pm\frac{1}{4},\cdots,
\end{gather*}
where $\phi$ is the left inverse of $\rho$. From this Longo derives the index-statistics relation, i.e., a relation between the Jones index Ind$(\rho)$ and the DHR statistical dimension $d(\rho)$:
\begin{gather*}
d(\rho):=|\lambda_\rho|^{-1}=\text{Ind}(\rho)^{1/2}.
\end{gather*}
In a final step the assumption of modular covariance bridges the gap between the local field algebra $\Fa(\O)$ (local net of observables $\Ma(\O)$) and the statistical parameter $\lambda_\rho$, because given this a representation of the Poincar\'e group with respect to a covariant irreducible superselection sector $\rho$ will be uniquely determined and this uniqueness connects them intrinsically.

\begin{theo}[Guido-Longo]\label{spin-statistics}
${}$\\
\vspace*{-0.4cm}
\begin{itemize}
\item[(i)] Let $\Fa(\O)$ be a local field algebra satisfying the conditions (1), (2) and (3) given above, then 
$$
\Gamma=U(2\pi)
$$
holds.
\item[(ii)] For an irreducible, modular-covariant local net of observables  $\Aa(\O)$ on Minkowski space and an irreducible, covariant and localized endomorphism $\rho$ with finite statistics one has
$$
U_\rho(2\pi)=\text{sign}(\lambda_\rho),
$$
where $U_\rho$ is the representation of $\tilde{\P}_+^\uparrow$ in the sector $\rho$.
\end{itemize}
\end{theo}
In the same paper \cite{Guido:1995fy} Guido and Longo also give a proof of the PCT theorem within the formalism of local field algebras. For this purpose they consider the intersection 
$$
\Fa^d(\O):=\bigcap_{\W\supset\O}\Fa(\W),
$$
which fulfills the Reeh-Schlieder property, locality and duality for convex, causally complete regions.

\begin{theo}[Guido-Longo]
 Let $\Fa(\O)$ be a local field algebra satisfying the conditions (1), (2) and (3) given above, then there exists an anti-unitary operator $\Theta$ implementing the PCT symmetry on the local field algebra $\Fa^d(\O)$:
\begin{gather*}
\Theta\Fa^d(\O)\Theta=\Fa^d(-\O).
\end{gather*}
\end{theo}

\subsection*{Modular Nuclearity Condition}

The axioms for the nets of local algebras do not yet implement one of the main concepts of high energy physics, namely the concept of particles. This problem is closely connected with the properties of the phase space volume associated with physical states of bounded energy which are localised in spacetime. In the algebraic formulation of quantum field theory there are several proposals how this volume should depend on the energy and localisation in order to allow for a particle interpretation.

Haag and Swieca were the first to give a characterisation of theories with a particle interpretation \cite{Haag:1965}. They have been motivated by the fact that in quantum mechanics, due to the uncertainty relation, only finitely many quantum states fit in to a bounded subset of phase space. Therefore, they formulate a compactness criterion for states in a phase space volume associated with energy-damped local excitations of the vacuum state. To be more precise, for given $\beta>0$ and bounded spacetime regions $\O$ they consider special maps defined as
\begin{align*}
\Theta_{\beta,\O}:\Aa(\O)&\longrightarrow\H\\
A&\mapsto\Theta_{\beta,\O}(A):=e^{-\beta H}A\Omega,
\end{align*}
where $H$ is the Hamiltonian and $\Omega\in\H$ the vacuum vector. They claim that in theories allowing for a particle interpretation the following condition holds true: 
\begin{flushleft}\textbf{Compactness condition (Haag-Swieca)}: The maps $\Theta_{\beta,\O}$ are compact in the norm topology for all $\beta>0$ and any bounded spacetime region $\O$.\end{flushleft}

Fredenhagen and Hertel \cite{Fredenhagen:1981zw} start from global states of limited energy and consider their restriction to local algebras. For each real number $\beta>0$  they first choose a subset of the space of normal linear functionals $\S_N$ on $\BH$, 
$$
\S_\beta:=\Big\{\o\in\S_N|\;e^{\beta H}\o e^{\beta H}\in\S_N\Big\},
$$
which is a Banach space of functionals of limited energy when equipped with the norm $\|\o\|_\beta:=\|e^{\beta H}\o e^{\beta H}\|$. Localisation in configuration space is introduced by restriction of these functionals to the local algebras $\Aa(\O)\subset\BH$. The map to be considered in this ansatz is defined as
\begin{align*}
\Pi_{\beta,\O}:\S_\beta&\longrightarrow\Aa(\O)^*\\
\o&\mapsto\Pi_{\beta,\O}(\o):=\o\arrowvert_{\Aa(\O)},
\end{align*}
and one postulates for quantum field theories with decent properties:
\begin{flushleft}\textbf{Compactness condition (Fredenhagen-Hertel)}: The maps $\Pi_{\beta,\O}$ are compact for all $\beta>0$ and any bounded spacetime region $\O$ in the norm topology.\end{flushleft}
In contradistinction to the approach of Haag and Swieca, the order of energy damping and localisation in configuration space is closer to the algebraic formulation and can be applied to all superselection sectors.

Buchholz and Wichmann were the first to introduce a quantitative description of the phase space in quantum field theory in terms of a nuclearity criterion. This does not only state the qualitative property of compactness of the phase space volume, but also requires a certain dependence on the available energy and localisation region \cite{Buchholz:1986dy}. They investigate the map
\begin{align*}
\tilde{\Theta}_{\beta,\O}:\Aa(\O)&\longrightarrow\H\\
A&\mapsto\tilde{\Theta}_{\beta,\O}(A):=e^{-\beta H}A\Omega
\end{align*}
and formulate their condition in terms of the notion of nuclearity.

\begin{defi}\label{nuclearity}
Let $\X$ and $\Y$ be Banach spaces. Then the linear operator $\Theta:\X\longrightarrow\Y$ is said to be nuclear if there exist a sequence of continuous linear functionals $\o_i\in\X^*$ and a sequence of vectors $Y_i\in\Ba$ such that 
\begin{gather*}
\Theta(X)=\sum_{i=1}^\infty \o_i(X)Y_i \quad\text{ for all}\quad X\in\X,\\
\text{with}\quad \sum_{i=1}^\infty\|\o_i\|\cdot\|Y_i\|<\infty.
\end{gather*}
Given this one defines the nuclearity index of $\Theta$ as
$$
\nu(\Theta):=\inf\left\{\sum_{i=1}^\infty\|\o_i\|\cdot\|Y_i\|\right\},
$$
where the infimum extends over all decompositions of $\Theta$ complying with the above properties.
\end{defi}

\begin{flushleft}\textbf{Nuclearity condition (Buchholz-Wichmann)}: The maps $\tilde{\Theta}_{\beta,\O}$ are nuclear for all $\beta>0$ and any bounded spacetime region $\O$, and their nuclearity index $\nu_{\beta,\O}(\Theta)$  is bounded by 
\begin{gather}\label{Buchholz-Wichmann}
\nu_{\beta,\O}(\Theta)<e^{cr^3\beta^{-n}},
\end{gather}
where $r$ is the spatial radius of $\O$ and $c$ and $n$ are constants.
\end{flushleft}

This formulation of the nuclearity condition is not completely local, since it makes use of the global Hamiltonian as well as of the global vacuum state. Exactly at this point the modular theory, to be more precise the Bisognano-Wichmann Theorem, plays a decisive role. Buchholz, D'Antoni and Longo give a local nuclearity condition in terms of the modular operator \cite{Buchholz:1989bj}, local in the sense that only information about the restriction of the vacuum state $\o_0$ to the local algebras $\Aa(\O)$ is needed. Moreover, this nuclearity condition is applicable to theories on any spacetime manifold, because the vacuum can be replaced by an arbitrary dense set of vectors. The idea for a modular formulation is that the Hamiltonian can be naturally connected with the modular operator $\Delta_{\W_\tau}=U\big(\Lambda(2\pi\tau),0\big)$ corresponding to the local algebra $\Aa(\W_\tau)$ on the wedge-shaped region
$$
\W_\tau:=\big\{x\in\mathbb{M}|\;|x^0|\leq x^1+\tau\big\}
$$
and the vacuum vector $\Omega$, where $U$ is the representation of the Poincar\'e group on the Hilbert space $\H$. In fact, one may prove the following relation:
$$
e^{-\sin 2\pi v\tau H}A\Omega=U\big(\Lambda(\mp iv),0\big)V_\pm(iv)U\big(1,\mp[1-\cos 2\pi v]\underline{\tau}\big)A\Omega
$$
with
$$
V_\pm(u):=U\big(1,\mp\underline{\tau}\big)U\big(\Lambda(\pm u),0\big)U\big(1,\mp\underline{\tau}\big)^{-1}
$$
for all $0\leq v<\frac{1}{4}$ and all operators 
$$
A\in\Aa\Big(\W_R+[1-\cos 2\pi v]\underline{\tau}\Big)\bigcap\Aa\Big(-\W_R+[1-\cos 2\pi v]\underline{\tau}\Big),
$$
where $\underline{\tau}$ denotes the spacelike vector, whose components are all zero apart from $\underline{\tau}^1=\tau>0$.

Now let $\O_1$ and $\O_2$, $\O_1\subset\O_2$, be two arbitrary bounded regions of Minkowski space with $\O_2\subset\W_\tau\cap(-\W_\tau)$, and consider the map
\begin{align*}
\Xi_{\O_1,\O_2}:\Aa(\O_1)&\longrightarrow\H\\
A&\mapsto\Xi_{\O_1,\O_2}(A):=\Big(1+\Delta_{\O_2}^{-1/2}\Big)A\Omega,
\end{align*}
where $\Delta_{\O_2}$ is the modular operator associated with the local algebra $\Aa(\O_2)$ and the vacuum vector $\Omega$. $N(\epsilon)$ is called the $\epsilon$-content of the map $\Xi_{\O_1,\O_2}$, if for a given $\epsilon>0$ $N(\epsilon)$ is the maximal number of elements $A_i\in\Aa_\1(\O_1)$, the unit ball of $\Aa(\O_1)$, $i=1,\cdots,N(\epsilon)$, such that $\|\Xi_{\O_1,\O_2}(A_i-A_j)\|>\epsilon$ for all $i\neq j$. The map $\Xi_{\O_1,\O_2}$ is compact if and only if its $\epsilon$-content $N(\epsilon)$ is finite for all $\epsilon>0$.

\begin{defi}
Let $\epsilon$, $N(\epsilon)$ and $\Xi_{\O_1,\O_2}$ be as aforementioned, then one defines, if existent, the order $q$ of the map $\Xi_{\O_1,\O_2}$ as 
$$
q:=\limsup_{\epsilon\;\searrow\; 0}\frac{\ln\ln N(\epsilon)}{\ln \epsilon^{-1}}.
$$
\end{defi}
With these notations we are now ready to formulate the local nuclearity criterion:
\begin{flushleft}\textbf{Modular nuclearity condition (Buchholz-D'Antoni-Longo)}: The maps $\Xi_{\O_1,\O_2}$ have to be of arbitrary small order if the inner distance between $\O_1$ and $\O_2$ is sufficiently large.
\end{flushleft}

The next task is to combine the compactness requirement of Fredenhagen and Hertel with the notion of $p$-nuclearity, a generalisation of the nuclearity given in Definition \ref{nuclearity}, in order to give a more precise description of the size of the phase space in quantum field theory. 

\begin{defi}\label{p-nuclearity}
Let $\X$ and $\Y$ be Banach spaces. Then the linear operator $\Theta:\X\longrightarrow\Y$ is said to be p-nuclear, where $0<p\leq 1$\footnote{The upper bound was mentioned in \cite{Fewster:2004} and will be discussed in the next page.}, if there exist a sequence of continuous, linear functionals $\o_i\in\X^*$ and a sequence of vectors $Y_i\in\Ba$ such that 
\begin{gather*}
\Theta(X)=\sum_{i=1}^\infty \o_i(X)Y_i \quad\text{for all }\quad X\in\X\\
\text{with}\quad \sum_{i=1}^\infty\|\o_i\|^p\cdot\|Y_i\|^p<\infty.
\end{gather*}
Combinations of such functional and vectors are called p-nuclear decompositions of $\Theta$. Furthermore, one defines the $p$-nuclearity index of $\Theta$ as
\begin{equation}\label{p-nuclearityind}
\nu_p(\Theta):=\underset{p-nucl. decom.}{\inf}\left(\sum_{i=1}^\infty\|\o_i\|^p\cdot\|Y_i\|^p\right)^{1/p},
\end{equation}
where the infimum extends over all $p$-nuclear decompositions of $\Theta$.
\end{defi}

Buchholz and Porrmann examine this problem and give the strongest of all phase space conditions, which is satisfied by the free field theory of massive and massless particles in four spacetime dimensions \cite{Buchholz:1989eb}.

\begin{flushleft}\textbf{Nuclearity condition (Buchholz-Porrmann)}: The maps $\Pi_{\beta,\O}$ are $p$-nuclear for all sufficiently large $\beta>0$ and any bounded spacetime region $\O$.
\end{flushleft}
To put it in a nutshell, one verifies the following diagram showing the relations between the different nuclearity and compactness conditions:\\

\begin{center}
\begin{tabular}{ccc}
Buchholz-Porrmann &$\Longrightarrow$& Buchholz-Wichmann\\
\\
$\Big\Downarrow$ && $\Big\Downarrow$\\
\\
Fredenhagen-Hertel &$\Longrightarrow$& Haag-Swieca\hspace{5mm},\\
\end{tabular}
\end{center}
\vspace{0,6cm}
where the arrows denote real implications. In this diagram neither an additional relation can be added, nor any of the existing ones can be reversed.

In a recent paper Fewster, Ojima and Porrmann, in the context of their investigation of a possible equivalence relation between nuclearity criteria and quantum energy inequalities, which restrict the violation of the classical energy conditions to the amount compatible with the uncertainty relations of quantum theory, mention the occurrence of two inconsistencies \cite{Fewster:2004}. First, one should be aware of the different definitions of $p$-nuclearity for $p\geq 1$ given by mathematicians, e.g. \cite{Jarchow:1981}, and physicists, which may provoke confusion.

 Their second concern is of greater importance for the whole concept, since they can show that the $p$-nuclearity index \eqref{p-nuclearityind} vanishes for any $p>1$. In order to save the Definition \ref{p-nuclearity}, they suggest to further restrict the set of admissible $p$-nuclear decompositions. One could demand, for example, linear independence of the vectors $(Y_i)_{i\in\mathbb{N}}$ of the sequence, which still does not suffice. In the special case of Hilbert spaces one could also restrict attention to $p$-nuclear decompositions in terms of orthonormal bases.

%%%%%%%%%%%%%%%%%%%%%%%%%%%%%%%%%%%%%%%%%%%%%%%%%%%%%%%%%%%%%%%%%%%%%%%%%%%%

\chapter{Modular Group on the Massive Algebra}

\begin{flushright}
\emph{Beweisen muss ich den K\"as'\\
sonst ist die Arbeit unseri\"os.}\\
\vspace{0,5cm}
F. Wille
\end{flushright}
\vspace{0,5cm}

We have seen in the previous chapter how the knowledge of the modular group has led to a variety of most important applications. Modular theory in general seems to become more and more a powerful tool for diverse problems and the natural formalism by which local quatum physics may be formulated. But the potential of modular theory will not be exhausted fully as long as the modular group $\sigma_m^t$ acting on the massive algebras $\Ma_m(\O)$ is not determined. With $\sigma_m^t$ one would obtain a deeper and easier accessible understanding of the dynamics of quantum systems.\\
Due to the result of Trebels, see Theorem \ref{Trebels}, $\sigma_m^t$ has to act non-locally on $\Ma_m(\O)$, otherwise it would coincide with $\sigma_0^t$ up to a scaling factor. Thus, examples of modular automorphism groups acting non-locally may serve as a testing ground for the aforementioned assumptions on their infinitesimal generators $\delta_m$, namely, they have to be of the form \eqref{assumption-gen}.

\begin{assump}\label{assumption-gen0}
 The massive infinitesimal generator $\delta_m$ has the following structure:
\begin{gather}\label{assumption-gen1}
\delta_m=\delta_0+\delta_r,
\end{gather}
where $\delta_0$ is the massless generator, satisfying the following properties:
\begin{itemize}
\item[(i)] the known massless generator $\delta_0$ is the principal term in $\delta_m$;
\item[(ii)] $\delta_r$ is a PsDO;
\item[(iii)] the order of $\delta_r$ is less than $1$.
\end{itemize}
\end{assump}

The first part of our investigation begins with the motivation of these assumptions. In Section $2$, we will then verify these assumptions for the first time explicitly for two concrete examples of modular groups with non local action. The first example is due to Yngvason from his analysis of essential duality \cite{Yngvason:1994nk}, and the second one is given by Borchers and Yngvason in the context of the formulation of modular groups with respect to general KMS states \cite{Borchers:1998ye}. In the subsequent Section we discuss the ansatz of Figliolini and Guido who treat the modular operator and its continuity properties in the second quantisation formalism \cite{Figliolini:1989vf}. The last two Sections contain two approaches of our own. The first one is based on the unitary equivalence of free local algebras and the second one relies on the cocycle theorem of Connes, Theorem \ref{Connes-theorem}.

\section{Why a Pseudo-Differential Operator?}

The assumption that the infinitesimal generator $\delta_m$ of the modular group $\sigma_m^t$ acting on the massive algebra $\Ma_m(\D)$, where $\D$ is a double cone, is a pseudo-differential operator, is mentioned explicitly first by Schroer and Wiesbrock \cite{Schroer:1998ax}. However, the authors restrict themselves to a few remarks on their strategy how to calculate $\delta_m$. This strategy supposes an intermediate step, namely the computation of the modular group $\sigma_{m,0}^t$ on $\Ma_m(\D)$ with respect to the `wrong' massless vacuum vector $\Omega_0$. They claim, without giving a proof, to have shown this to lead to a pseudo-differential operator for the infinitesimal generator $\delta_{m,0}$, whose principal symbol is identical to the infinitesimal generator $\delta_{0}$, a differential operator of first order, derived by Hislop and Longo for the massless algebra $\Ma_0(\D)$ with respect to $\Omega_0$, see Theorem \ref{Hislop-Longo}. Furthermore, they state that its action ``on a smaller massive subalgebra inside the original one is not describable in terms of the previous subgroup'', and ``the geometrical aspect of the action is wrecked by the breakdown of Huygens principle, which leads to a nonlocal reshuffling inside $\D$ but is still local in the sense of keeping the inside and its causal complement apart'' \cite{Schroer:1998ax}. They propose to derive, in a final step, the modular group $\sigma_m^t$ and its generator $\delta_m$ with respect to the `right' massive vacuum vector $\Omega_m$ via Connes' cocycle theorem,
\begin{gather*}
\sigma_m^t(A)=\Gamma_t\sigma_{m,0}^t(A)\Gamma_t^*,
\end{gather*}
for all $A\in\Ma_m(\D)$, where $\Gamma_t$ is the $\sigma$-strongly continuous family of unitaries defined in Theorem \ref{Connes-theorem}. Schroer and Wiesbrock assume that this procedure will not change the pseudo-differential nature of the infinitesimal generator.\\

\subsection*{$\mathbf{\delta_m}$ Has to Be Timelike}

It is well known that, due to the theorem of Noether, a conservation law is the consequence of a continuous symmetry group, and in our special case symmetry under time translations, i.e., the existence of a time-independent Hamiltonian, leads to the conservation of energy. The physically well-motivated demand of such a conservation condition determines the causal property of the Killing vector field $X^\mu$, or in our case of the infinitesimal generator $\delta_m$, see \cite{Komar:1958wp}, \cite{Wald:1984rg} and \cite{Stephani:2004ud} . \\
In general relativity, i.e., in a Riemannian spacetime, the stress-energy tensor satisfies the local conservation law
\begin{equation}\label{energy}
T^{\mu\nu}{}_{;\nu}=0,
\end{equation}
where the semicolon denotes the covariant derivative,
\begin{gather*}
A^{\mu}{}_{;\nu}:=\partial_\nu A^\mu-\Gamma^\mu{}_{\nu\rho}A^\rho\quad\text{with}\\
\Gamma^\mu{}_{\nu\rho}:=\frac{1}{2}g^{\mu\nu}\left(\frac{\partial}{\partial x^\nu} g_{\rho\lambda}+\frac{\partial}{\partial x^\rho} g_{\nu\lambda}+\frac{\partial}{\partial x^\lambda} g_{\nu\rho}\right).
\end{gather*}
But in general no genuine integral analogue can be found, because of the absence of a Gauss law for tensor fields of order greater than $1$. In the case of a spacetime with a Killing vector field $X^\mu$ one derives
$$
\big(X_\mu T^{\mu\nu}\big)_{;\nu}=X_{\mu;\nu}T^{\mu\nu}+X_\mu T^{\mu\nu}{}_{;\nu}=0.
$$
The first addend vanishes since $X_{\mu;\nu}$ is anti-symmetric and $T^{\mu\nu}$ is symmetric, while the second one is $0$ due to equation \eqref{energy}. Now we have a new situation, as the term on the left-hand side is a vector and the Gauss law may be applied to this relation. Finally, one obtains a conserved quantity, 
$$
T:=\int_{x^0=\text{const}}T^{\mu 0}X_\mu(-\det g_{\mu\nu})^{1/2}d^3\x.
$$
The interpretation of $T$ as the energy is physically meaningful, only if the vector field $X^\mu$ is timelike, i.e., $X_\mu X^\mu<0$. Thus, the energy of a gravitational system can be defined properly when a timelike vector field exists, and then it is automatically conserved. \\
In our case this means in particular that, if the infinitesimal generator $\delta_m$ is not timelike, then one would be able to find scalar fields, defined with respect to $X^\mu$, whose energy would not be bounded from below \cite{Kay:1988mu}.\\

\subsection*{Conformal Invariance of the Klein-Gordon Equation}

We begin by justifying the requirement of the infinitesimal generator of the modular group for the massive algebra to be a pseudo-differential operator. To this end we discuss the Klein-Gordon equation, in particular its maximal local symmetry. 

The invariance under a maximal symmetry operator $Q$, see \cite{Fushchich:1994} and \cite{Fulton:1962}, means that, if $\varphi$ is a solution of the Klein-Gordon equation, then so is the transformed scalar field $Q\varphi$:
$$
\big(\Box+m^2\big)\varphi(x)=:L\varphi(x)=0\quad\Longrightarrow\quad\big(\Box+m^2\big)Q\varphi(x)=0.
$$  
We are concerned here with linear differential operators of first order, which can be interpreted as generators of continuous transformation groups. If we express the symmetry operator by $Q:=A^\mu P_\mu+B$, where $A^\mu$ and $B$ are complex-valued functions and $P_\mu$ is the infinitesimal generator of translations, see Table \ref{generatortable}, then the implication given above is equivalent to the existence of a complex-valued function $\alpha_Q$ such that
$$
[Q,L]=\alpha_QL.
$$
The set of all symmetry operators for the Klein-Gordon equation forms a complex Lie algebra, since along with the symmetry operators $Q_1$ and $Q_2$ also their linear combinations and their commutators are symmetry operators. This so-called invariance algebra can be finite-dimensional or infinite-dimensional.\\
Writing the symmetry operator in terms of the anti-commutator $[A,B]_+:=AB+BA$,
$$
Q=\frac{1}{2}[A^\mu,P_\mu]_++C\quad\text{with}\quad C:=\frac{1}{2}[A^\mu,P_\mu]+B,
$$ 
we obtain
\begin{gather*}
\Big[\frac{1}{2}[A^\mu,P_\mu]_++C,L\Big]=\alpha_QL\quad\Longleftrightarrow\\
\frac{1}{2}\big[[\partial^\nu A^\mu,P_\mu]_+,P_\nu\big]_++\big[\partial^\nu C,P_\nu\big]_+=\\\frac{1}{4}\big[[\alpha_Q,P^\mu]_+,P_\mu\big]_++\frac{i}{2}\big[\partial^\mu\alpha_Q,P_\mu\big]_+-m^2\alpha_Q.
\end{gather*}
The operators on both sides are equivalent, if the coefficients of the same anti-commutators fulfill the following conditions: 
\begin{gather}\label{conformalinvariance}
\partial^\nu A^\mu+\partial^\mu A^\nu=\frac{1}{2}g^{\mu\nu}\alpha_Q,\notag\\
\partial^\mu C=\partial^\mu\alpha_Q,\qquad\text{and}\qquad m^2\alpha_Q=0.
\end{gather}
From now on we have to distinguish between two different possibilities. The first one is the massless case, namely the invariance algebra of the massless Klein-Gordon equation, and we conclude that the general solution of the first two equations is
$$
A^\mu=u^\mu(x)=a^\mu+g^{\mu\lambda}\o_{\lambda\nu}x^\nu+d\,x^\mu+2x^\mu c_\lambda x^\lambda-x_\lambda x^\lambda c^\mu,
$$
confer equation \eqref{conformaltrafo}. Thus the maximal invariance algebra of the massless Klein-Gordon equation is the whole conformal algebra generated by the translations, Lorentz transformations, dilations and the special conformal transformations.\\
Due to the implication $\alpha_Q=0$ in \eqref{conformalinvariance}, the second case, that is the massive Klein-Gordon equation, leads to the conditions 
$$
\partial^\nu A^\mu+\partial^\mu A^\nu=0\quad\text{and}\quad\partial^\mu C=0.
$$
The general solution of these equations is a linear combination of the infinitesimal generators of the translations and of the Lorentz transformations. Therefore, the massive Klein-Gordon equation is only Poincar\'e-covariant.\\

In the next Sections we will derive a group of automorphisms on the massive algebra $\Ma_m$ with respect to the massless vacuum state. The existence of the automorphism group on $\Ma_m$ for the massive vacuum state is then ensured by the cocycle theorem of Connes. Since Figliolini and Guido, see Theorem \ref{generatorcontinuity}, have shown continuity with respect to the strong toplogy of the mapping
$$
m\mapsto\delta_m \quad, m\geq 0,
$$
the generator $\delta_m$ of the massive group has to contain in a certain sense the massless one $\delta_0$, which is a differential operator of order one as derived by Hislop and Longo \cite{Hislop-Longo:1981uh}, as a special case, namely in the case $m=0$. But, as discussed above, $\delta_m$ cannot be maximally symmetric like $\delta_0$, because the massive Klein-Gordon equation is not conformally invariant. The `additional term' in $\delta_m$ seems to be responsible for the breakdown of conformal symmetry and, therefore, cannot be a differential operator. This circumstance is a hint for its pseudo-differential or even Fourier integral nature. In fact, this conjecture is confirmed by at least two examples due to Yngvason and Borchers, which will be discussed in the next Section in detail. The non-local action of these modular groups is reflected by their infinitesimal generators, which can be shown explicitly to be pseudo-differential and Fourier integral operators.

%\newpage

%\subsection{Cocycle}

%Since the infinitesimal generator $\delta_0$ of the modular group with respect to the massless vacuum state, $\sigma_{\o_0}^t$, is an ordinary differential operator of first order, and since the modular group with respect to the massive vacuum state, $\sigma_{\o_m}^t$, is connected to $\sigma_{\o_0}^t$ via a cocycle, 
%\begin{gather*}
%\sigma_{\o_m}^t(A)=\Gamma_t\sigma_{\o_0}^t(A)\Gamma_t^*,
%\end{gather*}
%one expects the massive generator to be ...

\section{Modular Groups with Nonlocal Action}

Before we try to calculate the modular automorphism group $\sigma_m^t$, we want to investigate the Assumption \ref{assumption-gen0} on its infinitesimal generator $\delta_m$, which is obviously not known in general. As aforementioned, the modular groups to be considered as constructive examples for our purpose are such with nonlocal action. To the best of our knowledge, there exist only two concrete examples for such modular groups in the literature. The first modular automorphism group is given by Yngvason in the context of his investigation on essential duality \cite{Yngvason:1994nk}, and the second one is introduced by Borchers and Yngvason who formulate modular groups in a more general setting, namely with respect to arbitrary KMS states instead of the vacuum state \cite{Borchers:1998ye}.

\subsection*{Yngvason's Counter-example}

In a Poincar\'e covariant Wightman framework, Bisognano and Wichmann identify the modular groups with the Lorentz boosts and, furthermore, show that wedge duality holds. Yngvason investigates the validity of these two properties for local nets \cite{Yngvason:1994nk}. He can give explicit examples for fields violating essential duality, an implication of wedge duality and major assumption in the superselection theory, or Lorentz covariance. We take one of his concrete examples as an opportunity to analyse the infinitesimal generator of a modular group with nonlocal action.\\
 He starts with a Hermitian Wightman field $\varphi$ which transforms covariantly under spacetime translations, but not necessarily under Lorentz transformations. A general two-point function, satisfying positivity, translation covariance, spectrum condition and locality, is of the following form in the Fourier space,
$$
\o_2(p)=\sum_{i=1}^n M_i(p)d\mu_i(p),
$$
where $M_i$ is a polynomial, which is positive on the support of $d\mu_i(p)$ in $\V_+$, $i=1,\cdots,n$, the positive, Lorentz-invariant measure. For the sake of simplicity let us consider a two-point function consisting of only one term,
$$
\o_2(p)=M(p)d\mu(p),
$$
whose polynomial factorises as
$$
M(p)=:F(p)F(-p)\quad\text{and}\quad F(p)^*=F(-p),
$$
where $F(p)$ (in general no polynomial) is analytic in a certain sense and has no zeros in the right wedge characterised by $x_+>0$ and $ x_-<0$. The existence of such polynomials is ensured by the following example,
$$
M(p):=\sum_{i=1}^n(p^i)^2+m^2,
$$
with
\begin{gather}\label{Yngvason-example}
F(p)=(\hat{p}\hat{p}+m^2)^{1/2}+ip^1=(\hat{p}\hat{p}+m^2)^{1/2}+\frac{i}{2}(p_++p_-),
\end{gather}
where we have used the notation $\hat{p}:=(p^2,\cdots,p^n)$ and $p_\pm:=p^0\pm p^1$. One obtains the generalised free field $\partial_t\varphi_m(x)$, where $\varphi_m$ is the free field of mass $m$, by setting $d\mu(p):=\Theta(p^0)\delta\big((p,p)_\Mi-m^2\big)$ and $M(p):=(p^0)^2$. For $\lambda>0$ one can now define the unitary operator $V_{\W_R}(\lambda)$ on the Fock space $\F$, first on the one-particle space $\H_1:=\L^2\big(\R^n,M(p)d\mu(p)\big)$ by
\begin{equation*}
V_{\W_R}(\lambda)\varphi(p):=\frac{F(-\lambda p_+,-\lambda^{-1} p_-,-\hat{p})}{F(-p_+,-p_-,-\hat{p})}\varphi(\lambda p_+,\lambda^{-1} p_-,\hat{p})
\end{equation*}
for $\varphi\in\H_1$, and then by canonical extension (second quantisation) to $\F$. One then introduces a one-parameter group of automorphisms on the von Neumann algebra $\mathfrak{M}(\W_R)$ over $\H$ generated by the Weyl operators $W(f):=e^{i\varphi[f]}$:
\begin{equation}\label{Yngvason}
\sigma_{\W_R}^t\big(W(f)\big):=V_{\W_R}\big(e^{-2\pi t}\big)W(f)V_{\W_R}\big(e^{2\pi t}\big).
\end{equation}
Yngvason identifies this group with the modular group with respect to the vacuum state on $\mathfrak{M}(\W_R)$ by proving, due to Theorem \ref{Takesaki-KMS}, the validity of the KMS condition, namely, he proves that the function 
\begin{gather*}
F(t):=\Big(\Omega,\sigma_{\W_R}^t\big(W(f)\big)W(g)\Omega\Big),
\end{gather*}
where $f$ and $g$ are test functions with support in $\R_+$, has an analytic continuation from the real axis into the half strip $\big\{t+is|\;0<s<1,\,t,s\in\R\big\}$ with
\begin{gather*}
\lim_{s\rightarrow 1}F(is)=\big(\Omega,W(g)W(f)\Omega\big).
\end{gather*}
But the operator $V_{\W_R}(\lambda)$ maps the Fourier transform $\tilde{f}$ of $f$ with $\supp f\subset \W_R$ into
\begin{equation*}
\tilde{f}_\lambda(p):=V_{\W_R}(\lambda)\tilde{f}(p)=\frac{(\hat{p}\hat{p}+m^2)^{1/2}-\frac{i}{2}(\lambda p_+-\lambda^{-1}p_-)}{(\hat{p}\hat{p}+m^2)^{1/2}-\frac{i}{2}(p_+-p_-)}\tilde{f}(\lambda p_+,\lambda^{-1}p_-,\hat{p}),
\end{equation*}
which is not analytic in $\hat{p}$ and therefore cannot be the Fourier transform of a function with compact support in the $\hat{x}$-direction, $\hat{x}:=(x^2,\cdots,x^n)$ . Consequently, $W(f_\lambda)$ cannot be an element of any wedge algebra unless the wedge is a translate of $\W_R$ or the left wedge $\W_L:=\big\{x\in\mathbb{M}|\;|x^0|<-x^3\big\}$. The operator $W(f_\lambda)$ is still localised only in the $x^0,x^1$-directions in the sense that it is an element of $\mathfrak{M}(\W_R+a)\cap\mathfrak{M}(\W_R+b)'$ for some $a,b\in\W_R$.\\
Correspondingly one derives for the left wedge $\W_L$,
\begin{equation*}
V_{\W_L}(\lambda)\varphi(p):=\frac{F(\lambda p_+,\lambda^{-1} p_-,\hat{p})}{F(p_+,p_-,\hat{p})}\varphi(\lambda p_+,\lambda^{-1} p_-,\hat{p}).
\end{equation*}
By comparing the modular conjugation of the two wedges,

\begin{gather*}
J_{\W_R}(\lambda)\varphi(p)=\frac{F(p_+,p_-,-\hat{p})}{F(-p_+,-p_-,-\hat{p})}\varphi(p_+,p_-,-\hat{p})^*\quad \text{and}\\
J_{\W_L}(\lambda)\varphi(p)=\frac{F(-p_+,-p_-,\hat{p})}{F(p_+,p_-,\hat{p})}\varphi(p_+,p_-,-\hat{p})^*,
\end{gather*}
one recognises that wedge duality, i.e., $\mathfrak{M}(\W_R)'=\mathfrak{M}(\W_L)$, is satisfied if and only if $F(p)=F(-p)$ holds on the support of $d\mu$. This condition is violated by our example mentioned above.\\

One may ask if the non-local behaviour of this example is reflected in some way by the infinitesimal generator of the group \eqref{Yngvason}. First, we derive the generator for the modular group,
\begin{align*}
\Delta_{\W_R}^{it}\varphi(p)&=\frac{F(-\lambda p_+,-\lambda^{-1}p_-,-\hat{p})}{F(-p_+,-p_-,-\hat{p})}\varphi(\lambda p_+,\lambda^{-1}p_-,\hat{p})\\
&=:\widehat{F}(\lambda,p_+,p_-,\hat{p})\varphi(\lambda p_+,\lambda^{-1}p_-,\hat{p}),
\end{align*}
where $\lambda=e^{-2\pi t}$, as
\begin{align*}
\delta_{\W_R}\varphi(p)&=\partial_t\Delta_{\W_R}^{it}\varphi(p)\big|_{t=0}\\
&=\partial_t\widehat{F}(\lambda,p_+,p_-,\hat{p})\big|_{t=0}\varphi(p_+,p_-,\hat{p})+\partial_t\varphi(\lambda p_+,\lambda^{-1}p_-,\hat{p})\big|_{t=0}\\
&=\bigg\{\frac{2\pi}{F(-p_+,-p_-,-\hat{p})}\big(p_+\partial_{p_+}-p_-\partial_{p_-}\big)F(-p_+,-p_-,-\hat{p})\\
&\qquad-2\pi p_+\partial_{p_+}+2\pi p_-\partial_{p_-}\bigg\}\varphi(p_+,p_-,\hat{p}).
\end{align*}
For our example \eqref{Yngvason-example} we obtain:
\begin{align*}
\delta_{\W_R}\varphi(p)&=\bigg\{\frac{i\pi(-p_++p_-)}{(\hat{p}\hat{p}+m^2)^{1/2}-\frac{i}{2}(p_++p_-)}-2\pi p_+\partial_{p_+}+2\pi p_-\partial_{p_-}\bigg\}\varphi(p_+,p_-,\hat{p})\\
&=\bigg\{\frac{-2i\pi p^1}{(\hat{p}\hat{p}+m^2)^{1/2}-ip^0}-4\pi\big(p^0\partial_{p^1}+p^1\partial_{p^0}\big)\bigg\}\varphi(p_+,p_-,\hat{p}).
\end{align*}
While the second term can be identified with the Bisognano-Wichmann infinitesimal generator \eqref{BW-generator}, the first term containing the mass $m$ is a PsDO of order zero. This additional part has to comprise the non local character of the modular group $\Delta_{\W_R}^{it}$. To put it in a nutshell, we have verified the Assumption \ref{assumption-gen0} with 
\begin{gather*}
\delta_r:=\frac{-2i\pi p^1}{(\hat{p}\hat{p}+m^2)^{1/2}-ip^0}.
\end{gather*}

\subsection*{Borchers-Yngvason's Counter-example}

In \cite{Borchers:1998ye} Borchers and Yngvason give other examples for modular automorphism groups which act non locally on the wedges, light cones and double cones. Whereas all investigations given so far have been concerned with modular groups with respect to the vacuum state, Borchers and Yngvason formulate the automorphism groups by means of KMS states.

They start with a general $C^*$-dynamical system $(\Aa,\alpha^t)$, an $\alpha^t$-invariant subalgebra $\Ba$, i.e., $\alpha^t(\Ba)\subseteq\Ba$, and an $(\alpha,\beta$)-KMS state $\o$. Due to the analyticity property of KMS states, $\Omega$ is separating, and also cyclic for $\Ma:=\pi_\o(\Aa)''$ and $\Na:=\pi_\o(\Ba)''$, if one assumes $\bigcup_{t\in\R}\alpha^t(\Ba)$ to be dense in $\Aa$ in the norm topology. Hence, the existence of the modular objects is ensured and one may determine the action of the modular automorphism group. Their main theorem reads as follows.

\begin{theo}[Borchers-Yngvason]\label{Borchers-Yngvason-theo1}
Let $T(t):=e^{itH}$ be the unitary group implementing the automrphism group $\alpha^t$ and $\Na(t):=T(t)\Na$, then one has:
\begin{gather*}
\Delta_\Na^{i\tau}\Na(t)\Delta_\Na^{-i\tau}=\Na\big(\nu^\tau(t)\big),
\end{gather*}
where
\begin{gather*}
\nu^\tau(t):=\frac{\beta}{2\pi}\log\Big(1+e^{-2\pi\tau}\big(e^{2\pi t/\beta}-1\big)\Big)
\end{gather*}
for all $t,\tau\in\R$ satisfying
\begin{gather*}
1+e^{-2\pi\tau}\big(e^{2\pi t/\beta}-1\big)>0.
\end{gather*}
Furthermore, 
\begin{gather*}
\Delta_\Na^{i\tau}\Ma\Delta_\Na^{-i\tau}\subset\Ma\quad\text{and}\quad\Na=\bigcap_{\tau\geq0}\Delta_\Na^{i\tau}\Ma\Delta_\Na^{-i\tau}
\end{gather*}
hold for all $\tau\geq0$.
\end{theo}

This result is then applied to quasi-local algebras $\Aa(\O)$ and $\Ba:=\Aa(\O_0)$, where $\O_0$ is invariant under half-sided translations in $t$-direction. The authors restrict themselves to two-dimensional theories which factorise in the light cone variables $x_+:=x^0+x^1$ and $x_-:=x^0-x^1$. In these cases one may first establish the modular group on the algebra $\Ma(\R_+)$ as
\begin{gather*}
\Delta_+^{i\tau}\Ma\big([x_\pm,\infty[\big)\Delta_+^{-i\tau}=\Ma\big([\nu_+^t(x_\pm),\infty[\big),
\end{gather*}
where
\begin{gather*}
\nu_+^t(x_\pm):=\frac{\beta}{2\pi}\log\Big(1+e^{-2\pi t}\big(e^{2\pi x_\pm/\beta}-1\big)\Big)
\end{gather*}
for all $t,x_\pm\in\R$ satisfying
\begin{gather}\label{Borchers-Yngvason2}
1+e^{-2\pi t}\big(e^{2\pi x_\pm/\beta}-1\big)>0.
\end{gather}
In the same manner one introduces the modular group on the algebra $\Ma(\R_-)$ as
\begin{gather*}
\Delta_+^{i\tau}\Ma\big(]-\infty,x_\pm]\big)\Delta_+^{-i\tau}=\Ma\big(]-\infty,\nu_-^t(x_\pm)]\big)
\end{gather*}
with
\begin{gather*}
\nu_-^t(x_\pm):=-\nu_+^{-t}(-x_\pm)
\end{gather*}
for all $t,x_\pm\in\R$ fulfilling
\begin{gather}\label{Borchers-Yngvason3}
1+e^{2\pi\tau}\big(e^{-2\pi x_\pm/\beta}-1\big)>0.
\end{gather}
Now, one can express the algebra for the two-dimensional space $I_+\times I_-\subseteq\R^2$ via the tensor product
\begin{gather*}
\Ma(I_+\times I_-)=\Ma(I_+)\otimes\Ma(I_-),
\end{gather*}
in particular, one obtains for the examples of our interest:
\begin{gather*}
\Ma(\W_R)=\Ma(\R_-)\otimes\Ma(\R_+),\\
\Ma(\V_+)=\Ma(\R_+)\otimes\Ma(\R_+),\quad\text{and}\\
\Ma(\O)=\Ma(I_-)\otimes\Ma(I_+).
\end{gather*}
The corresponding modular groups with respect to a factorising KMS state $\o\otimes\o$ are given as:
\begin{gather*}
\Delta_{\W_R}^{it}=\Delta_-^{it}\otimes\Delta_+^{it},\\
\Delta_{\V_+}^{it}=\Delta_+^{it}\otimes\Delta_+^{it},\quad\text{and}\\
\Delta_\O^{it}=\Delta_-^{it}\otimes\Delta_+^{it}.
\end{gather*}
Thus, Theorem \ref{Borchers-Yngvason-theo1} can be applied and one gets as a corollary

\begin{theo}[Borchers-Yngvason]\label{Borchers-Yngvason-theo2}
For the forward light cone one has 
\begin{gather*}
\Delta_{\V_+}^{it}\varphi[f]\Delta_{\V_+}^{-it}=\varphi\big[\nu_{\V_+}^tf\big]\quad\text{with}\\
\big(\nu_{\V_+}^tf\big)(x_-,x_+):=f\big(\nu_+^t(x_-),\nu_+^t(x_+)\big),
\end{gather*}
for all $t\in\R$ and $x_\pm\in\R$ satisfying \eqref{Borchers-Yngvason2} for $x=x_\pm$.\\
Analoguously, one has for the right wedge
\begin{gather*}
\Delta_{\W_R}^{i\tau}\varphi[f]\Delta_{\W_R}^{-i\tau}=\varphi\big[\nu_{\W_R}^tf\big]\quad\text{with}\\
\big(\nu_{\W_R}^tf\big)(x_-,x_+):=f\big(\nu_-^t(x_-),\nu_+^t(x_+)\big),
\end{gather*}
for all $t\in\R$ and $x_\pm\in\R$ satisfying \eqref{Borchers-Yngvason3} and \eqref{Borchers-Yngvason2} for $x=x_-$ and $x=x_+$, respectively.
\end{theo}
For more concrete calculations Borchers and Yngvason investigate the Weyl algebra of free Bose fields generated by elements $W(f)$, $f\in\D(\R)$, with
\begin{gather*}
W[f]^*=W[-f]\quad\text{and}\\
W[f]W[g]=e^{-K(f,g)/2}W[f+g],
\end{gather*}
and
\begin{gather*}
K(f,g):=\int_{-\infty}^\infty p\,Q(p^2)\tilde{f}(-p)\tilde{g}(p)dp,
\end{gather*}
where $Q(p^2)$ is a non-negative polynomial. They introduce for each scaling dimension $n\in\N$ and interval $I\subset\R$ the algebra $\Ma^{(n)}(I)$ which is generated by the Weyl operators $W^{(n)}[f]$ corresponding to $Q(p^2)=p^{2n}$. While the algebra is known to be independent of $n$ for unbounded  $I$, for bounded intervals one only has the inclusion 
\begin{gather}\label{Borchers-Yngvason-inclusion}
\Ma^{(m)}(I)\subset\Ma^{(n)}(I),
\end{gather}
whenever $m>n$. Thus the modular operators $\Delta_+$ and $\Delta_-$ corresponding to the positive real axis and the negative one, respectively, are independent of $n$. 
\begin{theo}[Borchers-Yngvason]\label{Borchers-Yngvason-theo3}
Let $\o$ be a quasi-free KMS state on the Weyl algebra $\Ma^{(0)}(\R_+)$ and $\pi$ the corresponding cyclic representation, then one has:
\begin{gather*}
\Delta_+^{it}\pi\big(W^{(0)}[f]\big)\Delta_+^{-it}=\pi\big(W^{(0)}[\eta_+^{t,(0)}f]\big),\\
\end{gather*}
with
\begin{gather*}
\big(\eta_+^{t,(0)}f\big)(x_\pm):=f\big(\nu_+^t(x_\pm)\big):=f\left(\frac{\beta}{2\pi}\log\left\{1+e^{-2\pi t}\big(e^{2\pi x_\pm/\beta}-1\big)\right\}\right),
\end{gather*}
and $\supp f\subset\R_+$.
\end{theo} 
Because of 
\begin{gather*}
W^{(n)}[f]=W^{(0)}\big[i^nf^{(n)}\big],
\end{gather*}
one may transform the action of the modular group to the case $n>0$,
\begin{align*}
\Delta_+^{it}\pi\big(W^{(n)}[f]\big)\Delta_+^{-it}&=\Delta_+^{it}\pi\big(W^{(0)}\big[i^nf^{(n)}\big]\big)\Delta_+^{-it}\\
&=\pi\big(W^{(0)}\big[\eta_+^{t,(0)}i^nf^{(n)}\big]\big)\\
&=:\pi\big(W^{(n)}\big[\eta_+^{t,(n)}f\big]\big),
\end{align*}
and obtains the following result.
\begin{theo}[Borchers-Yngvason]\label{Borchers-Yngvason-theo4}
Let $\o$ be a quasi-free KMS state on the Weyl algebra $\Ma^{(n)}(\R_+)$, $n>0$, and $\pi$ the corresponding cyclic representation, then the action of $ad\,\Delta_+^{it}$ reads as follows:
\begin{gather*}
\Delta_+^{it}\pi\big(W^{(n)}[f]\big)\Delta_+^{-it}=\pi\big(W^{(n)}[\eta_+^{t,(n)}f]\big),
\end{gather*}
where
\begin{gather*}
\big(\eta_+^{t,(n)}f\big)(x_\pm)=\int_0^{x_\pm}\int_0^{x^1}\cdots\int_0^{x^{n-1}}\eta_+^{t,(0)}f^{(n)}(x^n)dx^n\cdots dx^1,
\end{gather*}
and $f^{(n)}$ is the $n$-th derivative of the test function $f$ with $\supp f\subset\R_+$.
\end{theo}
The modular action on the negative axis is formulated as aforementioned, i.e., $\eta_-^{t,(0)}$ is defined via the transformation $\nu_-^t(x_\pm)$.

Borchers and Yngvason show that the modular group acts locally only in the case $n=0$. While the action on the field operator, which can be regained from the Weyl operators 
\begin{gather*}
\pi\big(W^{(n)}[f]\big)=:e^{i\int\varphi^{(n)}(x)f(x)dx}
\end{gather*}
through functional derivation, is
\begin{gather*}
\Delta_+^{it}\varphi^{(0)}(x_\pm)\Delta_+^{-it}=\partial_{x_\pm}\nu_+^t(x_\pm)\varphi^{(0)}\big(\nu_+^t(x_\pm)\big),
\end{gather*}
for $n=1$ one gets an additional term, e.g., at the origin
\begin{gather*}
\Delta_+^{it}\varphi^{(1)}(0)\Delta_+^{-it}=e^{-2\pi t}\varphi^{(1)}(0)-\frac{2\pi}{\beta}e^{-4\pi t}\int_0^\infty\varphi^{(1)}(x)dx.
\end{gather*}
In the case of double cones, namely where we are dealing with bounded intervals $I_\pm\subset\R_\pm$, fields of higher scaling dimension $\varphi^{(n)}$, $n\geq1$, are in general localised only in the algebra $\Ma^{(0)}(I_\pm)$ after the modular action, due to the inclusion \eqref{Borchers-Yngvason-inclusion}, but no longer in the original subalgebra $\Ma^{(0)}(I_\pm)$.\\

To be more precise, due to Theorem \ref{Borchers-Yngvason-theo2}, for the modular action on the forward light cone we obtain:
\begin{gather*}
\Delta_{\V_+}^{it}\varphi^{(0)}[f]\Delta_{\V_+}^{-it}=\varphi^{(0)}\big[\eta_{\V_+}^{t,(0)}f\big]\quad\text{with}\\
\big(\eta_{\V_+}^{t,(0)}f\big)(x_-,x_+):=f\big(\nu_+^t(x_-),\nu_+^t(x_+)\big).
\end{gather*}
Written in terms of the originial spacetime coordinates the transformated coordinates, 
\begin{gather}
\bar{x}^0=\frac{1}{2}\big(\nu_+^t(x_+)+\nu_-^t(x_-)\big)\quad\text{and}\notag\\
\bar{x}^1=\frac{1}{2}\big(\nu_+^t(x_+)-\nu_-^t(x_-)\big)\label{x0-x1-trafo}
\end{gather}
of $x^0$ and $x^1$, respectively, are
\begin{gather*}
\bar{x}^0=\frac{\beta}{4\pi}\log\left\{\big[1+e^{-2\pi t}\big(e^{2\pi x_+/\beta}-1\big)\big]\big[1+e^{-2\pi t}\big(e^{2\pi x_-/\beta}-1\big)\big]\right\},\\
\bar{x}^1=\frac{\beta}{4\pi}\log\left\{\frac{1+e^{-2\pi t}\big(e^{2\pi x_+/\beta}-1\big)}{1+e^{-2\pi t}\big(e^{2\pi x_-/\beta}-1\big)}\right\},
\end{gather*}
Close to the apex of $\V_+$, one obtains the known case $\beta=\infty$, i.e., dilations with the light cone coordinates $x_+$ and $x_-$ scaled by the factor $e^{-2\pi t}$.

In the case of the right wedge the same arguments lead to the following modular action:
\begin{gather*}
\Delta_{\W_R}^{i\tau}\varphi^{(0)}[f]\Delta_{\W_R}^{-i\tau}=\varphi^{(0)}\big[\eta_{\W_R}^{t,(0)}f\big]\quad\text{with}\\
\big(\eta_{\W_R}^{t,(0)}f\big)(x_-,x_+):=f\big(\nu_-^t(x_-),\nu_+^t(x_+)\big).
\end{gather*}
The transformed coordinates are:
\begin{gather*}
\bar{x}^0=\frac{\beta}{4\pi}\log\left\{\frac{1+e^{-2\pi t}\big(e^{2\pi x_+/\beta}-1\big)}{1+e^{2\pi t}\big(e^{-2\pi x_-/\beta}-1\big)}\right\},\\
\bar{x}^1=\frac{\beta}{4\pi}\log\left\{\big[1+e^{-2\pi t}\big(e^{2\pi x_+/\beta}-1\big)\big]\big[1+e^{2\pi t}\big(e^{-2\pi x_-/\beta}-1\big)\big]\right\}.
\end{gather*}
Here, near the edge of the the wedge, the action may be identified with the case $\beta=\infty$, i.e., with Lorentz boosts where the light cone coordinates $x_+$ and $x_-$ are scaled by the factors $e^{-2\pi t}$ and $e^{2\pi t}$, respectively.

Also in this case, we are interested in the infinitesimal generator $\delta^{(n)}$ of the modular automorphism group acting on wedges, forward light cones and double cones, since we expect to see this non local behaviour in the pseudo-differential structure of $\delta^{(n)}$. The generator corresponding to the positive real axis in the case of $n=0$ is
\begin{align*}
\delta_+^{(0)}\varphi^{(0)}[f]:&=\partial_t\Delta_+^{it}\varphi^{(0)}[f]\Delta_+^{-it}\big|_{t=0}\\
&=\varphi^{(0)}\big[\partial_t\eta_+^{t,(0)}f\big]\big|_{t=0},
\end{align*}
with
\begin{gather*}
\big(\partial_t\eta_+^{t,(0)}f\big)(x_\pm)\big|_{t=0}=\partial_tf\big(\nu_+^t(x_\pm)\big)\big|_{t=0}=-\beta\big(1-e^{-2\pi x_\pm/\beta}\big)\partial_{x_\pm}f(x_\pm),
\end{gather*}
while the counterpart with respect to the negative real axis reads
\begin{align*}
\delta_-^{(0)}\varphi^{(0)}[f]:&=\partial_t\Delta_-^{it}\varphi^{(0)}[f]\Delta_-^{-it}\big|_{t=0}\\
&=\varphi^{(0)}\big[\partial_t\eta_-^{t,(0)}f\big]\big|_{t=0},
\end{align*}
with
\begin{gather*}
\big(\partial_t\eta_-^{t,(0)}f\big)(x_\pm)\big|_{t=0}=\partial_tf\big(\nu_-^t(x_\pm)\big)\big|_{t=0}=-\beta\big(1-e^{2\pi x_\pm/\beta}\big)\partial_{x_\pm}f(x_\pm).
\end{gather*}
In terms of the original spacetime coordinates the infinitesimal generators have the following form:
\begin{align}
\delta_{\V_+}^{(0)}f(x^0,x^1)=\frac{\beta}{2}\Big[\big(&e^{-2\pi x_+/\beta}+e^{-2\pi x_-/\beta}-2\big)\partial_{x^0}\notag\\
&+\big(e^{-2\pi x_+/\beta}-e^{-2\pi x_-/\beta}\big)\partial_{x^1}\Big]f(x^0,x^1),\\
\delta_{\W_R}^{(0)}f(x^0,x^1)=\frac{\beta}{2}\Big[\big(&e^{-2\pi x_+/\beta}+e^{2\pi x_-/\beta}-2\big)\partial_{x^0}\notag\\
&+\big(e^{-2\pi x_+/\beta}-e^{2\pi x_-/\beta}\big)\partial_{x^1}\Big]f(x^0,x^1).
\end{align}
The generator for an arbitrary $n>0$ is given in the next 
\begin{theo} 
The infinitesimal generator of the modular group acting on the algebra $\Ma^{(n)}(\R_+)$ is for $n>0$
\begin{gather}
\delta_+^{(n)}f(x_\pm)=\delta_+^{(n-1)}f(x_\pm)+\delta_{+}^{(n,r)}f(x_\pm),
\end{gather}
where
\begin{align}
\delta_{+}^{(n,r)}f(x_\pm):=2\pi\int\frac{(i\xi)^n}{\big(i\xi-\frac{2\pi}{\beta}\big)^n}\tilde{f}(\xi)e^{ix_\pm(\xi+2\pi i/\beta)}d\xi.
\end{align}
The counterpart for $\Ma^{(n)}(\R_-)$ for $n>0$ reads 
\begin{gather}
\delta_-^{(n)}f(x_\pm)=\delta_-^{(n-1)}f(x_\pm)+\delta_{-}^{(n,r)}f(x_\pm)
\end{gather}
with
\begin{align}
\delta_{-}^{(n,r)}f(x_\pm):=-2\pi\int\frac{(i\xi)^n}{\big(i\xi+\frac{2\pi}{\beta}\big)^n}\tilde{f}(\xi)e^{ix_\pm(\xi-2\pi i/\beta)}d\xi.
\end{align}
\end{theo}

\proof By induction one obtains for the positive real axis:
\begin{align*}
\delta_+^{(n+1)}f(x_\pm)&=\partial_t\int_0^{x_\pm}\int_0^{x^1}\cdots\int_0^{x^n}\eta_t^{(0)}f^{(n+1)}(x^{n+1})dx^{n+1}\cdots dx^1\bigg|_{t=0}\\
&=\int_0^{x_\pm}\int_0^{x^1}\cdots\int_0^{x^n}\partial_tf^{(n+1)}\big(\nu_+^t(x^{n+1})\big)dx^{n+1}\cdots dx^1\bigg|_{t=0}\\
&=\int_0^{x_\pm}\int_0^{x^1}\cdots\int_0^{x^n}\partial_{x^{n+1}}^{n+2}f(x^{n+1})\partial_t\nu_+^t(x^{n+1})\big|_{t=0}dx^{n+1}\cdots dx^1\\
&=\int_0^{x_\pm}\int_0^{x^1}\cdots\int_0^{x^{n-1}}\partial_{x^n}^{n+1}f(x^n)\partial_t\nu_+^t(x^n)\big|_{t=0}dx^n\cdots dx^1\\
&\quad-\int_0^{x_\pm}\int_0^{x^1}\cdots\int_0^{x^n}\partial_{x^{n+1}}^{n+1}f(x^{n+1})\partial_{x^{n+1}}\partial_t\nu_+^t(x^{n+1})\big|_{t=0}\\
&\hspace{8,5cm}dx^{n+1}\cdots dx^1\\
&=\delta_+^{(n)}f(x_\pm)-\delta_{+}^{(n+1,r)}f(x_\pm).
\end{align*}
Due to the fact that $\supp f\subset\R_+$, we get for the additional term
\begin{align*}
\delta_{+}^{(n+1,r)}f(x_\pm)&=2\pi\int_0^{x_\pm}\int_0^{x^1}\cdots\int_0^{x^n}(i\xi)^{n+1}\tilde{f}(\xi)e^{ix_\pm^{n+1}\xi}e^{-2\pi x_\pm^{n+1}/\beta}d\xi dx^{n+1}\cdots dx^1\\
&=2\pi\int\frac{(i\xi)^{n+1}}{\big(i\xi-\frac{2\pi}{\beta}\big)^{n+1}}\tilde{f}(\xi)e^{ix_\pm(\xi+2\pi i/\beta)}d\xi.
\end{align*}
The expression for the generator $\delta_{-}^{(n+1,r)}$ corresponding to $\R_-$ is calculated in the same way.

\qed

What we have shown is that the infinitesimal generators $\delta_+^{(n)}$ and $\delta_-^{(n)}$ with scaling dimension $n\geq 1$ are no longer differential operators but Fourier integral operators instead, see Definition \ref{FIO}. To be more precise, the generators do have the following structure:
\begin{gather*}
\delta_\pm^{(n)}=\delta_\pm^{(0)}+\sum_{k=1}^n\delta_\pm^{(k,r)}=:\delta_\pm^{(0)}+\delta_{\pm,r}^{(n)}.
\end{gather*}
Whereas the principal symbol $\delta_\pm^{(0)}$ is still a differential operator of order one, the additional part $\delta_{\pm,r}^{(n)}$ is a Fourier integral operator of order zero with complex-valued symbol
\begin{gather*}
a_{\pm,r}^{(n)}(\xi):=\sum_{k=1}^n \frac{(i\xi)^k}{\big(i\xi\mp\frac{2\pi}{\beta}\big)^k}
\end{gather*}
and a complex-valued phase function
\begin{gather*}
\theta_\pm(x_\pm,\xi):=x_\pm\left(\xi\pm\frac{2\pi i}{\beta}\right),
\end{gather*}
which is independent of $n$. In H\"ormander's terminology, see Remark \ref{symbol-Hoermander}, $\delta_{\pm,r}^{(n)}$ is a PsDO of order zero with the symbol,
\begin{gather*}
p_{\pm,r}^{(0)}(x_\pm,\xi):=\sum_{k=1}^n \frac{(i\xi)^k}{\big(i\xi\mp\frac{2\pi}{\beta}\big)^k}\;e^{-2\pi x_\pm/\beta}.
\end{gather*}

The generators with respect to the spacetime coordinates can be computed via the equations \eqref{x0-x1-trafo}. We denote them by $\delta_{\W_R,r}^{(n)}$, $\delta_{\V_+,r}^{(n)}$ and $\delta_{\D,r}^{(n)}$. Thus, Assumption \ref{assumption-gen0} is proved with
\begin{gather*}
\delta_0:=\delta_{\W_R}^{(0)},\;\delta_{\V_+}^{(0)},\;\delta_{\D}^{(0)}\quad\text{and}\\
\delta_r:=\delta_{\W_R,r}^{(n)},\;\delta_{\V_+,r}^{(n)},\;\delta_{\D,r}^{(n)}
\end{gather*}
for all $n\in\N$.

\section{The Approach of Figliolini and Guido}\label{Guido-section}

The analysis of Figliolini and Guido \cite{Figliolini:1989vf} is based on the second quantisation formalism for free time-zero fields over a one-particle Hilbert space. Confer for a short introduction Appendix B. They consider the modular operator as the second quantisation of an operator $\check{\Delta}$, i.e.,
$$
\Delta\equiv d\Gamma(\check{\Delta}),
$$
for which they give an explicit expression. This procedure preserves properties like self-adjointness, positivity and unitarity, but not, for example, boundedness. Now consider a real-valued and closed subset $\mathcal{K}\subset\H$, denoted by $\mathcal{K}\subseteq_\R\H$, such that the vacuum $\Omega_0$ is cyclic and separating for the von Neumann algebra $\mathfrak{M}(\mathcal{K}):=\big\{W(f)|\;f\in\mathcal{K}\big\}''$ which is generated by the Weyl operators $W(f)$ . In this case one can start with modular theory and define the modular objects $S,J$ and $\Delta$ in the usual way.

Eckmann and Osterwalder \cite{Eckmann:1973} give important insight into the interplay of modular theory with quantum field theory of Bose fields and perform explicit calculations of fundamental quantities using their

\begin{theo}
If $\mathcal{K}$ is cyclic and separating for $\mathfrak{M}(\mathcal{K})$, then the Tomita operator $S$ is the second quantisation $d\Gamma(\tilde{S})$ of the closed, densely defined and conjugate-linear operator $\check{S}$ over $\H$ defined by
\begin{align*}
\check{S}:\mathcal{K}+i\mathcal{K}&\longrightarrow\mathcal{K}+i\mathcal{K}\\
f+ig&\mapsto f-ig.
\end{align*}
Furthermore, for the polar decomposition $\check{S}=\check{J}\check{\Delta}^{1/2}$ of $S$, one has 
$$
J=d\Gamma(\check{J}) \quad\text{and}\quad\Delta=d\Gamma(\check{\Delta}).
$$
\end{theo}
The investigation of Figliolini and Guido relies on the statements of this theorem and the results of Araki \cite{Araki:1963}, and of Leyland, Roberts and Testard \cite{Leyland:1978iv}.

Now let us consider a double cone $\D\subset\R^4$ whose basis $\O\subset\R^3$ is contained in the time-zero hyperplane. Then we can associate with $\O$ the von Neumann algebra 
\begin{gather*}
\mathfrak{M}(\O):=\big\{W(h)|\;h\in\mathcal{K}_m(\O)\big\}\\
\text{with}\quad\mathcal{K}_m(\O):=\big\{\o_m^{-1/2}f-i\o_m^{1/2}g|\;f,g\in\D(\O)\big\},
\end{gather*}
where the energy operator $\o_m$ is given as
\begin{align*}
(\o_mf)\,\tilde{}\;(\p)&:=\int(-\Delta+m^2)^{1/2}f(x)e^{-i\p\x}d^3\x\\
&=(\p^2+m^2)^{1/2}\tilde{f}(\p)
\end{align*}
for $m\geq 0$. The operator $\o_m$ is anti-local, which means
\begin{gather}\label{anti-local}
\supp f\subset\O\quad\text{and}\quad\supp(\o f)\subset\O\quad\Longrightarrow \quad f=0.
\end{gather}
The investigation of Figliolini and Guido \cite{Figliolini:1989vf} is mainly based on `local' Sobolev spaces,
$$
\Hs_m^{\pm 1/2}(\O):=\overline{\L^2(\O)\cap\D(\o_m^{\pm 1/2})}^{\|\o_m^{\pm 1/2}(\cdot)\|},
$$
in terms of which they reformulate 
$$
\mathcal{K}_m(\O)=\o_m^{-1/2}\Hs_{m,\R}^{-1/2}(\O)-i\o_m^{1/2}\Hs_{m,\R}^{1/2}(\O).
$$
For bounded regions $\O\subset\R^3$ and $\alpha>-\frac{3}{2}$ they show
\begin{gather*}
\Hs_m^\alpha(\O)\cong\Hs_0^\alpha(\O),
\end{gather*}
i.e., it is about the same vector spaces with equivalent norms. On these Sobolev spaces the important operator $A_\sigma$ is defined for $\sigma=\pm 1$ as
\begin{gather*}
A_\sigma:\Hs^{\sigma/2}(\O)\supset\F_\sigma\longrightarrow\Hs^{-\sigma/2}(\O),\quad A_\sigma:=P_{-\sigma}\o^\sigma\big\vert_{\Hs^{\sigma/2}(\O)},\\
\text{and}\quad\F_\sigma:=\o^{-\sigma}\big(\Hs^{-\sigma/2}(\O)+\Hs^{-\sigma/2}(\O^c)\big)\cap\Hs^{\sigma/2}(\O),
\end{gather*}
where we have skipped the indices $m$ and $\R$. $\F_\sigma$ is dense in $\Hs^{\sigma/2}(\O)$, and the indicator function $P_\sigma$ is given by
\begin{gather*}
P_\sigma:\Hs^{\sigma/2}(\R^3)\supset\big(\Hs^{\sigma/2}(\O)+\Hs^{\sigma/2}(\O^c)\big)\longrightarrow\Hs^{\sigma/2}(\R^3)\\
P_\sigma:=\begin{cases}\1&\text{ on  }\;\Hs^{\sigma/2}(\O),\\
0&\text{ on }\;\Hs^{\sigma/2}(\O^c),\end{cases}
\end{gather*}
where $\O^c$ represents the causal complement of $\O$. It can be shown that $A_\sigma$ is a densely defined and closed operator with the property
$$
(A_\sigma)^*=A_{-\sigma}.
$$
Now, Figliolini and Guido introduce an isommetric isomorphism between the domain of $\tilde{S}$ and a direct sum of Sobolev spaces via the unitary operator 
\begin{gather*}
T:\Hs^{-1/2}(\O)\oplus\Hs^{1/2}(\O)\longrightarrow\mathbf{D}(\check{S})\\
T:=2^{-1/2}\big(\o^{-1/2}\oplus -i\o^{1/2}\big),
\end{gather*}
and they identify herewith $\check{S}$ with the operator diag$(C,C)$, since this isomorphism maps the invariant elements with respect to diag$(C,C)$ onto the invariant ones of $\check{S}$. With this correspondence Figliolini and Guido give an explicit formula for the infinitesimal generator $\check{\Delta}_m$ of the modular operator with respect to the mass $m\geq 0$.

\begin{theo}\label{Guido-generator1}
Let $B$ be the self-adjoint operator defined as
\begin{gather*}
B:\Hs^{-1/2}(\O)\oplus\Hs^{1/2}(\O)\supset\F_{-1/2}\oplus\F_{1/2}\longrightarrow\Hs^{-1/2}(\O)\oplus\Hs^{1/2}(\O)\\
B:=\begin{pmatrix}0 & iA_{+1}\\ -iA_{-1} & 0 \end{pmatrix},
\end{gather*}
then for $m\geq 0$
\begin{gather}\label{Guido-generator2}
1\notin\mathbf{Spec}_p(B)\quad\text{and}\quad T^*\check{\Delta}_mT=\frac{B+\1}{B-\1}.
\end{gather}
\end{theo}
Consequently, $1$ is also not contained in the point spectrum of $\check{\Delta}_m$, but one can show that, since $A_\sigma$ and therefore $B$ are unbounded, $1$ has to be in the spectrum of $\check{\Delta}_m$. This means that the von Neumann algebra $\Ma(\O)$ is the unique hyperfinite factor of type $III_1$.\\   
The next result of Figliolini and Guido will also be of importance for us.

\begin{theo}\label{generatorcontinuity}
Let $\delta_m$ be given as aforementioned. Then:

\begin{itemize}
\item[(i)] The function $m\mapsto\check{\Delta}_m$ is continuous in the strong generalised sense.
\item[(ii)] The function $m\mapsto\check{\Delta}_m^{it}$ is strongly continuous, uniformly for every $t$ in any finite interval.
\item[(iii)] The function $m\mapsto\Delta_m^{it}$ is strongly continuous, uniformly for every $t$ in any finite interval.
\end{itemize}
\end{theo}
%\vspace{1cm}

Although with Theorem \ref{Guido-generator1} Figliolini and Guido seem to give an explicit expression for the massive generator, its real nature is not apparent. Even if one tries to calculate the special case $m=0$ directly via \eqref{Guido-generator2} in order to confirm the results of Bisognano and Wichmann, Theorem \ref{Bisognano}, Buchholz, Theorem \ref{Buchholz}, and Hislop and Longo, Theorem \ref{Hislop-Longo}, one faces some obstacles. Also the verification of Theorem \ref{Guido-generator1} by use of the known modular operators for $m=0$, i.e., by insertion of the known massless modular operators acting on the Cauchy data into equation \eqref{Guido-generator2}, is not free of difficulties. 

As mentioned above, see \eqref{anti-local}, the energy operator $\o$ reshuffles the support of the test functions also into the causal complement of $\O$, what is then corrected by the characteristic operator $P_\sigma$. Because of this anti-local property of $\o$, the term
\begin{gather*}
A_{-1}A_{+1}=P_{+1}\o^{-\sigma}\big\vert_{\Hs^{-1/2}(\O)}P_{-1}\o^\sigma\big\vert_{\Hs^{1/2}(\O)},
\end{gather*}
can hardly be further analysed directly, as long as one is not able to completely determine this reshuffling procedure. This lack of knowledge termitates the idea to approach the problem with the use of asymptotic expansions of $\o$ and $\o^{-1}$, see Example \ref{asymptotic0},
\begin{gather*}
\o f(x)=\left((-\Delta)^{1/2}+\frac{m^2}{2}(-\Delta)^{-1/2}+\sum_{k=2}^\infty\frac{1\cdot 3\cdots(2k-3)}{2\cdot 4\cdots 2k}(-\Delta)^{-k+\frac{1}{2}}m^{2k}\right)f(x),
\end{gather*}
and
\begin{align*}
\o^{-1} f(x)=\Bigg((-\Delta)^{-1/2}-\frac{m^2}{2}&(-\Delta)^{-3/2}\\
&+\sum_{k=2}^\infty\frac{1\cdot 3\cdots(2k-3)}{2\cdot 4\cdots 2k}(-\Delta)^{-k-\frac{1}{2}}m^{2k}\Bigg)f(x),
\end{align*}
where $f$ is a test function.

It is more probable that Theorem \ref{Guido-generator1} could contribute to the determination of the order of the infinitesimal generator, since the analysis of Figliolini and Guido is realised by means of Sobolev spaces and, for pseudo-differential operators, the order is linked up with the mapping property via Equation \eqref{Sobolev1}, namely via the implication:
\begin{equation*}
s<s'\quad\Longrightarrow\quad H^{s'}(\R^{n})\subset H^{s}(\R^{n}).
\end{equation*}
This relation could lead to an upper bound for the order and, if the upper bound is less than one, confirm the assertion of the infinitesimal generator for the massless theory being the leading part.

\section{Unitary Equivalence of Free Local Algebras}

In \cite{Eckmann:1974} Eckmann and Fr\"ohlich prove the unitary equivalence of local algebras $\Ma_m(\O)\equiv\mathfrak{M}(\O,m)$, where $\O$ is a bounded open subset of $\R^d$, in the quasifree representation for  $m\geq 0$. Their proof is based on the strategy that two representations of the CCR are isomorphic if the number operators from one of them can be described by means of the state which implements the other representation. In the sequel we want to summarize their result. Let $\R^d$ be the configuration space and $\H:=L^{2}(\R^d,dx)$ the Hilbert space of the one-particle wave functions. Furthermore we denote by 
\begin{equation*}
\F(\H):=\bigoplus_{m=0}^{\infty}L^{2}(\R^d,dx)^{\otimes_{s}m},\qquad\text{with}\quad L^{2}(\R^d,dx)^{\oplus_{s}0}:=\C,
\end{equation*}
the symmetric Fock space over $\H$ and the vacuum vector by $\Omega_{0}:=(1,0,0,\cdots)\in\H$.\\
The creation and annihilation operators $A^{*}(f)$ and $A(g)$, $f,g\in\H$, satisfy the canonical commutation relations (CCR),
\begin{gather*}
\big[A(f),A^{*}(g)\big]=(f,g)_{2}:=\int_{\R^d}\bar{f}(x)g(x)dx,\\
\big[A(f),A(g)\big]=\big[A^{*}(f),A^{*}(g)\big]=0,
\end{gather*}
and 
$$
A(f)\Omega_{0}=0\quad\forall f\in\H.
$$
Eckmann and Fr\"ohlich are considering  the Fourier transform of multiplication by $(k^2+m^2)^{1/2}$ on $\L^2(\R^d,dk)$,
$$
\mu f(x):=\int \big((\k^{2}+m^{2})^{1/2}\big)^{\sim}(x-y)f(y)dy
$$
for all $m\geq 0$. Then $\mu$ is a real, positive and self-adjoint operator on $\H$, and, for some dense set $D_{\mu}\subseteq L_{\text{real}}^{2}(\R^{d},dk)$, it fulfills the following inclusions:
\begin{equation*}
\D(\mu^{1/2})\supseteq\D(\mu):=D_{\mu}+iD_{\mu},\quad\text{and}\quad\D(\mu^{-1/2})\supseteq\D(\mu).
\end{equation*}
Further on they restrict themselves to the time-zero formulation, i.e., 
\begin{gather}\label{Eckmann1}
\varphi_{\mu}[f]=2^{-1/2}\big(A^{*}(\mu^{-1/2}f)+A(\mu^{-1/2}f)\big),\\
\pi_{\mu}[f]=2^{-1/2}\big(A^{*}(\mu^{+1/2}f)+A(\mu^{+1/2}f)\big)\notag
\end{gather}
for $f\in D_{\mu}$. These fields satisfy the Weyl relations,
\begin{equation*}
e^{i\varphi_{\mu}[f]}e^{i\pi_{\mu}[g]}=e^{i(f,g)_{2}}e^{i\pi_{\mu}[g]}e^{i\varphi_{\mu}[f]}.
\end{equation*}
For a bounded open region $\O\subseteq\R^d$ a complex ${}^{*}$-algebra
\begin{equation}\label{Eckmann2}
\overset{\circ}{\mathfrak{M}}_{\mu}(\O):=\Big\{e^{i\varphi_{\mu}[f]},\,e^{i\pi_{\mu}[f]}|\quad f\in D_{\mu},\supp f\subset \O\Big\}
\end{equation}
is constructed along with the von Neumann algebra $\mathfrak{M}_{\mu}(\O)$ as its weak closure in $\F(\H)$.\\
Since von Neumann algebras with respect to different masses are to be compared, Eckmann and Fr\"ohlich examine two different operators $\mu_{1}^{-1}$ and $\mu_{2}^{-1}$, which satisfy the  following statements for $\D:=D+iD$ with a dense subset $D\subseteq L_{\text{real}}^{2}(\R^{d},dk)$:
\begin{gather*}
\D\big(\mu_{1}^{+1/2}\big)\supseteq\D\cup\mu_{2}^{-1/2}\D,\quad \D\big(\mu_{2}^{+1/2}\big)\supseteq\D\cup\mu_{1}^{-1/2}\D,\\
\D\big(\mu_{1}^{-1/2}\big)\supseteq\D\cup\mu_{2}^{+1/2}\D,\quad \D\big(\mu_{2}^{-1/2}\big)\supseteq\D\cup\mu_{1}^{+1/2}\D.
\end{gather*}
The $\*$-algebras $\Ma_{\mu_{i}}(\O),\,i=1,2,$ are defined via \eqref{Eckmann2} by replacing $D_{\mu}$ with $D$.
The Bogoliubov transformation $\beta:=(\beta_{+}, \beta_{-})$  is introduced as
\begin{equation*}
\beta_{\pm}:=\frac{1}{2}\big(\mu_{2}^{-1/2}\mu_{1}^{1/2}\pm\mu_{2}^{1/2}\mu_{1}^{-1/2}\big).
\end{equation*}
If all conditions given above are fulfilled, then the mapping 
$$
(A^{*},A)\mapsto(A_{\beta}^{*},A_{\beta})
$$
with
\begin{gather*}
A_{\beta}^{*}:=A^{*}(\beta_{+}f)+A(\beta_{-}f),\\
A_{\beta}:=A^{*}(\beta_{-}f)+A(\beta_{+}f),
\end{gather*}
defines another one for the fields 
\begin{equation*}
\big(\varphi_{\mu_{1}}[f],\pi_{\mu_{1}}[g]\big)\mapsto\big(\varphi_{\mu_{2}}[f],\pi_{\mu_{2}}[g]\big),
\end{equation*}
via the relations \eqref{Eckmann1} to get herewith an invertible homomorphism
\begin{equation}\label{EckmannIso1}
\tau_\beta:\overset{\circ}{\mathfrak{M}}_{\mu_{1}}(\R^{d})\longrightarrow\overset{\circ}{\mathfrak{M}}_{\mu_{2}}(\R^{d}). 
\end{equation}
One obtains the explicit form
\begin{gather}\label{Eckmann3}
 \begin{split}
  \varphi_{\mu_{2}}[f]
  &=2^{-1/2}\big(A_{\beta}^{*}[\mu_{1}^{-1/2}f]+A_{\beta}[\mu_{1}^{-1/2}f]\big)\\
  &=2^{-1/2}\big(A^{*}[\beta_{+}\mu_{1}^{-1/2}f]+A[\beta_{-}\mu_{1}^{-1/2}f]+A^{*}[\beta_{-}\mu_{1}^{-1/2}f]+A[\beta_{+}\mu_{1}^{-1/2}f]\big)\\
  &=2^{-1/2}\big(A^{*}[(\beta_{+}+\beta_{-})\mu_{1}^{-1/2}f]+A[(\beta_{-}+\beta_{+})\mu_{1}^{-1/2}f]\big)\\
  %&=2^{-1/2}\big(A^{*}[\mu_{2}^{-1/2}\mu_{1}^{1/2}\mu_{1}^{-1/2}f]+A[\mu_{2}^{-1/2}\mu_{1}^{1/2}\mu_{1}^{-1/2}f]\big)\\
  &=2^{-1/2}\big(A^{*}[\mu_{2}^{-1/2}f]+A[\mu_{2}^{-1/2}f]\big).
 \end{split}
\end{gather}
For irreducible representations, $\tau_{\beta}$ according to \eqref{EckmannIso1} is unitarily implementable on $\F$ if and only if $\beta$ is a Hilbert-Schmidt operator.\\
The main result of Eckmann and Fr\"ohlich, namely the unitary equivalence of von Neumann algebras of local observables $\mathfrak{M}_{\mu_{i}}$, $i=1,2$, is stated in the following

\begin{theo}\label{EckmannIso2}
Let 
\begin{gather}\label{Eckmann-operator}
\mu_if(x):=\int \big((k^{2}+m_i^{2})^{1/2}\big)^{\sim}(x-y)f(y)dy,\quad i=1,2,
\end{gather}
then the factors $\mathfrak{M}_{\mu_{1}}(\O)$ and $\mathfrak{M}_{\mu_{2}}(\O)$ are unitarily equivalent for bounded open regions $\O\subset\R^d$ with piecewise smooth boundaries
\begin{itemize}
\item[(i)] in $d=2,3$ space dimensions if $m_i=0$ for one $\mu_i$, and
\item[(ii)] in $d=1$ space dimensions if $m_i\neq 0$ for both $i=1,2$. 
\end{itemize}
\end{theo}

We want to make use of this unitary equivalence of local algebras for the description of the massive fields by means of the massless ones in order to reduce the modular action in the massive to the known massless case.

\begin{cor}
For the massive $\varphi_m[f]\in\Ma_m(\O)$, $m>0$, and massless free scalar fields $\varphi_0[f]\in\Ma_0(\O)$, where $\O$ is a bounded open region in $\R^3$, the following statement holds: 
\begin{gather}\label{Eckmann4}
\varphi_m[f]=\varphi_0[f]+\varphi_0\big[f_{rest}\big]\\
\text{with}\quad f_{rest}(\x)=-4\pi\int\left(\frac{|r|^{3/2}}{(r^{2}+m^{2})^{1/4}}-r\right)\frac{\sin\big(r|\x-\y|\big)}{|\x-\y|}f(\y)drd\y.\notag
\end{gather}
\end{cor}

\proof First, we choose in \eqref{Eckmann-operator} operators $\mu_0$ and $\mu_m$ with the masses $m=0$ and $m>0$, respectively, and obtain for the product $\mu_{0}^{1/2}\mu_{m}^{-1/2}(f)$:

\begin{gather}\label{Eckmann6}
 \begin{split}
  \mu_{0}^{1/2}\mu_{m}^{-1/2}(f)
  &=\int\left(\frac{|\k|^{1/2}}{(\k^{2}+m^{2})^{1/4}}\right)^{\sim}(\x-\y)f(\y)d\y\\
  &=\int\left(\frac{|\k|^{1/2}}{(\k^{2}+m^{2})^{1/4}}-1+1\right)^{\sim}(\x-\y)f(\y)d\y\\
  &=\int\left\{\left(\frac{|\k|^{1/2}}{(\k^{2}+m^{2})^{1/4}}-1\right)^{\sim}(\x-\y)+\delta_{0}(\x-\y)\right\}f(\y)d\y\\
  &=f(\x)+\int\left(\frac{|\k|^{1/2}}{(\k^{2}+m^{2})^{1/4}}-1\right)^{\sim}(\x-\y)f(\y)d\y\\
&=f(\x)+\int\underbrace{\left(\frac{|\k|^{1/2}}{(\k^{2}+m^{2})^{1/4}}-1\right)}_{=:K(\k)}e^{-i\langle \k,(\x-\y)\rangle}f(\y)d\k d\y\\
  &=:f(\x)+f_r(\x).
 \end{split}
\end{gather}
Since $K(\k)$ and therefore $\widetilde{K}(\x-\y)$ are invariant under rotations we are allowed to simplify the right-hand side by choosing special coordinates $(0,0,|\x-\y|)^{t}$,
\begin{gather}
f_{rest}(\x)=  \int\left(\frac{|\k|^{1/2}}{(\k^{2}+m^{2})^{1/4}}-1\right)e^{-i\langle \k,(0,0,|\x-\y|)^{t}\rangle}f(\y)d\k d\y.
\end{gather}
Passing to spherical coordinates yields,
\begin{gather*}
 \begin{split}
  f_{rest}(\x)
  &=-\int\left(\frac{|r|^{1/2}}{(r^{2}+m^{2})^{1/4}}-1\right)e^{-ir\left\langle{\tiny\left(\begin{array}{c}\cos\theta\cos\phi\\ \cos\theta\sin\phi\\ \sin\theta\end{array}\right),\left(\begin{array}{c}0\\0\\|\x-\y|\end{array}\right)}\right\rangle}\cdot\\
  &\hspace{7cm}\cdot r^{2}\cos\theta f(\y)drd\phi d\theta d\y\\
  &=\int\left(\frac{|r|^{1/2}}{(r^{2}+m^{2})^{1/4}}-1\right)e^{-ir\sin\theta|\x-\y|}r^{2}\cos\theta f(\y)drd\phi d\theta d\y\\
  &=\frac{2\pi}{i}\int\left[\left(\frac{|r|^{3/2}}{(r^{2}+m^{2})^{1/4}}-r\right)\frac{e^{-ir\sin\theta|\x-\y|}}{|\x-\y|}\right]_{-\frac{\pi}{2}}^{\frac{\pi}{2}} f(\y)drd\y\\
  &=\frac{2 \pi}{i}\int\left(\frac{|r|^{3/2}}{(r^{2}+m^{2})^{1/4}}-r\right)\frac{e^{-ir|\x-\y|}-e^{ir|\x-\y|}}{|\x-\y|}f(\y)drd\y\\
  &=-4\pi\int\left(\frac{|r|^{3/2}}{(r^{2}+m^{2})^{1/4}}-r\right)\frac{\sin\big(r|\x-\y|\big)}{|\x-\y|}f(\y)drd\y\\
  &=-4\pi\int R(r)\frac{\sin\big(r|\x-\y|\big)}{|\x-\y|}f(\y)drd\y,
\end{split}
\end{gather*}
where
\begin{gather*}
R(r):=\frac{|r|^{3/2}}{(r^{2}+m^{2})^{1/4}}-r.
\end{gather*}
Because of relation \eqref{Eckmann3} we conclude:
\begin{gather}\label{Eckmann5}
 \begin{split}
  \varphi_m[f]
  &=2^{-1/2}\big(A^{*}[\mu_{2}^{-1/2}f]+A[\mu_{2}^{-1/2}f]\big)\\
  &=2^{-1/2}\big(A^{*}[\mu_{1}^{-1/2}\mu_{1}^{1/2}\mu_{2}^{-1/2}f]+A[\mu_{1}^{-1/2}\mu_{1}^{1/2}\mu_{2}^{-1/2}f]\big)\\
  &=\varphi_0\big[\mu_{1}^{1/2}\mu_{2}^{-1/2}(f)\big],
 \end{split}
\end{gather}
where we set $\varphi_0[f]:=\varphi_{\mu_{1}}[f]$ and $\varphi_m[f]:=\varphi_{\mu_{2}}[f]$. 
We complete the proof of the statement with the help of Theorem \ref{EckmannIso2} and the relations \eqref{Eckmann6} and \eqref{Eckmann5},
\begin{gather*}
 \varphi_m[f]=\varphi_0[f(\x)]+\varphi_0[f_{rest}(\x)].
\end{gather*}

\qed

Thus the term $\varphi_0[f_{rest}]$ represents the difference between the massless and massive algebra.\\
Evidently, the next step would be to derive the action of $\delta_m$ on the massive algebra via the action of $\delta_0$ on the massless algebra, i.e. via equation \eqref{Eckmann4}. But this strategy cannot be applied since the test functions $f$ and $f_{rest}$ do not have the same support. In general, the support of $f_{rest}$ does not have to lie in the same double cone as that of $f$. Also here, the reshuffling procedure cannot be specified further, what would be necessary for the the final step of this analysis.

In the cases where this non locality would not occur, the massive infinitesimal generator reads
\begin{gather*}
\delta_m=\delta_0+\delta_{rest}
\end{gather*} 
with
\begin{align*}
\delta_{rest}f(\x):=&\int R(r)\frac{\Big(r|\x-\y|\cos\big(r|\x-\y|\big)-\sin\big(r|\x-\y|\big)\Big)}{|\x-\y|^{3}}\cdot\\
&\hspace{7cm}\cdot y_0(1+\x^2)f(\y)drd\y.
\end{align*} 
Under the assumption of $\delta_{rest}$ being a pseudo-differential operator, the symbol of the infinitesimal generator has to be of the following form:
\begin{align*}
p(\x,\y,r):=R(r)e^{-i(\x-\y)r}&y^0(1+\x^2)\cdot\\
&\cdot\frac{\Big(r|\x-\y|\cos\big(r|\x-\y|\big)-\sin\big(r|\x-\y|\big)\Big)}{|\x-\y|^{3}}.
\end{align*}
One identifies $p(\x,\y,r)$ as a classical symbol with a singularity at the point $\x=\y$.

\section{Cocycle-Theorem}

One may use the unitary equivalence relation between the massive algebra $\mathfrak{M}_m(\O)$ and the massless algebra $\mathfrak{M}_0(\O)$ for the investigation of another approach. The idea is to formulate the modular automorphism group for the massive von Neumann algebra $\sigma_{m,0}^\tau$ with reference to the `wrong' massless vacuum vector $\Omega_0$ in an intermediate step. This strategy is allowed insofar as, due to the unitary equivalence, $\Omega_0$ is cyclic and separating for the massive algebra, too, and consequently one can formulate the modular objects in this case.

As a well-known fact the massive free scalar field $\varphi_m\in\mathfrak{M}_m(\O)$ can be described by means of the $\mathbf{\Delta}_m(x-y):=E(x,y)$ and the Cauchy data $\varphi_0(0,\y)$ and $\partial_t\varphi_0(0,\y)$,

\begin{align*}
\varphi_m(t,\x)=\int\big(\mathbf{\Delta}_m(t,\x-\y)\dot{\varphi}_0(0,\y)+\dot{\mathbf{\Delta}}_m(t,\x-\y)\varphi_0(0,\y)\big)d\y,
\end{align*}
where $\varphi_0\in\mathfrak{M}_0(\O)$. The modular group $\sigma_{m,0}^\tau$ acts on the smeared-out field according to:

\begin{align*}
\sigma_{m,0}^\tau\varphi_m[f]&=\sigma_0^\tau\left(\int\big[\mathbf{\Delta}_m(t,\x-\y)\dot{\varphi}_0(0,\y)+\dot{\mathbf{\Delta}}_m(t,\x-\y)\varphi_0(0,\y)\big]f(x)d\y d\x\right) \\
&=\int\Big(\mathbf{\Delta}_m(t,\x-\y)\dot{\varphi}_0\big(\nu_{-\tau}(0,\y)\big)\\
 &\qquad\quad+\partial_t\mathbf{\Delta}_m(t,\x-\y)\varphi_0\big(\nu_{-\tau}(0,\y)\big)\Big)\gamma_\tau f\big(\nu_{-\tau}(x)\big)d\y d\x,
\end{align*}
where $\sigma_0^\tau$ denotes the modular group in Theorem \ref{Hislop-Longo}, which has been derived by Hislop and Longo for the massless algebra and double cones. Also for the definition of $\gamma_\tau\equiv\gamma(x^0,x^3,\tau)$ and $\nu_{-\tau}(x)$ consult Theorem \ref{Hislop-Longo}.The infinitesimal generator $\delta_{m,0}$ of $\sigma_{m,0}^\tau$ is then given by the derivative at the point $\tau=0$.

\begin{align}\label{cocycle1}
\partial_\tau\sigma_{m,0}^\tau\varphi_m[f]&=\int\partial_\tau\Big\{\Big(\mathbf{\Delta}_m(t,\x-\y)\dot{\varphi}_0\big(\nu_{-\tau}(0,\y)\big)\notag\\
 &\qquad\quad+\dot{\mathbf{\Delta}}_m(t,\x-\y)\varphi_0\big(\nu_{-\tau}(0,\y)\big)\Big)\gamma_\tau f\big(\nu_{-\tau}(x)\big)\Big\}d\y d\x\notag\\
&=\int\Big(\mathbf{\Delta}_m(t,\x-\y)\partial_\tau\dot{\varphi}_0\big(\nu_{-\tau}(0,\y)\big)\notag\\
 &\qquad\quad+\dot{\mathbf{\Delta}}_m(t,\x-\y)\partial_\tau\varphi_0\big(\nu_{-\tau}(0,\y)\big)\Big)\gamma_\tau f\big(\nu_{-\tau}(x)\big)d\y d\x\notag\\
&+\int\Big(\mathbf{\Delta}_m(t,\x-\y)\dot{\varphi}_0\big(\nu_{-\tau}(0,\y)\big)\notag\\
 &\qquad\quad+\dot{\mathbf{\Delta}}_m(t,\x-\y)\varphi_0\big(\nu_{-\tau}(0,\y)\big)\Big)\partial_\tau\big(\gamma_\tau\big) f\big(\nu_{-\tau}(x)\big)d\y d\x\notag\\
&+\int\Big(\mathbf{\Delta}_m(t,\x-\y)\dot{\varphi}_0\big(\nu_{-\tau}(0,\y)\big)\notag\\
 &\qquad\quad+\dot{\mathbf{\Delta}}_m(t,\x-\y)\varphi_0\big(\nu_{-\tau}(0,\y)\big)\Big)\gamma_\tau\partial_\tau f\big(\nu_{-\tau}(x)\big)d\y d\x.\notag\\
\end{align}
Let us have a closer look on the first addend on the right-hand side of equation \eqref{cocycle1},
\begin{align*}
\partial_\tau\partial_t\varphi_0\big(\nu_{-\tau}(0,\y)\big)&=\partial_t^2\varphi_0\big(\nu_{-\tau}(0,\y)\big)\partial_\tau\nu_{-\tau}(0,\y)\\
&\quad+\sum_{i=1}^3\partial_{y_i}\partial_t\varphi_0\big(\nu_{-\tau}(0,\y)\big)\partial_\tau\nu_{-\tau}(0,\y)\\
&=\big(\Delta-m^2\big)\varphi_0\big(\nu_{-\tau}(0,\y)\big)\partial_\tau\nu_{-\tau}(0,\y)\\
&\quad+\sum_{i=1}^3\partial_{y_i}\partial_t\varphi_0\big(\nu_{-\tau}(0,\y)\big)\partial_\tau\nu_{-\tau}(0,\y),\\
\partial_\tau\varphi_0\big(\nu_{-\tau}(0,\y)\big)&=\partial_t\varphi_0\big(\nu_{-\tau}(0,\y)\big)\partial_\tau\nu_{-\tau}(0,\y)\\
&\quad+\sum_{i=1}^3\partial_{y_i}\varphi_0\big(\nu_{-\tau}(0,\y)\big)\partial_\tau\nu_{-\tau}(0,\y)
\end{align*}
By plugging this into \eqref{cocycle1} at the origin $\tau=0$ we get,
\begin{align}\label{cocycle2}
\delta_{m,0}\varphi_m[f]&=\partial_\tau\sigma_{m,0}^\tau\varphi_m[f]\big|_{\tau=0}\notag\\
&=\int\Big\{\mathbf{\Delta}_m(t,\x-\y)\Big[\big(\Delta-m^2\big)\varphi_0(0,\y)+\sum_{i=1}^3\partial_{y^i}\dot{\varphi}_0(0,\y)\Big]\partial_\tau\nu_{-\tau}(0,\y)\big|_{\tau=0}\notag\\
 &\qquad\quad+\dot{\mathbf{\Delta}}_m(t,\x-\y)\sum_{i=0}^3\partial_{y^i}\varphi_0(0,\y)\partial_\tau\nu_{-\tau}(0,\y)\big|_{\tau=0}\Big\}f(x)d\y d\x\notag\\
&\quad+\int\Big\{\mathbf{\Delta}_m(t,\x-\y)\partial_t\varphi_0(0,\y)\\
 &\qquad\quad+\dot{\mathbf{\Delta}}_m(t,\x-\y)\varphi_0(0,\y)\Big\}\partial_\tau\gamma_\tau\big|_{\tau=0} f(x)d\y d\x\notag\\
&\quad+\int\Big\{\mathbf{\Delta}_m(t,\x-\y)\dot{\varphi_0}(0,\y)\notag\\
 &\qquad\quad+\dot{\mathbf{\Delta}}_m(t,\x-\y)\varphi_0(0,\y)\Big\}\partial_\tau f\big(\nu_{-\tau}(x)\big|_{\tau=0}\big)d\y d\x,\notag\\
&=\int\Big\{\mathbf{\Delta}_m(t,\x-\y)\Big[(\Delta-m^2)\varphi_0(0,\y)+\notag\\
&\hspace{5,5cm}\sum_{i=1}^3\partial_{y^i}\dot{\varphi}_0(0,\y)+\dot{\varphi}_0(0,\y)\partial_t\Big]f(x)\notag\\
 &\qquad\quad+\dot{\mathbf{\Delta}}_m(t,\x-\y)\Big[\sum_{i=1}^3\partial_{y^i}\varphi_0(0,\y)+\varphi_0(0,\y)\partial_t\Big]\Big\}\notag\\
&\hspace{8cm}f(x)(1+\x^2)d\y d\x,\notag
\end{align}
where we have used the relations
\begin{gather*}
\lim_{\tau\rightarrow 0}\gamma(x^0,x^3,\tau)=\lim_{\tau\rightarrow 0}\nu_{-\tau}(x)=1,\\
\partial_\tau\nu_{-\tau}(x^0)\big|_{\tau=0}=-\frac{1}{2}(1+\x^2)+(x^0)^2,\\
\text{and}\quad\partial_\tau\nu_{-\tau}(\x)\big|_{\tau=0}=\x x^0.
\end{gather*}
To put it in a nutshell we have shown

\begin{theo}
Let $\mathfrak{M}_m(\O)$ be the von Neumann algebra generated by the massive free scalar field $\varphi_m[f]$, then $\delta_{m,0}$ given in \eqref{cocycle2} is the infinitesimal generator of the modular automorphism group $\sigma_{m,0}^\tau$ on $\mathfrak{M}_m(\O)$ with respect to the massless vacuum vector $\Omega_0$.
\end{theo}

Now one hopes to bridge the gap between the `wrong' massless vacuum and the massive vacuum via the cocycle theorem. For this purpose we have to thoroughly investigate the proof of Connes' Theorem \ref{Connes-theorem}.\\
First of all, we define a new faithful, semifinite and normal state on $\mathfrak{N}_\rho:=\mathfrak{M}_m\bigotimes\mathbf{M_2}$ with reference to the massive and massless states on $\mathfrak{M}_m$ by
\begin{equation*}
\rho\begin{pmatrix}A_{11} & A_{12}\\ A_{21} & A_{22}\end{pmatrix}:=(\o_m\oplus\o_0)\begin{pmatrix}A_{11} & A_{12}\\ A_{21} & A_{22}\end{pmatrix}:=\o_m(A_{11})+\o_0(A_{22}).
\end{equation*}
Because of the properties of $\rho$ mentioned above, we may introduce the corresponding modular objects. One finds that the representation space can be expressed as a direct sum of mutually orthogonal subspaces,
$$
\H_\rho=\H_{\o_m,\o_m}\oplus\H_{\o_m,\o_0}\oplus\H_{\o_0,\o_m}\oplus\H_{\o_0,\o_0},
$$
with
\begin{align*}
\H_{\o_m,\o_m}&:=\Big[\big((\mathfrak{n}_{\o_m}\cap\mathfrak{n}_{\o_m}^*)\otimes e_{11}\big)+\mathbf{N}_\rho\Big],\\
\H_{\o_m,\o_0}&:=\Big[\big((\mathfrak{n}_{\o_m}^*\cap\mathfrak{n}_{\o_0})\otimes e_{12}\big)+\mathbf{N}_\rho\Big],\\
\H_{\o_0,\o_m}&:=\Big[\big((\mathfrak{n}_{\o_m}\cap\mathfrak{n}_{\o_0}^*)\otimes e_{21}\big)+\mathbf{N}_\rho\Big],\\
\H_{\o_0,\o_0}&:=\Big[\big((\mathfrak{n}_{\o_0}\cap\mathfrak{n}_{\o_0}^*)\otimes e_{22}\big)+\mathbf{N}_\rho\Big],
\end{align*}
where $\mathbf{N}_\rho:=\big\{A\in\mathfrak{N}_\rho|\;\rho(A^*A)=0\big\}$ is a left ideal of $\mathfrak{N}_\rho$, $\mathfrak{n}_{\o_m}:=\big\{A\in\Ma_m|\;\o_m(A^*A)<\infty\big\}$, $\mathfrak{n}_{\o_0}:=\big\{A\in\Ma_m|\;\o_0(A^*A)<\infty\big\}$
and 
\begin{align*}
e_{11}:=\begin{pmatrix}1 & 0\\ 0 & 0\end{pmatrix},\;e_{12}:=\begin{pmatrix}0 & 1\\ 0 & 0\end{pmatrix},\;e_{21}:=\begin{pmatrix}0 & 0\\ 1 & 0\end{pmatrix},\;e_{22}:=\begin{pmatrix}0 & 0\\ 0 & 1\end{pmatrix}
\end{align*}
are the matrix units. The brackets `$[\;\cdot\;]$' denote the closure in the Hilbert space $\H_\rho$. The von Neumann algebra $\mathfrak{N}_\rho$ inherits its involution from $\mathfrak{M}$. Since we are concerned with states, not weights, the sets simplify to $\mathfrak{n}_{\o_m}=\mathfrak{n}_{\o_0}=\Ma_m$. Furthermore, because  $\rho$ is faithful, we obtain $\mathbf{N}_\rho=\{0\}$. For the domain of the $\*$-operation on has,
$$
\D^*=\big(\D^*\cap\H_{\o_m,\o_m}\big)\oplus\big(\D^*\cap\H_{\o_m,\o_0}\big)\oplus\big(\D^*\cap\H_{\o_0,\o_m}\big)\oplus\big(\D^*\cap\H_{\o_0,\o_0}\big),
$$
which determines the Tomita operator as
\begin{align*}
S\big(\D^*\cap\H_{\o_m,\o_m}\big)&=\D^*\cap\H_{\o_m,\o_m},\\
S\big(\D^*\cap\H_{\o_m,\o_0}\big)&=\D^*\cap\H_{\o_0,\o_m},\\
S\big(\D^*\cap\H_{\o_0,\o_m}\big)&=\D^*\cap\H_{\o_m,\o_0},\\
S\big(\D^*\cap\H_{\o_0,\o_0}\big)&=\D^*\cap\H_{\o_0,\o_0}.
\end{align*}
Written in a more compact form, we obtain for the Tomita operator the following matrix:
$$
S=
\begin{pmatrix}
S_{11} & 0 & 0 & 0\\ 0 & 0 & S_{23} & 0\\ 0 & S_{32} & 0 & 0\\ 0 & 0 & 0 & S_{44}
\end{pmatrix}.
$$
Its adjoint is given by
$$
F=
\begin{pmatrix}
F_{11} & 0 & 0 & 0\\ 0 & 0 & F_{23} & 0\\ 0 & F_{32} & 0 & 0\\ 0 & 0 & 0 & F_{44}
\end{pmatrix},
$$
where the components are fixed as
$$
F_{11}:=S_{11} ^*,\quad F_{23}:=S_{32}^*,\quad F_{32}:=S_{23}^*\quad\text{and}\quad F_{44}:=S_{44}.
$$
Finally, via polar decomposition
$$
S=J\Delta^{1/2}=\Delta^{-1/2}J,
$$
which is equivalent to
\begin{align*}
S_{11}&=J_{11}\Delta_{11}^{1/2}=\Delta_{11}^{-1/2}J_{11},\\
S_{23}&=J_{23}\Delta_{33}^{1/2}=\Delta_{22}^{-1/2}J_{23},\\
S_{32}&=J_{32}\Delta_{22}^{1/2}=\Delta_{33}^{-1/2}J_{32},\\
S_{44}&=J_{44}\Delta_{44}^{1/2}=\Delta_{44}^{-1/2}J_{44},
\end{align*}
we arrive at the modular conjugation and, most important for us, the modular operator:
\begin{gather*}
J=
\begin{pmatrix}
J_{11} & 0 & 0 & 0\\ 0 & 0 & J_{23} & 0\\ 0 & J_{32} & 0 & 0\\ 0 & 0 & 0 & J_{44}
\end{pmatrix}
\quad\text{and}\quad
\Delta=
\begin{pmatrix}
\Delta_{11} & 0 & 0 & 0\\ 0 & \Delta_{22} & 0 & 0\\ 0 & 0 & \Delta_{33} & 0\\ 0 & 0 & 0 & \Delta_{44}
\end{pmatrix}.
\end{gather*}
Let $\mathfrak{M}_m$ act on the Hilbert space $\H$, then one can find linear maps 
\begin{gather*}
\xi:\mathfrak{n}_{\o_m}\ni A\mapsto\xi(A)\in\H\quad\text{and}\\
\eta:\mathfrak{n}_{\o_0}\ni A\mapsto\eta(A)\in\H
\end{gather*}
such that for all $A_\mathfrak{M}\in\mathfrak{M}_m$, $\;A_m,B_m\in\mathfrak{n}_{\o_m}$ and $\;A_0,B_0\in\mathfrak{n}_{\o_0}$
\begin{align*}
A_\mathfrak{M}\xi(A_m)&=\xi(A_\mathfrak{M}A_m),&\quad\o_m(A_m^*B_m)&=\big(\xi(A_m),\xi(B_m)\big),\\
A_\mathfrak{M}\eta(A_0)&=\eta(A_\mathfrak{M}A_0),&\quad\o_0(A_0^*B_0)&=\big(\eta(A_0),\eta(B_0)\big],
\end{align*}
\begin{align*}
\text{and}\quad\H=\big[\xi(\mathfrak{n}_{\o_m})\big]=\big[\eta(\mathfrak{n}_{\o_0})\big].
\end{align*}
The form of the representation space with respect to $\rho$ is reduced to
\begin{gather*}
\H_\rho=\H\oplus\H\oplus\H\oplus\H,
\end{gather*}
and the representation itself to:
\begin{gather*}
\pi_\rho\begin{pmatrix} A_{11} & A_{12}\\ A_{21} & A_{22}\end{pmatrix}=\begin{pmatrix}A_{11} & A_{12} & 0 & 0\\ A_{21} & A_{22} & 0 & 0\\ 0 & 0 & A_{11} & A_{12}\\ 0 & 0 & A_{21} & A_{22}\end{pmatrix}.
\end{gather*}
Since the modular automorphism group on $\mathfrak{N}_\rho$ should leave the representation space invariant, i.e., 
\begin{gather*}
\sigma_\rho^t\big(\pi_\rho(\mathfrak{N}_\rho)\big)=\Delta^{it}\pi_\rho(\mathfrak{N}_\rho)\Delta^{-it}=\pi_\rho(\mathfrak{N}_\rho),
\end{gather*}
we conclude
\begin{align*}
\Delta_{11}^{it}A_{11}\Delta_{11}^{-it}&=\Delta_{33}^{it}A_{11}\Delta_{33}^{-it},&\quad\Delta_{11}^{it}A_{12}\Delta_{22}^{-it}&=\Delta_{33}^{it}A_{12}\Delta_{44}^{-it},\\
\Delta_{22}^{it}A_{21}\Delta_{11}^{-it}&=\Delta_{44}^{it}A_{21}\Delta_{33}^{-it},&\quad\Delta_{22}^{it}A_{22}\Delta_{22}^{-it}&=\Delta_{44}^{it}A_{22}\Delta_{44}^{-it}.
\end{align*}
The identification of $\Delta_{11}$ and $\Delta_{44}$ with $\Delta_{\o_m}$ and $\Delta_{\o_0}$, respectively, yields the following expressions for the modular automorphism groups,
\begin{align*}
\sigma_{\o_m}^t(A)&=\Delta_{11}^{it}A\Delta_{11}^{-it}\in\mathfrak{M}_m,&\quad\sigma_{\o_m,\o_0}^t(A)&=\Delta_{11}^{it}A\Delta_{22}^{-it}\in\mathfrak{M}_m,\\
\sigma_{\o_0,\o_m}^t(A)&=\Delta_{22}^{it}A\Delta_{11}^{-it}\in\mathfrak{M}_m,&\quad\sigma_{\o_0}^t(A)&=\Delta_{22}^{it}A\Delta_{22}^{-it}\in\mathfrak{M}_m,
\end{align*}
for all $A\in\Ma_m$. Finally, we can write down the form of the modular group with respect to $\rho$ explicitly:
\begin{equation}
\sigma_\rho^t\left[\begin{pmatrix} A_{11} & A_{12}\\ A_{21} & A_{22}\end{pmatrix}\right]=\begin{pmatrix} \sigma_{\o_m}^t(A_{11})& \sigma_{\o_m,\o_0}^t(A_{12})\\ \sigma_{\o_0,\o_m}^t(A_{21}) & \sigma_{\o_0}^t(A_{22})\end{pmatrix}.
\end{equation}
As already mentioned, the unitary cocycle appearing in Connes' Theorem is defined by
\begin{equation*}
\begin{pmatrix}0 & \Gamma_{t}\\ 0 & 0\end{pmatrix}:=\sigma_{\rho}^t\left[\begin{pmatrix} 0 & \1\\ 0 & 0\end{pmatrix}\right],
\end{equation*}
and therefore one has its explicit structure
$$
\Gamma_{t}=\sigma_{\o_m,\o_0}^t(\1)
$$
for all $t\in\R$.

Thus, the modular automorphism group with respect to the massive vacuum on the massive algebra is determined up to a cocycle, i.e., up to a perturbation term.

%%%%%%%%%%%%%%%%%%%%%%%%%%%%%%%%%%%%%%%%%%%%%%%%%%%%%%%%%%%%%%%%%%%%%%%%%%%%

%\input{introdissp}
%\input{micro}
%\input{mathpre}
    %%\input{modudissp}
%\input{modqftdissp}
    %%\input{pseudissp}
%\input{massivegroup}
   %\input{eckdissp}
    %%\input{guidodissp}
   %\input{cocycledissp}

%%%%%%%%%%%%%%%%%%%%%%%%%%%%%%%%%%%%%%%%%%%%%%%%%%%%%%%%%%%%%%%%%%%%%%%%%%%%

\chapter{Summary and Outlook}

\begin{flushright}
  \emph{La\ss{} uns das Leben genie\ss en,\\
solange wir es nicht begreifen.}\\
\vspace{0,5cm}
K. Tucholsky
\end{flushright}
\vspace{0,5cm}

In this thesis, the interplay between modular theory and quantum field theory has been investigated extensively. It has been shown that modular theory not only is an excellent tool for the description of quantum field theory, especially for its most fundamental pillars like the PCT theorem, the spin-statistics theorem, and the concept of particles, but also discloses totally new and conceptually groundbreaking insights, for example the classification of quantum systems as hyperfinite factors of type $III_1$, the formulation of equilibrium states, the KMS states, and, last but not least, the introduction of the modular action as a geometric transformation. It has also been proven that another increasingly important mathematical discipline, namely micro-local analysis, apart from Radzikowki's characterisation of Hadamard states by wave front sets, enters quantum field theory a second time via the determination of the infinitesimal generator of modular groups with non local action as a pseudo-differential operator.

The well-known Assumption \ref{assumption-gen0} on the qualitative structure of the infinitesimal generator of modular groups acting non locally has been confirmed explicitly for the first time in two concrete examples. The first example, given by Yngvason \cite{Yngvason:1994nk}, is concerned with the algebra of massive Hermitian fields defined on the left or right wedges and transforming covariantly under translations, but not in general under Lorentz boosts. The second one, formulated by Borchers and Yngvason \cite{Borchers:1998ye}, introduces modular automorphism groups for the wedges, forward light cones, and double cones with respect not only to the vacuum state but also to general KMS states, which turn out to act locally only in the limit $\beta=T^{-1}\rightarrow\infty$, i.e., in the limit of ground states. One may pose the question if an automorphism group acting non locally on the von Neumann algebra always implies a pseudo-differential form for its infinitesimal generator.

The ansatz of Figliolini and Guido \cite{Figliolini:1989vf} for the derivation of the modular operator in the case of $m\geq 0$ has been discussed, and the conceptual and technical difficulties occurring there have been explored. Even the verification of the known results in the massless case can hardly be mastered as long as one is not able to determine the reshuffling property of the operators on which their analysis is based. It is possible, that the analysis of Figliolini and Guido could contribute to the identification of the order of the massive infinitesimal generator through the mapping property of the operator $B$, defined in Theorem \ref{Guido-generator1}, and the implication \eqref{Sobolev1} for Sobolev spaces,
\begin{equation*}
s<s'\quad\Longrightarrow\quad H^{s'}(\R^{n})\subset H^{s}(\R^{n}).
\end{equation*}

Furthermore, two new approaches to the calculation of the modular automorphism group $\big(\sigma_m^t\big)_{t\in\R}$ have been investigated. The first one, the ansatz via the unitary equivalence of local algebras, is shown to lead to non-local transformations, an obstacle which has been faced many times in various settings and which leads to uncontrollable domain problems. The second one gives a partial result, namely an automorphism group on the massive algebra, but with respect to the `wrong' massless vacuum state $\big(\sigma_{m,0}^t\big)_{t\in\R}$. The desire to bridge the gap to our main target $\big(\sigma_m^t\big)_{t\in\R}$, the modular automorphism group with reference to the `right' massive vacuum, via Connes' cocycle theorem remains unaccomplished, since this theorem is not constructive, but only ensures the existence of the cocycle. The perturbation formalism, which is used in the analysis of the interplay between KMS states and stability criteria, could provide assistance with this problem.

One may replace in future investigations the two-point function of the massice vacuum,
\begin{gather*}
\mathbf{\Delta}_m^+(x)=\frac{1}{(2\pi)^3}\int\Theta(p^0)\delta(p^2-m^2)e^{ipx}dp,
\end{gather*}
by the modified version, proposed in \cite{Duetsch:2004dd},
\begin{gather*}
H_m^\mu(x):=\mathbf{\Delta}_m^+(x)-\log\left(\frac{m^2}{\mu^2}\right)m^2h(m^2x^2), \\
\text{with}\qquad H_0^\mu(x)=\mathbf{\Delta}_0^+(x),
\end{gather*}
where $\mu>0$ is a new fixed mass parameter and $h$ is an analytic function such that $H_m^\mu(x)$ is smooth in $m\geq 0$. This ansatz is based on the idea that, since the second term comprises all logarithmic singularities, the new two-point function could be more compatible with conformal transformations.

%%%%%%%%%%%%%%%%%%%%%%%%%%%%%%%%%%%%%%%%%%%%%%%%%%%%%%%%%%%%%%%%%%%%%%%%%%%%

\begin{appendix}

\chapter{$C^*$-Algebras, States, Representations}

\begin{flushright}
\emph{Was auch immer geschieht:\\
Nie d\"urft ihr so tief sinken, von dem Kakao,\\
durch den man euch zieht, auch noch zu trinken!}\\
\vspace{0,5cm}
E. K\"astner 
\end{flushright}
\vspace{0,5cm}

%\begin{flushright}
%  \emph{Archimedes will be remembered \\
%when Aeschylus is forgotten.}\\
%\vspace{0,5cm}
%G. H. Hardy
%\end{flushright}
%\vspace{0,5cm}

In this chapter we want to give in a nutshell the basic terminology and some concepts of operator algebras which is used throughout this thesis. In the sequel we present in the following a number of definitions and statements collected mainly from \cite{Takesaki:2001}, \cite{Bratteli:1979tw} and \cite{Stratila:1981} without giving their proofs.  

\begin{defi}
A C$\*$-algebra $\Aa$ is a Banach$\*$-algebra, i.e., $\Aa$ is a Banach algebra with involution and $\|A^*\|=\|A\|$, such that $\|A^{*}A\|=\|A\|^{2}$ for all $A\in\Aa$. $\Aa$ is said to be simple, if the only closed ideals are $\{0\}$ and $\Aa$ itself.
\end{defi}

\begin{theo}
Let $\Aa$ be an abelian and unital C$\*$-algebra, then there exists a compact Hausdorff space $\X$ such that $\Aa$ is (isometrically) isomorphic to $\mathcal{C}(\X)$. In the case of a non-unital C$\*$-algebra, $\Aa$ is (isometrically) isomorphic to $\mathcal{C}_0(\X)$, if $\X$ is locally compact. In both cases the space $\X$ is uniquely determined up to homomorphisms.
\end{theo}

\begin{defi}
A C$\*$-algebra $\Ma$ is called $W^*$-algebra, if it is the dual space of some Banach space.
\end{defi}
If one is concerned with operator algebras, then it is usual to call $W^*$-algebras von Neumann algebras. Sakai has proved that the abstract characterisation of von Neumann algebras given above  is equivalent to the more usual definition: Let $\H$ be a Hilbert space and $\LH$ the algebra of linear bounded operators
\begin{equation*}
A:\H\longrightarrow\H
\end{equation*}
equipped with the norm
\begin{equation*}
\|A\|:=\underset{\psi\in\H, \|\psi\|\leq1}{\sup}\|A\psi\|<\infty.
\end{equation*}
Define the commutant $\Aa'$ of $\Aa$ as the algebra
\begin{equation*}
\Aa':=\{X\in\LH:\;XA=AX,\,A\in\Aa\}
\end{equation*}
with the properties  
\begin{equation*}
\begin{split}
\Aa\subseteq\Aa''=\Aa^{(4)}=\Aa^{(6)}=\cdots,\\
\Aa'=\Aa'''=\Aa^{(5)}=\Aa^{(7)}=\cdots,\\
\text{where}\quad\Aa^{(n+1)}:=\big(\Aa^{(n)}\big)'.
\end{split}
\end{equation*}
Then one verifies the following 

\begin{theo}[Sakai]
Every $\*$-algebra $\Ma\subseteq\LH$ is a von Neumann algebra if and only if $\Ma=\Ma''$.
\end{theo}

\begin{examps}\emph{
\begin{itemize}
\item[(i)] $\LH$ is not only a von Neumann algebra, but even a factor since  $\LH'=\C\1$.
\item[(ii)] The $C^*$-algebra of compact operators $\LCH$ on $\H$ cannot be a von Neumann algebra, because $\LCH''=(\C\1)'=\LH\neq\LCH.$
\end{itemize}}
\end{examps}

For further discussions on von Neumann algebras, additional topologies apart from the uniform topology are needed. For the sake of completeness, we introduce the most important ones although we will not make use of all of them explicitly.

\begin{defi}
On $\LH$ we distinguish between the locally convex operator topologies defined by the following sets of seminorms:

\begin{itemize}
\item [(i)] $\sigma$-weak: $p_{(u_n),(v_n)}(A):=\big|\underset{n}{\sum}(u_{n},Av_{n})\big|$ for all $u_n,v_n\in\H$ \\
with $\underset{n}{\sum}\big(\|u_{n}\|^{2}+\|v_{n}\|^{2}\big)<\infty$.
\item [(ii)] weak: $p_{u,v}(A):=|(u,Av)|$ for all $u,v\in\H$.
\item [(iii)] strong: $p_{u}(A):=\|Au\|$ for all $u\in\H$.
\item [(iv)] $\sigma$-strong: $p_{(u_n)}(A):=\underset{n}{\sum}\|Au_{n}\|^{2}$ for all $u_n\in\H$ with $\underset{n}{\sum}\|u_{n}\|^{2}<\infty$.
\item [(v)] $\*$-strong: $A\rightarrow p_{u}(A)+p_{u}(A^{*})$ for all $u\in\H$.
\item [(vi)] $\sigma\*$-strong: $A\rightarrow\left\{\underset{n}{\sum}\|Au_{n}\|^{2}+\underset{n}{\sum}\|A^{*}u_{n}\|^{2}\right\}^{\frac{1}{2}}$ for all $u_n\in\H$ \\
with $\underset{n}{\sum}\|u_{n}\|^{2}<\infty$.
\end{itemize}
\end{defi} 

For our purposes the $\sigma$-weak and $\sigma$-strong topologies will be of particular interest, because the modular group of automorphisms is contiuous with respect to them. If we denote by ``$<$'' the relation ``finer than'', then the following diagram shows the relation between the various topologies:\\

%\vspace{0.5cm}
\begin{center}
uniform $<\;\sigma\*$-strong$\;<\;\sigma$-strong$\;<\;\sigma$-weak\\
${}\hspace{2,3cm}\wedge\hspace{1.7cm}\wedge\hspace{1.5cm}\wedge$\\
${}\hspace{2cm}\*$-strong\hspace{0.2cm}$<\hspace{0.4cm}$strong$\hspace{0.3cm}<\hspace{0.5cm}$weak\quad.\\
\end{center}

%\vspace{0.5cm}

The $\sigma\*$-strong, $\sigma$-strong and $\sigma$-weak topologies allow for the same continuous linear functionals. This statement remains true if one drops the $\sigma$-. The main difference between the $\*$-strong and the $\sigma\*$-strong topology is the fact that the involution $A\mapsto A^{*}$ is only $\*$-strongly continuous.\\
With the help of these topologies we formulate the next 

\begin{theo}[Bicommutant -]
Let $\Aa\subset\LH$ be a nondegenerate $\*$-algebra, i.e., $[\Aa\H]:=\overline{\{A\xi|\;A\in\Aa,\xi\in\H\}}=\H$, then the following statements are equivalent:

\begin{itemize}
\item[(i)] $\Aa=\Aa''$.
\item[(ii)]$\Aa$ is weakly closed.
\item[(iii)]$\Aa$ is strongly closed.
\item[(iv)]$\Aa$ is $\*$-strongly closed.
\item[(v)]$\Aa$ is $\sigma$-weakly closed.
\item[(vi)]$\Aa$ is $\sigma$-strongly closed.
\item[(vii)]$\Aa$ is $\sigma\*$-strongly closed.
\end{itemize}
\end{theo} 
Therefore, a von Neumann algebra is a weakly closed $C^*$-subalgebra of $\LH$. The bicommutant theorem also states that the closure of the $\*$-algebra is independent of the choice of a particular topology. An immediate consequence is the next  

\begin{cor}[von Neumann density -]
The nondegenerate $\*$-algebra of operators $\Aa$ on $\H$ is weakly, strongly, $\*$-strongly, $\sigma$-weakly, $\sigma$-strongly and $\sigma\*$-strongly dense in $\Aa''$.
\end{cor}
This statement can be tightened to a stronger version.

\begin{theo}[Kaplanskys density -]
The unit ball of  the $\*$-algebra of operators $\Aa$ on $\H$ is $\sigma\*$-strongly dense in the unit ball of the weak closure of $\Aa$.
\end{theo}

\begin{defi}
Let $\Aa,\Ba$ be $C^*$-algebras, then a linear map $\phi:\Aa\longrightarrow\Ba$ is called a $\*$-homomorphism if

\begin{itemize}
\item[(i)] $\phi(AB)=\phi(A)\phi(B)$, and
\item[(ii)]$\phi(A^{*})=\big(\phi(A)\big)^{*}$
\end{itemize}
hold for all $A,B\in\Aa$.
\end{defi}
The notions mono-, epi-, iso-, endo- and automorphism are introduced in the usual manner. We want to keep hold of some fundamental properties of $\*$-homomorphisms $\phi$ in the following

\begin{lem}
Let $\Aa$ and $\Ba$ be C$\*$-algebras and $\phi:\Aa\longrightarrow\Ba$ a $\*$-homomorphism. Then the following statements are valid:

\begin{itemize}
\item[(i)] $\phi$ preserves positivity: $A\geq 0\Longrightarrow\phi(A)\geq 0$.
\item[(ii)]$\phi$ is continuous, and $\|\phi(A)\|\leq\|A\|$ for all $A\in\Aa$, thus $\phi$ can only decrease the norm.
\item[(iii)] $\phi$ is a $\*$-isomorphism if and only if $\ker\phi:=\{A\in\Aa:\,\phi(A)=0\}=\{0\}$.
\item[(iv)] A $\*$-isomorphism is automatically isometrical, i.e., norm preserving: $\|\phi(A)\|=\|A\|$ for all $A\in\Aa$.
\item[(iv)] The image $\phi(\Aa)$ of a $C^*$-algebra $\Aa$ is again a $C^*$-algebra.
\end{itemize}
\end{lem} 
In the case of von Neumann algebras we may state stricter properties of $\*$-homomorphism.

\begin{theo}\label{weakstrongtopology}
$\*$-homomorphisms $\phi:\Ma\longrightarrow\Na$ between von Neumann algebras $\Ma$ and $\Na$ are $\sigma$-weakly and $\sigma$-strongly continuous.
\end{theo}

\begin{defi}
A representation of a $C^*$-algebra $\Aa$ is a pair $(\H,\pi)$, consisting of a complex Hilbert space $\H$ and a $\*$-homomorphism $\pi:\Aa\longrightarrow\LH$, and it is said to be faithful if $\pi$ is a $\*$-isomorphism, and nondegenerate if $\{v\in\H:\;\pi(A)v=0,\,A\in\Aa\}=\{0\}$. A subspace $\F\subset\H$ is called invariant under $\pi(\Aa)$ if $\pi(A)\F\subseteq\F$ for all $A\in\Aa$. Whenever $\{0\}$ and $\H$ are the only invariant and closed subspaces, we call the representation $(\H,\pi)$  irreducible, otherwise reducible. Two representations $(\H_{1},\pi_{1})$ and $(\H_{2},\pi_{2})$ are said to be unitarily equivalent if there exists a unitary operator $U:\H_{1}\longrightarrow\H_{2}$ such that $\pi_{2}(A)=U\pi_{1}(A)U^{*}$ for all $A\in\Aa$. If the two Hilbert spaces are connected via a $\*$-isomorphism instead, then we call $\pi_1$ and $\pi_2$ quasi-equivalent.
\end{defi}

\begin{cor}
The representation $(\H,\pi)$ of a C$\*$-algebra $\Aa$ is faithful if and only if one of the following (equivalent) conditions is satisfied:

\begin{itemize}
\item[(i)] $\ker\pi=\{0\}$.
\item[(ii)] $\|\pi(A)\|=\|A\|\qquad\forall A\in\Aa$.
\item[(iii)] $\pi(A)>0\qquad\forall A\in\Aa_{+}$.
\end{itemize}

\end{cor}
Each representation $(\H,\pi)$ of a $C^*$-algebra $\Aa$ defines a faithful representation of the quotient algebra $\Aa_{\pi}:=\Aa/\ker\pi$. The representation of a simple algebra is always faithful.

\begin{defi}
Let $\Ma$ be a von Neumann algebra on a Hilbert space $\H$, then a subspace $\H'\subseteq\H$ is said to be separating for $\Ma$ if and only if $A\xi=0$ implies $A=0$ for all $A\in\Ma$ and $\xi\in\H'$.
\end{defi}

\begin{defi}
A cyclic  representation is a triple $(\H,\pi,\Omega)$, where $(\H,\pi)$ is a representation of $\Aa$ and $\Omega\in\H$ is a cyclic vector for $\pi$ in $\H$, i.e., $\{\pi(A)\Omega:\;A\in\Aa\}$ is dense in $\H$.
\end{defi}

\begin{cor}
For a von Neumann algebra $\Ma$ on a Hilbert space $\H$ and $\mathcal{K}\subseteq\H$ the following statements are equivalent:

\begin{itemize}
\item[(i)] $\mathcal{K}$ is cyclic for $\Ma$.
\item[(ii)] $\mathcal{K}$ is separating for $\Ma'$.
\end{itemize}
\end{cor}
If the vector $\Omega\in\H$ is cyclic and separating for the von Neumann algebra $\Ma$, then it has these properties also for its commutant $\Ma'$. Every nondegenerate representation of a $C^*$-algebra can be described as a direct sum of cyclic sub-representations. Therefore, the discussion of such representations can be restricted to the investigation of the cyclic ones.

\begin{defi}
Let $\Aa$ be a unital C$\*$-algebra and $\o:\Aa\rightarrow\C$ a linear functional on $\Aa$, then $\o$ is said to be

\begin{itemize}
\item[(i)] hermitian, if $\o(A^{*})=\overline{\o(A)}\quad\forall A\in\Aa$.
\item[(ii)] positive, if $\o(A)\geq 0\quad\forall A\in\Aa,\;A\geq 0$.
\item[(iii)] a weight, if $\o$ is positive.
\item[(iv)] a state, if $\o$ is positive and normalized, i.e., $\o(\1)=1$.
\item[(v)] a faithful state if $\o$ is a state and $\o(A)>0$ for all $A\in\Aa_+$, the set of positive elements of $\Aa$.
%\item[(vi)] a pure state, if $\o$ is not describable as a convex linear combination of other states.
\item[(vi)] a trace, if $\o(AB)=\o(BA)\quad\forall A,B\in\Aa$.
\item[(vii)] a vector state, if $\o(A)\equiv\o_{\Omega}(A):=\big(\Omega,\pi(A)\Omega\big)$ for a non-degenerate representation $(\H,\pi)$ of $\Aa$ and some vector $\Omega\in\H$ with $\|\Omega\|=1$.
\end{itemize} 
\end{defi}
In the case of an abelian $C^*$-algebra $\Aa$, $\o$ is a pure state if and only if $\o(AB)=\o(A)\o(B)$ holds for all $A,B\in\Aa$. If $\Aa$ does not possess a unit $\1$, then the norm property $(iv)$ is replaced by the condition $\|\o\|:=\sup\big\{|\o(A)|\,|\;A\in\Aa\text{ and }\|A\|=1\big\}=1$. 

\begin{defi}
If $\Ma$ is a von Neumann algebra, $\o$ a positive linear functional on $\Ma$ and $\o\big(\text{l.u.b.}_{\alpha}(A_{\alpha})\big)=\text{l.u.b.}_{\alpha}\big(\omega(A_{\alpha})\big)$ holds for all increasing bounded nets $(A_{\alpha})$ in $\Ma_{+}$, where `$l.u.b.$' denotes the least upper bound, then $\o$ is said to be normal.
\end{defi} 

\begin{theo}\label{normalstate}
Let the von Neumann algebra $\Ma$ operate on the Hilbert space $\H$ and let $\o$ be a state on $\Ma$, then the next statements are equivalent:

\begin{itemize}
\item[(i)] $\o$ is normal.
\item[(ii)] $\o$ is $\sigma$-weakly continuous.
\item[(iii)] There exists a density matrix $\rho$, i.e., a positive trace-class operator on $\H$ with $\Tr(\rho)=1$, satisfying $\omega(A)=\Tr(\rho A)$ for all $A\in\Aa$.
\end{itemize}
\end{theo}

\begin{defi}
A trace $\o$ on a von Neumann algebra $\Ma$ is said to be semifinite, if the set
$$
\Ma_+^\o:=\big\{A\in\Ma_+|\;\o(A)<\infty\big\}
$$
is dense in $\Ma$. A von Neumann algebra $\Ma$ is called semifinite, if there exists a faithful, normal and semifinite weight on $\Ma$.
 
\end{defi}

\begin{theo}
The commutant $\Ma'$ of a semifinite von Neumann algebra $\Ma$ on a separable Hilbert space is also semifinite. 
\end{theo}

\begin{defi}
Let us consider an involutive Banach algebra $\Aa$, then $P\in\Aa$ is called a projection if $P^{2}=P$ and $P^{*}=P$. Two projections $P,Q\in\Ma$, where $\Ma$ is a von Neumann algebra, are said to be equivalent, abbreviated by $P\sim Q$, if there exists a $W\in\Ma$, such that $W^{*}W=P$ and $WW^{*}=Q$. The projection $P$ is said to be finite if $Q\leq P$ and $Q\sim P$ imply $Q=P$, otherwise it is called infinite. If there is no nonzero finite projection $Q\leq P$, $Q\in\Ma$, then $P$ is called purely infinite. If $ZP\neq 0$ is infinite for every central projection $Z\in\Ma$, i.e., for every projection on $\Ma\cap\Ma'$, then $P$ is called properly infinite. $P$ is said to be $abelian$, if $P\Ma P$ is a commutative subalgebra of $\Ma$. Two projections $P_1$ and $P_2$ are said to be centrally orthogonal, if the smallest central projections $Z_{P_1}$ and $Z_{P_2}$ majorizing $P_1$ and $P_2$, respectively, are orthogonal. 

\end{defi}

Every projection $P\in\Ma$ can be uniquely described as the sum of two centrally orthogonal projections $P_1,P_2\in\Ma$, such that $P_1$ is finite and $P_2$ is properly infinite. Since the set spanned by the projections is dense in the von Neumann algebra $\Ma$, the property of the projections can be used to characterise their algebras.

\begin{defi}
A von Neumann algebra $\Ma$ is called finite, infinite, properly infinite or purely infinite if the identity projection $\1$ possesses these properties. The von Neumann algebra is said to be of\\
\\
\begin{tabular}{l l}

Type $I$,  & if every nonzero central projection in $\Ma$ majorises a nonzero\\
         & abelian projection in $\Ma$.\\
Type $II$, & if $\Ma$ has no nonzero abelian projection and every nonzero central\\
         & projection in $\Ma$ majorises a nonzero finite projection in $\Ma$.\\
Type $II_{1}$, & if $\Ma$ is of type II and finite.\\
Type $II_{\infty}$, & if $\Ma$ is of type II and has no nonzero central finite projection.\\
Type $III$, & if $\Ma$ is purely infinite.\\
L
\end{tabular}
\\
\end{defi}
If $\Ma$ is of type $I,II$ or $III$, then so is its commutant $\Ma'$ and contrariwise. For von Neumann algebras of type $II_{1}$ and $II_{\infty}$ this statement is in general not valid. The aforementioned characterisation allows one to decompose all von Neumann algebras completely in terms of these different types.

\begin{theo}
Every von Neumann algebra $\Ma$ is uniquely decomposable into the direct sum
$$
\Ma=\Ma_I\oplus\Ma_{II_1}\oplus\Ma_{II_\infty}\oplus\Ma_{III}.
$$
\end{theo}
A von Neumann algebra is semifinite if and only if the type $III$ part in the former equation is vanishing.

\begin{defi}
A von Neumann algebra $\Ma$ is called a factor, if it possesses a trivial center, i.e., $\Za(\Ma)=\Ma\cap\Ma'=\C\1$.
\end{defi}
\begin{defi}
 A state $\o$ on a C$\*$-algebra $\Aa$ is said to be a factor state or primary state, if $\pi_\o(\Aa)''$ is a factor, where $\pi_\o$ is the corresponding cyclic representation.
\end{defi}
A factor is either of type $I,II_{1},II_{\infty}$ or $III$. The table below illustrates the resulting type of the tensor product $\Ma\overline{\otimes}\Na$ of two von Neumann algebras $\Ma$ and $\Na$. 
\vspace{1cm}

\begin{center}
\begin{tabular}{c|ccccc}
$\Ma\diagdown\Na$ & $I_n$ & $I_\infty$ & $II_1$ & $II_\infty$ & $III$\\
\hline
$I_m$ & $I_{mn}$ & $I_\infty$ & $II_1$ & $II_\infty$ & $III$\\
$I_\infty$ & $I_\infty$ & $I_\infty$ & $II_\infty$ & $II_\infty$ & $III$\\
$II_1$ & $II_1$ & $II_\infty$ & $II_1$ & $II_\infty$ & $III$\\
$II_\infty$ & $II_\infty$ & $II_\infty$ & $II_\infty$ & $II_\infty$ & $III$\\
$III$ & $III$ & $II$ & $III$ & $III$ & $III$
\end{tabular}
\end{center}

\vspace{1cm}

\begin{defi}
A state $\o$ is said to be pure, if the only positive linear functionals majorised by $\o$ are of the form $\lambda\o$ with $0\leq\lambda\leq 1$.
\end{defi}
Pure states are extremal points of the set of states $\EA$ on $\Aa$, which means that a pure state $\o$ is not describable as a convex linear combination 
$$
\o=\lambda\o_1+(1-\lambda)\o_2,\quad 0<\lambda<1,
$$
of different states $\o_1$ and $\o_2$.\\
To each arbitrary nondegenerate representation of a $C^*$-algebra and a vector $\Omega\in\H$ with $\|\Omega\|=1$ one can assign a state, the so-called vector state. The construction in the opposite direction is ensured by the next

\begin{theo}
For an arbitrary state $\o$ on a C$\*$-algebra $\Aa$ there exists a (up to unitary equivalence) unique cyclic representation $(\H_\o,\pi_\o,\Omega_\o)$ of $\Aa$, the so-called canonical cyclic representation of $\Aa$ with respect to $\o$, such that 
\begin{equation*}
\o(A)=\big(\Omega_{\o},\pi_{\o}(A)\Omega_{\o}\big)
\end{equation*}
holds for all $A\in\Aa$. 
\end{theo}
The next theorem ensures that every $C^*$-algebra can be represented by a $C^*$-subalgebra of $\LH$ leading to the so-called GNS construction.

\begin{theo}[GNS-; Gelfand, Naimark, Segal]
For every C$\*$-algebra $\Aa$ there exists a Hilbert space $\H$, such that $\Aa$ is $\*$-isomorphic to a C$\*$-subalgebra of $\LH$.
\end{theo}

\begin{cor}\label{pure1}
Let $\o$ be a state on the unital C$\*$-algebra, then the following two statements are equivalent:
\begin{itemize}
\item[(i)] The state $\o$ is pure. 
\item[(ii)] The cyclic representation $(\H_\o,\pi_\o,\Omega_\o)$ with respect to $\o$ is irreducible.
\end{itemize} 
\end{cor}

\begin{cor}\label{pure2}
For a representation $(\H,\pi,\Omega)$ of a C$\*$-algebra $\Aa$ the next two conditions are equivalent:
\begin{itemize}
\item[(i)] $(\H,\pi,\Omega)$ is irreducible. 
\item[(ii)] $\pi(\Aa)$ is a factor.
\end{itemize} 
\end{cor}
Thus, each pure state is factorial, and, contrariwise, a factorial state is pure if it is normal, i.e., if it admits, due to Theorem \ref{normalstate}, a cyclic vector in the representation space.

The following theorem from the theory of $n$-dimensional complex-valued functions is used in the proof of the Reeh-Schlieder Theorem \ref{Reeh-Schlieder}.

\begin{theo}[Edge of the wedge]\label{Edge of the wedge}
Let $\mathbf{K}:=\big\{z\in\C|\;|z|<1\big\}$ be the unit ball in $\C$, $\mathbf{C}\subset\R^n$ an open convex cone with $\mathbf{C}\cap(-\mathbf{C})\neq\emptyset$, and $\mathbf{G}:=\big\{z=a+ib\in\mathbf{K}|\;b\in\mathbf{C},a\in\R^n\big\}\subset\C^n$. If the in $\mathbf{G}$ complex-valued, holomorphic function $f$ with $\lim_{\mathbf{C}\ni b\rightarrow 0}f(a+ib)$ existing for all open subsets $\mathbf{U}:=\big\{x\in\R^n|\;|x|<r\big\}\subset\R^n$, where the limit is independent of the chosen sequence, then $f$ is analytically continuable into an open region $\mathbf{G}\cup\mathbf{G}_0$, where $\mathbf{G}_0:=\bigcup_{x\in \mathbf{U}}\big\{z\in\C|\;|z-x|<\theta\cdot\text{dist}(x,\partial \mathbf{U}),\;0<\theta<1\big\}$ is a complex neighbourhood of $\mathbf{U}$ and $\theta$ is independent of $x,r,\mathbf{U}$ and $f$.
\end{theo}

\chapter{Free Quantum Fields}

Let $H_m$ be the mass hyperboloid in $\R^4$ consisting of those $p\in\R^4$ which satisfy the equation $p\cdot p-m^2=0$, $m\geq 0$, where $\H:=\L^2(H_m,d\Omega_m)$ is the Hilbert space of square integrable functions on $\H_m$ with respect to the invariant Lebesgue measure $d\Omega_m$, and define the symmetric Fock space $\F(\H)$ over $\H$ as
$$
\F(\H):=\bigoplus_{n=0}^\infty\H^{\bigotimes_s n}\quad\text{with}\quad\H^{\bigotimes_s n}:=\bigotimes_{k=1}^n\H,\quad\H^{\bigotimes_k 0}:=\C,
$$
where $\H^{\bigotimes_s n}$ is the so-called $n$-particle subspace of $\F(\H)$ and $\bigotimes_s$ denotes the symmetric tensor product. With the help of the second quantisation defined by
$$
d\Gamma(A):=A\otimes\1\otimes\cdots\otimes\1+\1\otimes A\otimes\1\otimes\cdots\otimes\1+\cdots+\1\otimes\cdots\otimes\1\otimes A
$$
on $\mathbf{D}_A\cap\H^{\bigotimes_s n}$, where $\mathbf{D}$ is the domain of $A$ and
\begin{gather*}
\mathbf{D}_A:=\Big\{(\psi_n)_{n\in\N}\in\F_0(\H)|\;\psi_n\in\bigotimes_{k=1}^n\mathbf{D}\;\text{ for each } n\Big\},\\
\F_0(\H):=\big\{(\psi_n)_{n\in\N}\in\F(\H)|\;\psi_n=0\;\text{ for all but finitely many } n\big\},
\end{gather*}
one introduces the Segal operator
$$
\varphi_S[f]:=\frac{1}{\sqrt{2}}\big(A[f]+A^*[f]\big),
$$
where $A^*[f]$ is the creation operator and
\begin{gather*}
A[f](\psi_1\otimes\cdots\otimes\psi_n):=\sqrt{N+1}(f,\psi_1)(\psi_2\otimes\cdots\otimes\psi_n),\\
N\psi:=n\psi\quad\forall\psi\in\H^{\bigotimes_s n},
\end{gather*}
are the so-called annihilation operator and the number operator, respectively.
Let $\Gamma(U)$, where $U$ is a unitary operator on the Hilbert space, be the unitary operator on the Fock space, defined as 
\begin{gather*}
\Gamma(U):=\begin{cases}\bigotimes_{k=1}^nU &\text{ on  }\;\H^{\otimes_sn}\text{ for  }n>0,\\
\1&\text{ on }\;\H^{\otimes_s0}.\end{cases} 
\end{gather*}
A continuous unitary groups $e^{itA}$ on $\H$ is then generated by the second quantisation $d\Gamma(A)$ and one has the identity:
\begin{gather*}
\Gamma\big(e^{itA}\big)=e^{itd\Gamma(A)}.
\end{gather*}
Furthermore, the following statements hold:

\begin{itemize}
\item[(i)] Self-adjointness: $\varphi_S[f]$ is essentially self-adjoint on the set of finite particle vectors $\F_0:=\big\{\psi\in\F(\H)|\;\psi^{\bigotimes_s n}=0\; \text{for all but finitely many $n$}\big\}$ for each $f\in\H$. 
\item[(ii)] Cyclicity of the vacuum: $\Omega_0:=(1,0,0,\cdots)$ is in the domain of all finite products $\varphi_S[f_1]\cdots\varphi_S[f_n]$ and the set $\big\{\varphi_S[f_1]\cdots\varphi_S[f_n]\Omega_0|\;\forall f\in\H\,\forall n\in\N\big\}$ is dense in $\F(\H)$.
\item[(iii)] Commutation relations: For every $\psi\in\F_0$ and $f,g\in\H$ one has 
\begin{gather*}
\big(\varphi_S[f]\varphi_S[g]-\varphi_S[g]\varphi_S[f]\big)\psi=i\Im(f,g)_\H\psi, \quad\text{and}\\
W(f+g):=e^{i\varphi_S[f+g]}=e^{-\frac{i}{2}\Im(f,g)_\H}W(f)W(g).
\end{gather*}
\item[(iv)] Continuity: $f_n\longrightarrow f$, $f,f_n\in\H$, implies
$$
W(f_n)\psi\longrightarrow W(f)\psi\quad\text{and}\quad\varphi_S[f_n]\tilde{\psi}\longrightarrow\varphi_S[f]\tilde{\psi}
$$
for all $\psi\in\F_s(\H)$, $\tilde{\psi}\in\F_0$ and $n\in\N$.
\item[(v)] For each unitary operator $U$ on $\H$, $\Gamma(U):\D(\overline{\varphi_S[f]})\longrightarrow\D(\overline{\varphi_S[Uf]})$, and every $\psi\in\D(\overline{\varphi_S[Uf]})$ the relation
$$
\Gamma(U)\overline{\varphi_S[f]}\Gamma(U)^{-1}\psi=\overline{\varphi_S[Uf]}\psi
$$
holds.
\end{itemize}
With the aid of the Segal operator we can now introduce the canonical free scalar field and the canonical conjugate momentum of mass $m\geq 0$,
\begin{gather*}
\varphi_m[f]:=\varphi[Ef]:=\varphi_S[Ef],\\
\pi_m[f]:=\pi[\mu Ef]:=\varphi_S[i\mu Ef],\\
\text{with}\quad Ef(x):=\sqrt{2\pi}\widehat{f(p)},\quad\mu:=(\p^2+m^2)^{1/2},
\end{gather*}
and in terms of the creation and annihilation operators,
\begin{gather*}
\varphi_m[f]=\frac{1}{\sqrt{2}}\big(A^*[Ef]+A[Ef]\big),\\
\pi_m[f]=\frac{i}{\sqrt{2}}\big(A^*[\mu Ef]+A[\mu Ef]\big).
\end{gather*}
The fundamental characteristics of the free scalar field and of the conjugate momentum are inherited from the Segal operator:

\begin{itemize}
\item[(a)]
\begin{itemize}
\item[(i)] The field $\varphi_m[f]$ is essentially self-adjoint on $\F_0$ for all $f\in\H$.
\item[(ii)] $\big\{\varphi_m[f]|\;f\in\H\big\}$ is a commuting family of self-adjoint operators.
\item[(iii)] $\Omega_0\in\H$ is cyclic for $\big\{\varphi_m[f]|\;f\in\H\big\}$.
\item[(iv)] $f_n\longrightarrow f$, $f,f_n\in\H$, implies
$$
e^{i\varphi_m[f_n]}\psi\longrightarrow e^{i\varphi_m[f]}\psi\quad\text{and}\quad\varphi_m[f_n]\tilde{\psi}\longrightarrow\varphi_m[f]\tilde{\psi}
$$
for all $\psi\in\F_s(\H)$, $\tilde{\psi}\in\F_0$ and $n\in\N$.
\end{itemize}
\item[(b)] The properties given in (a) hold also for the conjugate momentum $\pi_m[f]$.
\item[(c)] For all $f,g\in\H$ one has 
\begin{gather*}
\big[\varphi_m[f],\pi_m[g]\big]\psi=i(f,g)_\H\psi\quad\forall\psi\in\F_0\quad\text{and}\\
e^{i\varphi_m[f]}e^{i\pi_m[g]}=e^{i(f,g)_\H}e^{i\pi_m[g]}e^{i\varphi_m[f]}.
\end{gather*}
\end{itemize}
Although not relativistically invariant, the framework of free fields and their conjugate momentum at particular fixed times has turned out to be fruitful. One deals here with the mappings
$$
\S(\R^4)\ni f\mapsto \delta(t-t_0)f(\x)\in\S(\R^3).
$$
In this thesis we will be confronted several times with the so-called time-zero fields, i.e., maps defined by
$$
\S(\R^3)\ni f\mapsto\varphi\big[\delta(t)f(\x)\big]\quad\text{and}\quad S(\R^3)\ni f\mapsto\pi\big[\delta(t)f(\x)\big].
$$
Transferring the scalar field and its conjugate momentum from the Fock space $\F\big(\L^2(H_m,d\Omega_m)\big)$ to that based on the configuration space $\F\big(\L^2(\R^3,dk)\big)$, one gets the following form:
\begin{gather*}
\tilde{\varphi}_m[f]=\frac{1}{\sqrt{2}}\big(A^*[\mu^{-1/2}Ef]+A[\mu^{-1/2}Ef]\big),\\
\tilde{\pi}_m[f]=\frac{i}{\sqrt{2}}\big(A^*[\mu^{1/2}Ef]+A[\mu^{1/2}Ef]\big).
\end{gather*}
%In the following we will omit the sign '$\;\tilde{}\;$' and denote the time-zero field and its conjugate momentum by $\varphi_m[f]$ and $\pi_m[f]$, respectively.

\end{appendix}

%\begin{appendices}
%\input{appendix}
%\end{appendices}

%%%%%%%%%%%%%%%%%%%%%%%%%%%%%%%%%%%%%%%%%%%%%%%%%%%%%%%%%%%%%%%%%%%%%%%%%%%%

\cleardoublepage
\addcontentsline{toc}{chapter}{\protect\numberline{Notation}}
\chapter*{Notation}

\begin{tabular}{ll}
$x$ & $=(x^{1},\cdots,x^{n})$\\
$\xi$ & $=(\xi^{1},\cdots,\xi^{n})$\\
$\alpha$ & $=(\alpha_{1},\cdots,\alpha_{n})$\\
$|\alpha|$ & $=\Sigma_{i=1}^{n}\alpha_{i}$\\
$\alpha!$ & $=\Pi_{i=1}^{n}\alpha_{i}$\\
$x^{\alpha}$ & $=\Pi_{i=1}^{n}x_{i}^{\alpha_{i}}$\\
$\partial_{x}^{\alpha}$ & $=(\partial/\partial x^{1})^{\alpha_{1}}\cdots(\partial/\partial x^{n})^{\alpha_{n}}$\\
$D_{x}^{\alpha}$ & $=(-1)^{|\alpha|}\partial_{x}^{\alpha}$\\
\\
$\mathcal{C}^{l}(\Omega)$ & vector space of $l$-times continuously differentiable \\
& functions on $\Omega$\\
$\mathcal{C}(\Omega)$ & $=\mathcal{C}^{0}(\Omega)$ continuous functions on $\Omega$\\
$\mathcal{C}^{\infty}(\Omega)$ & $=\cap\{\mathcal{C}^{l}(\Omega)|\;l\in\N_{0}\}$\\
$\mathcal{B}(\R^n)$ & $=\big\{f\in\Cg(\R^n)|\;\forall\alpha\in\N_0^n:\;\sup\{|D^\alpha f(x)|\,|x\in\R^n\}<\infty\big\}$\\
$\mathcal{E}(\Omega)$ & $=\mathcal{C}^{\infty}(\Omega)$\\
$\mathcal{D}(\Omega)$ & $=\mathcal{C}_{0}^{\infty}(\Omega)$\\
$\mathcal{E}'(\Omega)$ & dual space of $\mathcal{E}(\Omega)$\\
$\mathcal{D}'(\Omega)$ & dual space of $\mathcal{D}(\Omega)$\\
$\mathcal{S}'(\R^{n})$ & dual space of $\mathcal{D}(\R^{n})$\\
\\
$\langle u,\phi\rangle$ & Application of the distribution $u$ on $\phi\in\mathcal{X}\quad\big(\mathcal{X}\in\{\mathcal{E}(\Omega),\mathcal{D}(\Omega),\mathcal{S}(\R^{n})\}\big)$\\
$(u,\phi)$ & $=\langle u,\overline{\phi}\rangle$\\
singsupp($u$) & singular support of $u$\\
$\hat{u}$ & Fourier transform of $u$\\
\\
\\
$S_{\rho,\delta}^{m}(\R^{n})=S_{\rho,\delta}^{m}$ & vector space of (Kumano-go's) symbols of order $m\in\R$\\
& and of type $(\rho,\delta)$\\
$S^{m}$ & $=S_{0,1}^{m}$\\
$S^{\infty}$ & $=\cup\{S^{m}|\quad m\in\R\}$\\
$S^{-\infty}$ & $=\cap\{S^{m}|\quad m\in\R\}$\\

\end{tabular}

\begin{tabular}{ll}

$\EuScript{S}_{\rho,\delta}^{m}(\R^{n})=\EuScript{S}_{\rho,\delta}^{m}$ & vector space of (Kumano-go's) pseudo-differential operators\\ 
& (PsDO) of order $m\in\R$ and of type $(\rho,\delta)$\\
$\EuScript{S}^{m}$ & $=\EuScript{S}_{0,1}^{m}$\\
$\EuScript{S}^{\infty}$ & $=\cup\{\EuScript{S}^{m}|\quad m\in\R\}$\\
$\EuScript{S}^{-\infty}$ & $=\cap\{\EuScript{S}^{m}|\quad m\in\R\}$\\
$P\equiv p(X,D_{x})$ & PsDO generated by the symbol $p\in S_{\rho,\delta}^{m}$\\
$A\equiv a(X,D_{x})$ & FIO generated by the symbol $a\in S_{\rho,\delta}^{m}$\\
$S_{\rho,\delta}^{m,m'}(\R^{n})=S_{\rho,\delta}^{m,m'}$ & vector space of (Kumano-go's) double symbols \\
&of order $(m,m')$ and of type $(\rho,\delta)$\\ 
$S^{m,m'}$ & $S_{0,1}^{m,m'}$\\
$p(X,X',D_{x},D_{x'})$ & PsDO generated by the double symbol $p\in S_{\rho,\delta}^{m,m'}$\\
$H^{s}(\R^{n})=H^{s}$ & Sobolev space\\
$H^{-\infty}$ & $\bigcup\{H^{s}|\quad s\in\R\}$\\
$H^{\infty}$ & $\bigcap\{H^{s}|\quad s\in\R\}$\\
%$\EuFrak{K}(\xi_{0})$ & Set of conical neighbourhood of $\xi_{0}\in\R_{*}^{n}$\\
%$\EuFrak{K}(x_{0},\xi_{0})$ & set of the conical neighbourhood (with respect to $\xi$)\\
& of $(x_{0},\xi_{0})\in\R^{n_{1}}\times\R_{*}^{n_{2}}$\\
$\Sigma(u)$ & set of the `direction with singular property' of $u\in\mathcal{E}'$\\
$WF(u)$ & wave front set  of $u\in\mathcal{D}'$\\

$\mu Tr(p)=\mu Tr(P)$ & micro-support of the symbol $p$ or the PsDO $P$, respectively\\
\\
\\
\\

$x$ & $(x^0,x^1,x^2,x^3)$\\
$\x$ & $(x^1,x^2,x^3)$\\
$x_\pm$ & $:=x^0\pm|\x|$\\
$\xi^\alpha$ & pseudo-orthogonal coordinates $\alpha=0,1,2,\cdots,d+2$\\
$\xi_\pm$ & $:=\xi_{d+2}\pm\xi_{d+1}$\\
$T_{\alpha\beta}$ & conformal transformation\\

$\mathcal{M}$ & differentiable manifold\\
$\mathbb{M}$ & four-dimensional Minkowski space\\
$\Sigma$ & Cauchy surface\\
$\O^c$ & $=\big\{x\in\M|\;\text{ the vector }x-y\text{ is spacelike for all }y\in\O\big\}$\\
&  causal complement of $\O\subset\M$\\
$\W_R$ & $:=\big\{x\in\mathbb{M}|\;|x^0|<x^1\big\}$ right Rindler wedge\\
$\W_L$ & $:=\big\{x\in\mathbb{M}|\;|x^0|<-x^1\big\}$ left Rindler wedge\\
$\V_+$ & $:=\big\{x\in\mathbb{M}|\;x\cdot x>0\text{ and } x^0>0\big\}$ forward light cone\\
$\V_-$ & $:=\big\{x\in\mathbb{M}|\;x\cdot x>0\text{ and } x^0<0\big\}$ backward light cone\\
$\D$ & $:=\V_+\cap\V_-$ double cone\\
$\D_1$ & $:=(\V_+-e_0)\cap(\V_-+e_0)$ double cone with radius one\\
$\Wa$, $\Va$, $\Da$ & set of wedges, light cones, double cones\\

\end{tabular}

\begin{tabular}{ll}
$\H$ & Hilbert sapce\\
$\LH$ & set of bounded operators on $\H$\\
$\Aa,\Ba$ & $C^{*}$-algebras\\
$\Ma,\Na$ & von Neumann algebras\\
$\Aa(\O)$ & $C^{*}$-algebra of local observables localised in $\mathcal{O}\in\mathcal{M}$\\
$\Ma(\O)$ & von Neumann algebra of local observables localised in $\mathcal{O}\in\mathcal{M}$\\
$\Ma_m(\O)$ & local algebra of fields with mass $m$\\
$\Fa(\O)$ & field algebra\\
$\triangle(x_{1},x_{2})$ & van Vleck-Morette determinant\\
$\sigma(x,y)$ & geodesic distance between $x$ and $y$ in $\mathcal{M}$\\
$\omega_{n}(f_{1},\cdots,f_{n})$ & $n$-point-function, $f_{i}\in\mathcal{D}(\mathcal{M})$\\
$\Lambda(f,g)$ & $=\omega_{2}(f,g)$\\
$\mathcal{H}$ & Hilbert space\\
$\mathcal{L}(\mathcal{H})$ & set of bounded linear operators on $\mathcal{H}$\\
$\mathcal{E}_\Aa$ & set of states on $\Aa$\\
$\mathcal{K}_{\beta}$ & set of KMS-states on $\Aa$\\
$\alpha^t$, $\sigma^t$, $\tau^t$ & one-parameter groups of automorphisms\\
\\
$\Omega\in\H$ & vacuum vector\\
$(\H_\o,\pi_\o,\Omega_\o)$ & cyclic representation with respect to $\o$\\
$\o_m$ & vacuum state with respect to $\Ma_m$, $m\geq 0$\\
$S$, $J$, $\Delta$ & Tomita oerator, modular conjugation, modular operator\\
$\sigma_m^t(A)$ & $:=\Delta_m^{it}A\Delta_m^{-it}$ modular group of automorphisms on $\Ma$\\
$\sigma_{m,0}^t$ & modular group on $\Ma_m(\O)$ with respect to $\o_0$\\
 & with respect to the mass $m\geq 0$\\
$\sigma_{\o_m}$ & modular group of automorphisms with respect to\\
 & the faithful, normal and semifinite (vacuum) state $\o_m$\\
$\Gamma_t$ & Connes' cocycle\\
\\
$(\Aa,\mathbf{G},\alpha)$ & $C^{*}$-dynamical system\\
$\delta$ & derivation\\
$\delta(A)$ & $=\lim_{t\rightarrow 0}\big(\tau^{t}(A)-A\big)$ infinitesimal generator of $\tau$\\
$\delta_m$ & infinitesimal generator of  $\sigma_m^t$\\
$\delta_{m,0}$ & infinitesimal generator of  $\sigma_{m,0}^t$\\
$P_\mu$, $M_{\mu\nu}$, $D$, $K_\mu$ & infinitesimal generators of translations, Lorentz transformations,\\
 & dilations, (special) conformal transformations\\
%$\delta_m$ & infinitesimal generator of $\sigma_m^\tau$ with respect to the mass $m\geq 0$\\
$P$ & perturbation, $P=P^{*}\in\Aa$\\
$\Theta$ & PCT operator\\

\end{tabular}

%%%%%%%%%%%%%%%%%%%%%%%%%%%%%%%%%%%%%%%%%%%%%%%%%%%%%%%%%%%%%%%%%%%%%%%%%%%%

\cleardoublepage

\addcontentsline{toc}{chapter}{\protect\numberline{Bibliography}}
%\addcontentsline{toc}{chapter}{Bibliography}
%\bibliographystyle{cmp}
%\bibliography{dissp}

\nocite{*}

%%%%%%%%%%%%%%%%%%%%%%%%%%%%%%%%%%%%%%%%%%%%%%%%%%%%%%%%%%%%%%%%%%%%%%%%%%%%

\clearpage

\addcontentsline{toc}{chapter}{\protect\numberline{Acknowledgements}}
\chapter*{Acknowledgements}

\thispagestyle{empty}

First and foremost, I want to express my sincere gratitude to my supervisor Prof Dr K. Fredenhagen for encouraging me to investigate the interplay of modular theory and local quantum physics, his guidance through the years of my studies and his patience during our discussions.\\
I am also deeply grateful to Prof Dr G. Mack for composing the co-report.\\
This thesis, which has been written mainly at the Helmut-Schmidt-Universit\"at, Universit\"at der Bundeswehr Hamburg, would not have been possible without the support of Prof Dr W. Krumbholz and Prof Dr H. Hebbel, whose confidence I have enjoyed and who have given me financial security as well as a maximum of freedom in my scientific activities. I do not take this for granted.  \\
In times of need, lucky are who can rely on someone's constant presence and skilled help. Thank you very much, Dr M. Porrmann. \\
Furthermore, I am indebted to all friends and colleagues from the F\"achergruppe f\"ur Statistik, Helmut-Schmidt-Universit\"at, and from the AQFT group of the II. Institut f\"ur Theoretische Physik, Universit\"at Hamburg. \\ \\
Last but not least, to my family, scattered all over the world, but still an entity and always present.

\end{document}